\documentclass[twocolumn]{aastex62}

\newcommand{\sna}{SN~1987A}
\newcommand{\sleak}{Sanduleak~-69$^\circ$~202}

\newcommand{\ks}{\textit{K}$_\mathrm{s}$}
\newcommand{\kmps}{~km~s$^{-1}$}
\newcommand{\ergps}{~erg~s$^{-1}$}
\newcommand{\ergscmhz}{~erg~s$^{-1}$~cm$^{-2}$~Hz$^{-1}$}
\newcommand{\Msun}{~M$_{\sun}$}
\newcommand{\Lsun}{~L$_{\sun}$}

\usepackage{amsmath}

\received{May 11, 2018}
\revised{July 27, 2018}
\accepted{July 28, 2018}
\submitjournal{ApJ}
\shorttitle{The Compact Object in SN~1987A}
\shortauthors{Alp et al.}

\begin{document}
\title{The 30-Year Search for the Compact Object in SN~1987A}

\correspondingauthor{Dennis Alp}
\email{dalp@kth.se}

\author[0000-0002-0427-5592]{Dennis Alp}
\affil{Department of Physics, KTH Royal Institute of Technology, 
  The Oskar Klein Centre, AlbaNova, SE\nobreakdash{-}106\nobreakspace{ }91 Stockholm, Sweden}

\author[0000-0003-0065-2933]{Josefin Larsson}
\affil{Department of Physics, KTH Royal Institute of Technology, 
  The Oskar Klein Centre, AlbaNova, SE\nobreakdash{-}106\nobreakspace{ }91 Stockholm, Sweden}

\author[0000-0001-8532-3594]{Claes Fransson}
\affil{Department of Astronomy, Stockholm University, 
  The Oskar Klein Centre, AlbaNova, SE\nobreakdash{-}106\nobreakspace{ }91 Stockholm, Sweden}

\author{Remy Indebetouw}
\affil{National Radio Astronomy Observatory and University of Virginia, 520 Edgemont Rd, Charlottesville, VA 22903, USA}

\author[0000-0001-8005-4030]{Anders Jerkstrand}
\affil{Max Planck Institute for Astrophysics, 
       Karl-Schwarzschild-Str.~1, D\nobreakdash{-}85748 Garching,
       Germany}

\author[0000-0002-9722-9009]{Antero Ahola}
\affil{Tuorla observatory, Department of Physics and Astronomy, 
University of Turku, V\"{a}is\"{a}l\"{a}ntie 20, FI\nobreakdash{-}21500 Piikki\"{o}, Finland}

\author{David Burrows}
\affil{Department of Astronomy \& Astrophysics, The Pennsylvania State University, University Park, PA 16802, USA}

\author{Peter Challis}
\affil{Harvard-Smithsonian Center for Astrophysics, 60 Garden Street, Cambridge, MA 02138, USA}

\author[0000-0002-8736-2463]{Phil Cigan}
\affil{School of Physics and Astrophysics, Cardiff University, Queens buildings, The Parade, Cardiff CF24 3AA, UK}

\author[0000-0001-7101-9831]{Aleksandar Cikota}
\affil{European Southern Observatory, Karl-Schwarzschild-Strasse 2,
  D\nobreakdash{-}85748 Garching, Germany}

\author[0000-0002-1966-3942]{Robert P. Kirshner}
\affil{Harvard-Smithsonian Center for Astrophysics, 60 Garden Street, Cambridge, MA 02138, USA}
\affil{Gordon and Betty Moore Foundation, 1661 Page Mill Road, Palo Alto, CA 94304  USA }

\author[0000-0002-1272-3017]{Jacco Th. van Loon}
\affil{Lennard-Jones Laboratories, Keele University, ST5 5BG, UK}

\author[0000-0001-7497-2994]{Seppo Mattila}
\affil{Tuorla observatory, Department of Physics and Astronomy, 
University of Turku, V\"{a}is\"{a}l\"{a}ntie 20,
FI\nobreakdash{-}21500 Piikki\"{o}, Finland}

\author[0000-0002-5847-2612]{C.-Y.~Ng}
\affiliation{Department of Physics, The University of Hong Kong, Pokfulam
Road, Hong Kong}

\author[0000-0003-3900-7739]{Sangwook Park}
\affil{Department of Physics, Box 19059, 108 Science Hall, University
  of Texas at Arlington, Arlington, TX 76019}

\author[0000-0001-6815-4055]{Jason Spyromilio}
\affil{European Southern Observatory, Karl-Schwarzschild-Strasse 2,
  D\nobreakdash{-}85748 Garching, Germany}

\author[0000-0002-3352-7437]{Stan Woosley}
\affil{Department of Astronomy and Astrophysics, University of California, Santa Cruz, California, 95064}

\author[0000-0002-3930-2757]{Maarten Baes}
\affil{Sterrenkundig Observatorium, Universiteit Gent,
  Krijgslaan 281 S9, B\nobreakdash{-}9000 Gent, Belgium}

\author{Patrice Bouchet}
\affil{IRFU, CEA, Universit\'e Paris-Saclay, F\nobreakdash{-}91191
  Gif-sur-Yvette, France}
\affil{CNRS/AIM, Universit\'e Paris Diderot, F\nobreakdash{-}91191
  Gif-sur-Yvette, France}

\author[0000-0002-9117-7244]{Roger Chevalier}
\affil{Department of Astronomy, University of Virginia, P.O. Box
  400325, Charlottesville, VA 22904-4325, USA}

\author[0000-0003-0570-9951]{Kari A. Frank}
\affil{Center for Interdisciplinary Exploration and Research in Astrophysics (CIERA), Northwestern University, 2145 Sheridan Road, Evanston, IL 60208, USA}

\author[0000-0002-3382-9558]{B.~M. Gaensler}
\affil{Dunlap Institute for Astronomy and Astrophysics, 50 St. George
  Street, Toronto, ON M5S 3H4, Canada}

\author[0000-0003-3398-0052]{Haley Gomez}
\affil{School of Physics and Astrophysics, Cardiff University, Queens buildings, The Parade, Cardiff CF24 3AA, UK}

\author[0000-0002-0831-3330]{Hans-Thomas Janka}
\affil{Max Planck Institute for Astrophysics, 
       Karl-Schwarzschild-Str.~1, D\nobreakdash{-}85748 Garching, Germany}

\author[0000-0002-4413-7733]{Bruno Leibundgut}
\affil{European Southern Observatory, Karl-Schwarzschild-Strasse 2,
  D\nobreakdash{-}85748 Garching, Germany}

\author[0000-0002-3664-8082]{Peter Lundqvist}
\affil{Department of Astronomy, Stockholm University, 
  The Oskar Klein Centre, AlbaNova, SE\nobreakdash{-}106\nobreakspace{ }91 Stockholm, Sweden}

\author{Jon Marcaide}
\affil{Departmento de Astronomia y Astrofisica, Universidad de Valencia, 
  Dr. Moliner 50, Burjassot-Valencia, E\nobreakdash{-}46100, Spain}

\author[0000-0002-5529-5593]{Mikako Matsuura}
\affil{School of Physics and Astrophysics, Cardiff University, Queens
  buildings, The Parade, Cardiff CF24 3AA, UK}

\author[0000-0003-1546-6615]{Jesper Sollerman}
\affil{Department of Astronomy, Stockholm University, 
  The Oskar Klein Centre, AlbaNova, SE\nobreakdash{-}106\nobreakspace{ }91 Stockholm, Sweden}

\author[0000-0003-1440-9897]{George Sonneborn}
\affil{Laboratory for Observational Cosmology, Code 665, 
NASA Goddard Space Flight Center, Greenbelt, MD 20771, USA}

\author[0000-0002-8057-0294]{Lister Staveley-Smith}
\affil{International Centre for Radio Astronomy Research (ICRAR), University of Western Australia, 35 Stirling Hwy,
Crawley, WA 6009, Australia}

\author[0000-0003-2742-771X]{Giovanna Zanardo}
\affil{International Centre for Radio Astronomy Research (ICRAR), University of Western Australia, 35 Stirling Hwy,
Crawley, WA 6009, Australia}

\author[0000-0002-1663-4513]{Michael Gabler}
\affil{Max Planck Institute for Astrophysics, 
       Karl-Schwarzschild-Str.~1, D\nobreakdash{-}85748 Garching, Germany}

\author{Francesco Taddia}
\affil{Department of Astronomy, Stockholm University, 
  The Oskar Klein Centre, AlbaNova, SE\nobreakdash{-}106\nobreakspace{ }91 Stockholm, Sweden}

\author[0000-0003-1349-6538]{J. Craig Wheeler}
\affil{Department of Astronomy and McDonald Observatory, The University of Texas at Austin, Austin, TX 78712, USA}

\begin{abstract}
  Despite more than 30 years of searches, the compact object in
  Supernova (SN) 1987A has not yet been detected.  We present new
  limits on the compact object in \sna{} using millimeter,
  near-infrared, optical, ultraviolet, and X-ray observations from
  ALMA, VLT, \textit{HST}, and \textit{Chandra}.  The limits are
  approximately 0.1~mJy ($0.1\times 10^{-26}$\ergscmhz{}) at 213~GHz,
  1\Lsun{} ($6\times 10^{-29}$\ergscmhz{}) in optical if our
  line-of-sight is free of ejecta dust, and $10^{36}$\ergps{}
  ($2\times 10^{-30}$\ergscmhz{}) in 2--10~keV X-rays. Our X-ray
  limits are an order of magnitude less constraining than previous
  limits because we use a more realistic ejecta absorption model based
  on three-dimensional neutrino-driven SN explosion
  models~\citep{alp18b}. The allowed bolometric luminosity of the
  compact object is 22\Lsun{} if our line-of-sight is free of ejecta
  dust, or 138\Lsun{} if dust-obscured. Depending on assumptions,
  these values limit the effective temperature of a neutron star to
  $< 4$--8~MK and do not exclude models, which typically are in the
  range 3--4~MK. For the simplest accretion model, the accretion rate
  for an efficiency $\eta$ is limited to
  $< 10^{-11}\,\eta^{-1}$\Msun{}~yr$^{-1}$, which excludes most
  predictions. For pulsar activity modeled by a rotating magnetic
  dipole in vacuum, the limit on the magnetic field strength ($B$) for
  a given spin period ($P$) is $B \lesssim 10^{14}\,P^2$~G~s$^{-2}$,
  which firmly excludes pulsars comparable to the Crab. By combining
  information about radiation reprocessing and geometry, it is likely
  that the compact object is a dust-obscured thermally-emitting
  neutron star, which may appear as a region of higher-temperature
  ejecta dust emission.
\end{abstract}

\section{Introduction}\label{sec:intro}
Supernova (SN) 1987A provides a unique opportunity to observe the
development of a SN and subsequent early evolution of a very young SN
remnant~\citep[for reviews of \sna{}, see][]{arnett89, mccray93,
  mccray16}. \sna{} is expected to have created a compact object. The
existence of the compact object is supported by the detection of the
initial neutrino burst, which was observed by Kamiokande
II~\citep{hirata87, hirata88} and the Irvine-Michigan-Brookhaven
detector~\citep{bionta87, bratton88}, with a possible supporting
detection by the Baksan Neutrino Observatory~\citep{alekseev87,
  alexeyev88}. While the prompt neutrino emission is attributed to the
formation of a compact object, more than 30 years of diligent searches
across the electromagnetic spectrum have failed to observe it. Being
able to observe the compact object in \sna{} would provide valuable
insight into the explosion mechanisms of SNe, the connection between
SN progenitors and compact objects, and the early evolution of neutron
stars (NSs). This has implications for our description of fundamental
physics in the strong-gravity regime.

Previous studies have been able to indirectly infer some properties of
the compact object in \sna{}. The progenitor star,
\sleak{}~\citep{sanduleak70}, was identified as a B3 Ia blue
supergiant~\citep{west87, white87, kirshner87, walborn87}. The
zero-age main-sequence mass of the progenitor is estimated to be in
the range 16--22\Msun{}, and the progenitor mass 14\Msun{} at the time
of explosion~\citep{fransson02, smartt09b, utrobin15,
  sukhbold16}. Most studies predict that the collapse of a star like
\sleak{} would create a NS, which is supported by the prompt neutrino
burst~\citep{burrows88} and SN simulations~\citep{fryer99, perego15,
  ertl16, sukhbold16}. However, some authors advance the hypothesis
that a black hole (BH) was created in \sna{}~\citep{brown92, blum16}.
The mass estimates of the possible NS are only loosely
constraining. Early estimates based on the neutrino signal predicts a
baryonic NS mass in the range 1.2--1.7\Msun{}~\citep{burrows88},
explosion simulations calibrated to \sna{} estimates a baryonic mass
of 1.66\Msun{}~\citep{perego15}, and a lower limit on the baryonic
mass of 1.7\Msun{} has been placed through constraints on explosive
silicon burning by measuring $\mathrm{Ni}/\mathrm{Fe}$
ratios~\citep{jerkstrand15}.

The detection of the compact object is made difficult by the light
from the ejecta and surrounding circumstellar medium (CSM). The CSM is
in the form of a triple-ring structure, possibly created by a binary
merger 20\,000 years before explosion~\citep{blondin93, morris07,
  morris09} or a rapidly rotating progenitor~\citep{chita08}. The
brightest of the three rings is the inner equatorial ring (ER), which
is seen in Figure~\ref{fig:pos}. It appears elliptical because it is
inclined by ${\sim}43$\arcdeg{}~\citep{tziamtzis11}. The outer parts
of the SN ejecta reached the ER by 1995. The interactions gave rise to
the first hotspot~\citep{lawrence00} and the ER subsequently
brightened during several years across the entire electromagnetic
spectrum~\citep{ng13, fransson15, arendt16, frank16}. However, the
mid-infrared emission from the ER started decreasing in
2010~\citep[MIR,][]{arendt16}, the optical emission started decreasing
in 2009~\citep{fransson15}, and the soft X-ray luminosity flattened
around 2013~\citep{frank16}. The radiation from central ejecta is also
affected by the ER. \citet{larsson11} showed that the brightening of
the SN ejecta in optical most likely is explained by the increase of
X-ray emission from the ER. However, the decay of $^{44}$Ti is still
expected to be the dominant energy source in the innermost parts of
the ejecta, where the compact object is expected to
reside~\citep{fransson13, larsson13}.

In this paper, we place limits on the compact object in \sna{} using
observations from the Atacama Large Millimeter/submillimeter Array
(ALMA) at millimeter wavelengths, the Very Large Telescope (VLT) in
near-infrared (NIR), \textit{Hubble Space Telescope} (\textit{HST}) in
optical and ultraviolet (UV), and \textit{Chandra} in X-rays. We then
discuss the implications of the limits on physical properties of the
compact object and prospects for future observations.

This paper is organized as follows. The observations and data
reduction are presented in Section~\ref{sec:obs} and the analysis
methods in Section~\ref{sec:methods}. We present our compact object
limits in Section~\ref{sec:results} and discuss the implications of
our results in Section~\ref{sec:discussion}. Finally, we summarize and
list the main conclusions in Section~\ref{sec:conclusions}. In an
accompanying paper~\citep{alp18b}, we estimate the X-ray absorption in
SN ejecta using three-dimensional (3D) neutrino-driven explosion
models.

\section{Observations and Data Reduction}\label{sec:obs}
\begin{deluxetable*}{lccccccccc}
  \tablecaption{Observations of \sna{}\label{tab:obs_all}}
  \tablewidth{0pt}
  \tablehead{\colhead{Instrument} & \colhead{Epoch} &
    \colhead{Exposure} & \colhead{Band/Filter/Grating} & \colhead{Frequency/Wavelength/Energy} \\
    \colhead{} & \colhead{(YYYY-mm-dd)} & \colhead{(s)} & \colhead{} } \startdata
  ALMA                  & 2014-09-02 &  1800--2040\tablenotemark{a} &                     6 & 230~GHz           \\
  VLT/NACO              & 2010-10-26 &  2160\tablenotemark{b}       &            \textit{H} & 1.7~\micron{}     \\
  VLT/NACO              & 2012-12-14 &  2070\tablenotemark{c}       &                 \ks{} & 2.2~\micron{}     \\
  VLT/SINFONI           & 2014-10-10 &  2400\tablenotemark{d}       &            \textit{H} & 1.7~\micron{}     \\
  VLT/SINFONI           & 2014-10-12 &  2400\tablenotemark{d}       &            \textit{K} & 2.2~\micron{}     \\
  \textit{HST}/WFC3     & 2009-12-13 &   800                        &                 F438W & 4300~\AA{}        \\
  \textit{HST}/WFC3     & 2009-12-13 &  3000                        &                 F625W & 6300~\AA{}        \\
  \textit{HST}/WFC3     & 2009-12-12 &   800                        &                 F225W & 2400~\AA{}        \\
  \textit{HST}/WFC3     & 2009-12-13 &   800                        &                 F336W & 3400~\AA{}        \\
  \textit{HST}/WFC3     & 2009-12-12 &   400                        &                 F555W & 5300~\AA{}        \\
  \textit{HST}/WFC3     & 2009-12-13 &   400                        &                 F841W & 8100~\AA{}        \\
  \textit{HST}/WFC3     & 2011-01-05 &   403                        &                 F110W & 1.2~\micron{}     \\
  \textit{HST}/WFC3     & 2011-01-05 &   805                        &                 F160W & 1.5~\micron{}     \\
  \textit{HST}/WFC3     & 2015-05-24 &  1200                        &                 F438W & 4300~\AA{}        \\
  \textit{HST}/WFC3     & 2015-05-24 &  1200                        &                 F625W & 6300~\AA{}        \\
  \textit{HST}/STIS     & 2014-08-16 & 40490\tablenotemark{e}       &                 G750L & 5300--10000~\AA{} \\
  \textit{Chandra}/ACIS & 2015-09-17 & 66598                        & HETG\tablenotemark{f} & 0.3--10~keV       \\
  \enddata
  \tablenotetext{a}{Different exposures for individual segments}
  \tablenotetext{b}{Integrated from single exposures of 120~s.}
  \tablenotetext{c}{Integrated from single exposures of 90~s.}
  \tablenotetext{d}{Integrated from single exposures of 600~s.}
  \tablenotetext{e}{Sum of 8098~s for each of the five slits.}
  \tablenotetext{f}{Only zeroth-order image and CCD spectrum used for the analysis.}
\end{deluxetable*}
All observations are summarized in Table~\ref{tab:obs_all}. We use the
most recent observations available at the time of analysis, unless
previous ones have better quality. It is possible that older
observations place more stringent constraints if the compact object
was brighter in the past. We briefly inspect an X-ray observation from
2000 (Section~\ref{sec:obs_cha}) besides the detailed study of the
2015 X-ray observation, but investigating all observations of \sna{}
is beyond the scope of this paper. Some consequences of the temporal
evolution of the compact object are discussed in
Section~\ref{sec:birth}.

We also analyze a circular polarimetric observation using the FOcal
Reducer and low dispersion Spectrograph 2 (FORS2) at the VLT, which
did not detect any significant polarization. The polarization
observation is presented in Appendix~\ref{app:pol}.

\subsection{ALMA}\label{sec:obs_alm}
ALMA observations of \sna{} (Table~\ref{tab:obs_all}) at 1.3~mm (Band
6, 211--275~GHz) were performed on two different epochs: Cycle 2
modest angular resolution data (2013.1.00280; 23--770~k$\lambda$) were
obtained on 2 September 2014. The quasar J0519-4546 (05:19:49.72,
\mbox{-45:46:43.85}; 0.75~Jy at 234~GHz) was the absolute flux
calibrator, which is monitored regularly and calibrated against solar
system objects by the observatory. The quasar J0635-7516 (06:35:46.51,
\mbox{-75:16:16.82}; 0.68~Jy at 234~GHz) was the phase calibrator.

Cycle 3 high angular resolution data (2015.1.00631;
190--8600~k$\lambda$) were obtained from 1 to 15 November 2015, using
J0601-7036 (06:01:11.25, -70:36:08.79; 0.70~Jy at 224~GHz; 0.58~Jy at
253~GHz) as the phase calibrator. Data from 211--213~GHz used
J0519-4546 (0.75~Jy at 224~GHz) as the absolute flux calibrator. Data
at 247~GHz used J0519-4546 for one execution, and J0334-4008
(03:34:13.65, -40:08:25.10; 0.44~Jy at 253~GHz) for a second
execution. All of these quasar calibrators are observed regularly as
part of the observatory calibration network, so we can evaluate the
temporal evolution of each to estimate the uncertainty in the absolute
flux calibrator due to quasar variability.  We can then compare the
derived flux densities of the phase calibrator to the monitoring
observations to estimate the uncertainty in fluxscale transfer to the
science target. These combined yield an estimated absolute flux
calibration uncertainty of better than 7\,\%.

We use the Common Astronomy Software
Application\footnote{\url{https://casa.nrao.edu}}~\citep[CASA,][]{mcmullin07}
to calibrate and image the interferometric data into 3 images with
spectral ranges deemed to be largely free of line emission:
211.83--213.25~GHz, 232.55--233.52~GHz, and
245.95--247.20~GHz~\citep[see Figure~2 of][]{matsuura17}. For imaging
we use the task \texttt{tclean} with multiscale deconvolution with
scales of 0 and 7 times the $6^2$\nobreakdash{-}mas$^{2}$ pixel
size. The FWHM of the restoring beam of the 213\nobreakdash{-}GHz
image is $57\times40$~mas$^{2}$ (major and minor axis), the
233\nobreakdash{-}GHz image $49\times30$~mas$^{2}$, and the
247\nobreakdash{-}GHz image $40\times34$~mas$^{2}$. Analysis of the
phase RMS during the observation with knowledge of the ALMA
calibration efficacy~\citep{asaki14} leads us to conclude that the
astrometric accuracy is better than 10~mas. After image
reconstruction, the real-space images are transformed by a linear
mapping such that the beams are circular for optimal performance with
the finding algorithm, which is described in Appendix~\ref{app:find}.

\subsection{VLT/NACO}
\sna{} has been observed for a total of six epochs between 2006 and
2017 with the Nasmyth Adaptive Optics System Near-Infrared Imager and
Spectrograph (NACO) at the VLT~\citep{rousset03, lenzen03}. Full
details of the observations and data reductions are given in
\citet{ahola18}. For the present work we selected only a single epoch
of \textit{H} and \ks{} band imaging of the highest image quality
(Table~\ref{tab:obs_all}). The \textit{H}-band observation is from
October 2010 with a total on-source integration time of 2160~s. The
\ks{}-band observation is from December 2012 with a total on-source
integration time of 2070~s.

​The images are reduced using standard recipes from the ESO
pipeline~\citep{schreiber04,modigliani07} and IRAF. A horizontal
striping pattern present in the images is removed by a custom script
that creates a one-dimensional (1D) image from the medians ​of the
image pixels along the detector rows. This 1D image is then subtracted
from the rows of the original image. A running sky subtraction is
performed for sets of three stripe-removed exposures at a time using
an ESO pipeline recipe. The three sky-subtracted exposures are aligned
and stacked by the recipe yielding one sky-subtracted image per
running set of three exposures. This process is repeated ​until all of
the sets of three exposures had been sky-subtracted such that $N$
exposures resulted in $N-2$ sky-subtracted images. The sky-subtracted
images are finally aligned based on the centroid coordinates of a
bright star manually selected in each image and subsequently median
averaged. The stripe removal script is run once more for the stacked
image to remove any remaining stripes and bands left by the first
stripe removal step.

Flux calibrations of the NACO observations are made using
star~2\footnote{2MASS J05352761-6916089} as a reference. We confirmed
that star 2 is not variable using observations from 1997 to 2006 from
the Near Infrared Camera and Multi-Object Spectrometer at the VLT in
the F160W and F205W filters. The NACO \textit{H} and \ks{} fluxes of
star~2 are obtained by
converting\footnote{\url{http://www.ipac.caltech.edu/2mass/releases/allsky/doc/sec6_4b.html}}
the \textit{H} and \ks{} fluxes from
2MASS~\citep{cohen03,skrutskie06}. The accuracy of this zeropoint
construction is checked by repeating the 2MASS--ESO comparison for
star 3\footnote{2MASS J05352822-6916118}. The relative difference in
the flux is a factor of 1.06 in the \textit{H} band and 0.82 in the
\ks{} band. We note that star 3 shows variability within a factor of 2
in optical~\citep{walborn93}, which is confirmed by regular
\textit{HST} observations in the \textit{R} and \textit{B} bands over
the past two decades.

\subsection{VLT/SINFONI}\label{sec:obs_snf}
The SINgle Faint Object Near-IR Investigation (SINFONI) Integral Field
Spectrograph at the VLT~\citep{eisenhauer03, bonnet04} observed \sna{}
in the \textit{H} and \textit{K} bands between October and December
2014 (Table~\ref{tab:obs_all}). SINFONI provides moderate angular
resolution and high spectral resolution in a small field of view
(FOV). We only use the spatial resolution to extract spectra from a
search region (Section~\ref{sec:sea_reg}). The limits are then
constructed from the extracted spectra. The data are reduced using the
standard ESO pipeline~\citep{schreiber04,modigliani07} with the
improved subtraction of the OH airglow emission following
\citet{davies07}. A more detailed presentation of the processing of
these particular observations is provided by \citet{larsson16} and a
comprehensive description of SINFONI data reduction can be found in
\citet{kjaer10}.

Contaminating light from the ER is subtracted from the spectra of the
central region. The lines from the ER have a FWHM of
${\sim}$300\kmps{} and the lines from the central ejecta a FWHM of
${\sim}$2500\kmps{}~\citep{fransson15, larsson16}. Even though the
ejecta are clumpy and illumination is non-uniform, the line profiles
of the ejecta are relatively smooth and much broader than the sharp
narrow lines from the ER. The difference allows us to subtract the ER
spectra from the central spectra by scaling the ER spectra such that
they cancel the narrow components of the central spectra. Backgrounds
are constructed from the cleanest available regions in the relatively
small FOV of SINFONI. These are then subtracted from the extracted
central spectra. It is verified that different choices of background
regions do not significantly alter the results. The
signal-to-background ratio (S/B) for the low continuum level is 0.88
in the \textit{H} band and 2.31 in the \textit{K} band.

\subsection{\textit{HST}/WFC3}\label{sec:obs_wfc3}
\sna{} was observed using Wide Field Camera 3 (WFC3) in December 2009
in six filters; F225W, F336W, F438W, F555W, F625W, and F814W
(Table~\ref{tab:obs_all}). We choose these observations
from 2009 because they provide the most complete wavelength coverage
at a recent epoch. Together, the six filters provide coverage over the
2150--8860~\AA{} wavelength interval. The latest WFC3 NIR observations
are from January 2011 in the F110W and F160W filters. The latest
wide-filter observations with high quality at the time of analysis are
from May 2015 in the F438W and F625W filters.

All observations were performed using the four-point box dither
pattern and drizzled~\citep{fruchter02} onto a final pixel size of
$25^2$~mas$^2$ using a value for the DrizzlePac~\citep{gonzaga12}
parameter \texttt{pixfrac} of either 0.6 or 0.7. Cosmic ray rejection
is also performed when drizzling to combine the dithered exposures.
The flux zeropoints for all WFC3 images are taken from the IRAF/STSDAS
package Synphot~\citep[calibration database updated on 17 January
2017]{bushouse94}.

\subsection{\textit{HST}/STIS}\label{sec:obs_sts}
Between 16--20 August 2014, \textit{HST}/Space Telescope Imaging
Spectrograph (STIS) observed \sna{} using the G750L grating, which
covers the wavelength interval 5300--10\,000~\AA{}
(Table~\ref{tab:obs_all}).  The STIS observations have been described
in detail in \citet{larsson16}. Here, we provide additional
information on astrometry and background subtraction because of their
importance to the compact object limit. The observations were made at
five adjacent slit positions, shown in Figure~\ref{fig:pos}. Each slit
is 100~mas wide and oriented in the north-south direction. The
position of \sna{} that is presented in Section~\ref{sec:pos} is very
close to the dividing line between the second and third slit.

Contaminating light from the ring is subtracted from the spectrum of
the central region using the same method as for SINFONI
(Section~\ref{sec:obs_snf}). In addition, a background is extracted
from regions north and south of the SN and subtracted from the ejecta
spectrum. The S/B is 0.74 for the low continuum level. The background
is extracted from 75 pixel rows (50~mas~pixel$^{-1}$) in three regions
both north and south of the \sna{}.

\subsection{Chandra}\label{sec:obs_cha}
\textit{Chandra} observed \sna{} on 17 September 2015 (Obs.~ID 16756,
Table~\ref{tab:obs_all}), utilizing the Advanced CCD Imaging
Spectrometer~\citep[ACIS,][]{garmire03} S-array equipped with the
High-Energy Transmission Grating~\citep[HETG,][]{canizares05}. The
ACIS detector provides imaging capabilities with an angular resolution
of 700~mas (FWHM) and a moderate energy resolution of ${\sim}$100~eV
at 2~keV. The spatial resolution is just enough to resolve \sna{},
which allows us to extract a spectrum of the central region that can
be used to set an upper limit on the compact object. Among many
\textit{Chandra} observations~\citep[e.g.][]{frank16}, we choose this
particular observation because it was the latest observation at the
time of analysis, which implies that the absorption toward the center
is the lowest (Appendix~\ref{app:xray_abs}).

Contamination on the ACIS optical blocking
filter~\citep[OBF,][]{odell13} has previously led to inaccurate flux
measurements~\citep[e.g.][]{park11, helder13}. \citet{frank16} have
verified that the OBF contamination is now well-modeled using the High
Resolution Camera/Low Energy Transmission Grating observation of
\sna{} from 14 March 2015 (Obs.~ID 16757), which does not suffer from
the OBF contamination. Only data from the zeroth-order image in energy
range 0.3--10~keV of the 17 September 2015 ACIS/HETG observation are
used in this analysis.

The data are reduced following standard procedures using CIAO 4.9 and
CALDB 4.7.7~\citep{fruscione06}. No background flares are observed
resulting in a total exposure of 66~ks with ${\sim}$11\,000 source
counts. XSPEC 12.9.1p~\citep{arnaud96} is used for the spectral
analysis and all extracted spectra are binned with a minimum of 20
counts per bin. A background is extracted from an annulus with an
inner radius of 15\arcsec{} and outer radius of 30\arcsec{}. The
background is negligible for the spectra of \sna{} because of the high
source count rates and small spatial region of interest.

The HETG provides dispersed spectra and also reduces pileup in the
zeroth-order image~\citep{helder13, frank16}. The dispersed spectra
are used to verify that pileup is not significant in the zeroth-order
CCD spectrum and we find that the bad grades 1, 5, and 7 combine to be
${\sim}$3\,\% of the total level 1 source
counts\footnote{\url{http://cxc.harvard.edu/proposer/POG/html/chap6.html}}.
This indicates that the level of pileup is low enough to not
significantly affect our analysis. A more detailed treatment of pileup
is difficult because \sna{} is marginally resolved and we primarily
use the spectrum from the region inside the ER, which is smaller than
a single ACIS pixel. Pileup properties could be different for the ER
and the ejecta because the count rate is significantly higher in the
pixels neighboring the few central pixels. However, the grades are
assigned based on $3\times 3$ pixel islands. How all these effects
combine require custom methods, which would be excessive for our
analysis.

\section{Methods}\label{sec:methods}
Below, we describe the methods used to determine upper limits on the
compact object in \sna{}. The position of \sna{} and the spatial
regions in which we search for the compact object are described in
Sections~\ref{sec:pos} and~\ref{sec:sea_reg}
Sections~\ref{sec:reddening}--\ref{sec:met_spe} present how the image
and spectral limits at millimeter, UV, optical, and NIR (UVOIR)
wavelengths are determined. The spread light in the X-ray observation,
which complicates the computation of the X-ray limits, is described in
Section~\ref{sec:cha_lim}. Finally, the X-ray ejecta absorption model
based on 3D neutrino-driven SN explosion models~\citep{alp18b} is
described in Appendix~\ref{app:xray_abs}, and the spatial alignment of
the images is described in Appendix~\ref{app:spa_ali}.

The source luminosity limits based on the imaging analysis rely on the
assumption that the compact object is a point source. The limits from
the spectral analysis in UVOIR are based on the assumption that the
compact object emission is a continuum. Finally, the X-ray limits are
constructed by assuming certain spectra for the compact object. The
distance to \sna{} is taken to be 51.2~kpc~\citep{panagia91, gould98,
  panagia99, mitchell02}. All two-sided confidence intervals are
1$\sigma$ and all one-sided upper limits are 3$\sigma$ unless
otherwise stated.

\subsection{Position of \sna{}}\label{sec:pos}
\begin{figure*}
  \centering
  \raisebox{-0.5\height}{\includegraphics[width=.525\textwidth]{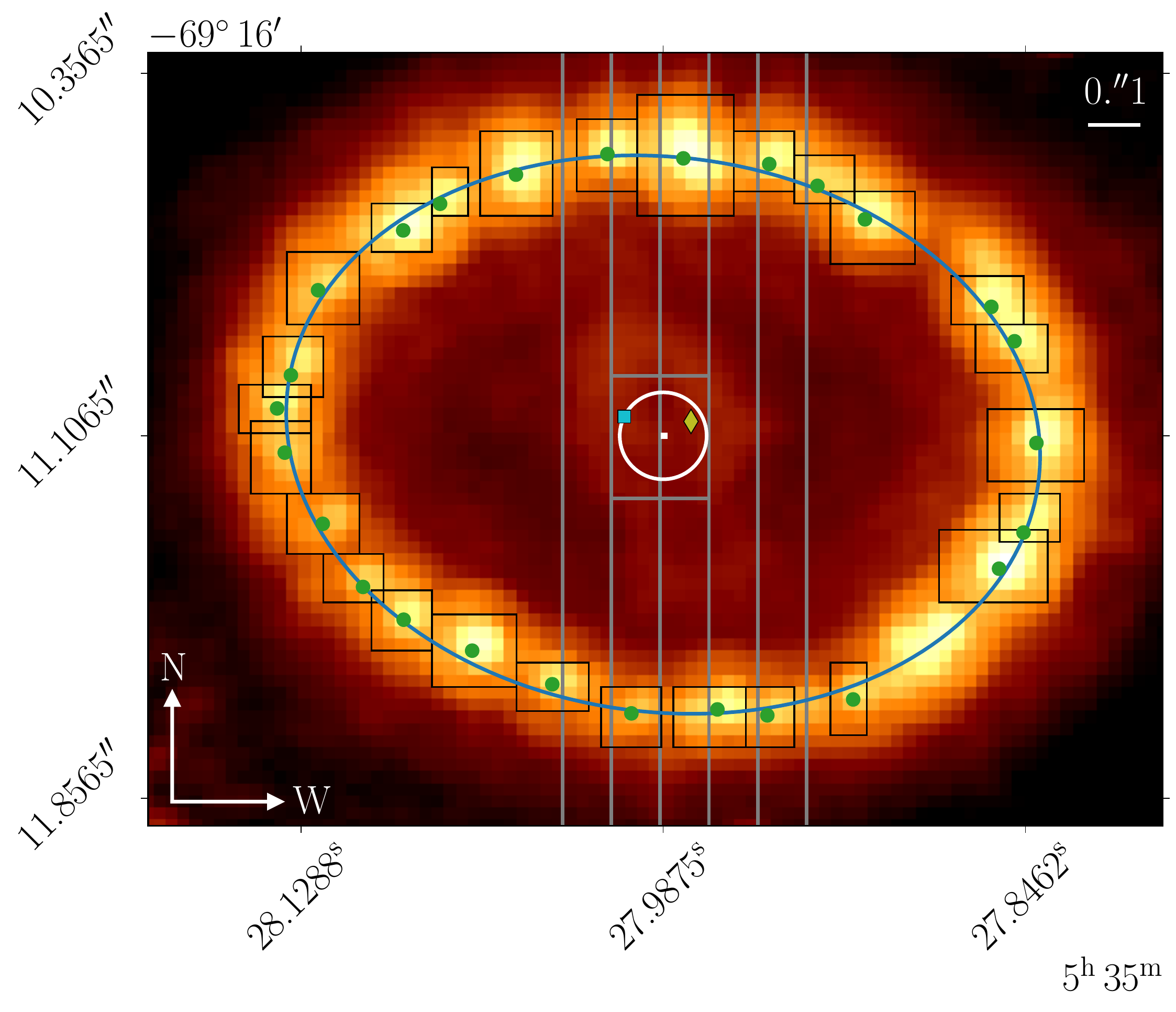}}
  \raisebox{-0.5\height}{\includegraphics[width=.462\textwidth]{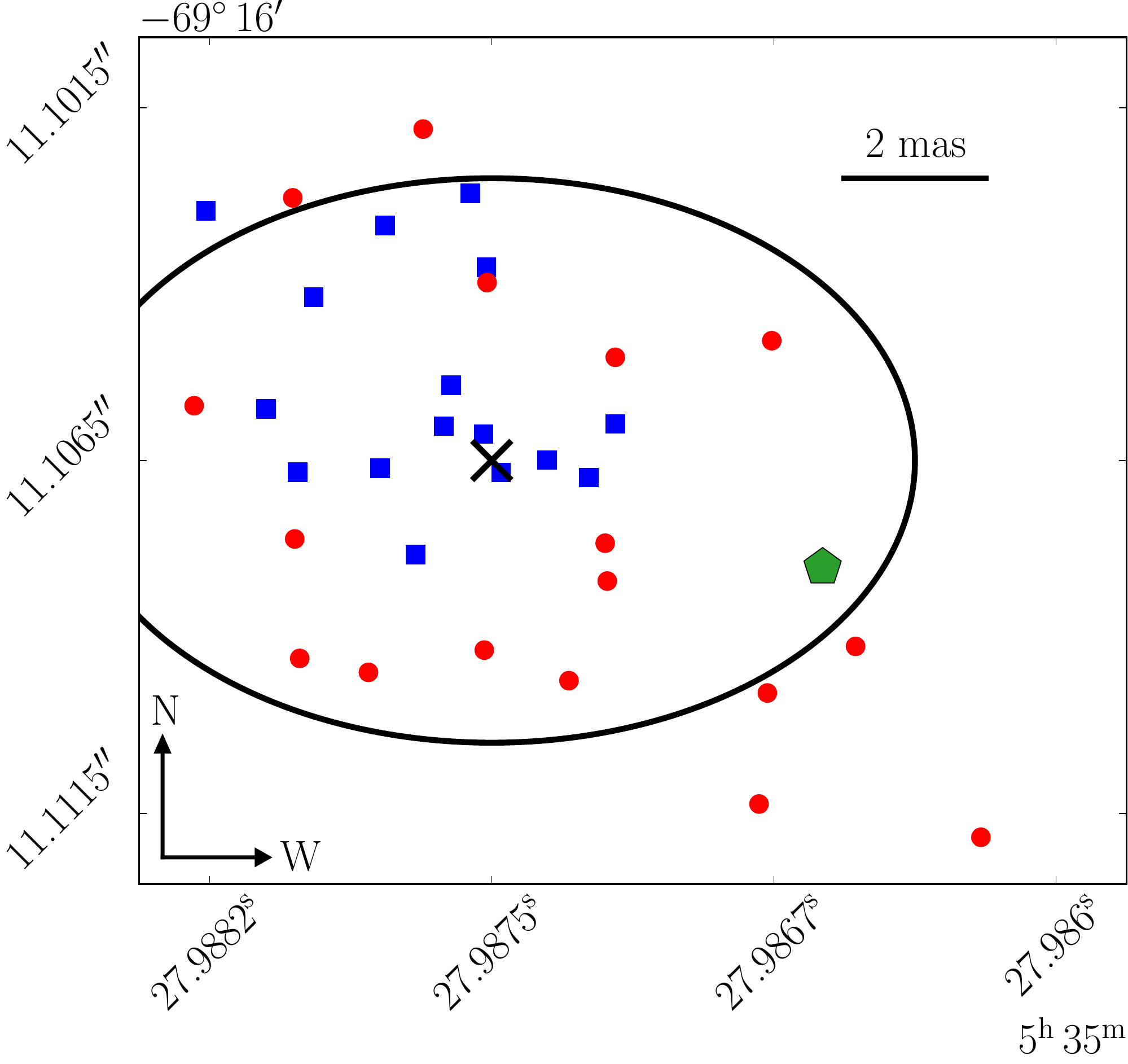}}
  \caption{Position estimates of \sna{} plotted on the
    \textit{HST}/ACS F625W 6 December 2006 observation. The left panel
    shows the fitted hotspot locations as green dots, which are
    determined by fitting 2D Gaussians within the black rectangles.
    The blue ellipse is the best-fit ellipse to the green dots and the
    (very small) white rectangle is the entire region shown in the
    right panel. The white circle is the search region corresponding
    to a kick velocity of 800\kmps{} at 10\,000~days (11 July
    2014). The cyan square is the radio centroid position (fit of a
    spherical shell to radio data) and the yellow diamond is the
    progenitor position reported by \citet{reynolds95} after
    correcting for the proper motion of the position of \sna{} in the
    LMC~\citep{kallivayalil13, van_der_marel14, van_der_marel16}. The
    vertical gray lines show the area covered by the five slits of the
    \textit{HST}/STIS observations and the horizontal gray lines
    represent the STIS extraction region. The right panel shows the
    estimated positions of \sna{} based on the 33 \textit{HST}
    observations from 2003 to 2016 by fitting ellipses to the
    hotspots. The size of the right panel is $14\times 12$~mas. Blue
    points denote positions from \textit{B} band images and red points
    positions from \textit{R} band images. The black solid ellipse is
    the 1-$\sigma$ confidence contour of the best estimate (black
    cross):
    $\alpha = 05^\mathrm{h}\,35^\mathrm{m}\,27^\mathrm{s}\!.9875(11),
    \delta = -69^\circ\,16'\,11\farcs{}107(4)$ (ICRF J2015.0). The
    green pentagon is the position (5~mas from the favored position)
    based on fits using an elliptic annulus as described in the
    text.\label{fig:pos}}
\end{figure*}
We need an accurate position estimate of \sna{} because we only search
for the compact object in a kick region that has a radius of
${\sim}$100~mas (Section~\ref{sec:sea_reg}). Therefore, it is
important that the position estimate of \sna{} is accurate to
${\sim}$10~mas. The ``position of \sna{}'' refers to the projected
position of the progenitor star at current epochs. The position is
determined by fitting an ellipse to the hotspots in the ER in
\textit{HST} images. This assumes that the progenitor is located at
the center of the ring of hotspots. The first step is to tie the
\textit{HST} images to \textit{Gaia} data release
1~\citep[DR1,][]{gaia16b, gaia16} because the error of the absolute
astrometry of \textit{HST} is relatively large. This is done by
mapping two unsaturated stars\footnote{\textit{Gaia} source ID:
  4657668007091797248, 4657668075811272704}. The uncertainties in the
positions of the two reference stars in the \textit{Gaia} archive are
$<1$~mas and the accuracy of the localization of the stars in the
\textit{HST} images are a few mas.

All 33 observations from 2003 to 2016 in the \textit{B} (F435W, F438W,
and F439W) and \textit{R} (F625W and F675W) bands (presented in
Appendix~\ref{app:pos_obs}) are used for determining the
position. Observations from before 2003 are excluded because the low
number of hotspots present before this time make the fits inaccurate.
The hotspots are defined using the F625W 6 December 2006 observation
because it provides the best spatial resolution at an epoch when most
hotspots were bright~\citep{fransson15}. Two-dimensional (2D)
Gaussians are fitted to the 26 hotspots, shown in
Figure~\ref{fig:pos}. The hotspots are located in the other
observations by fitting a radial, 1D Gaussian along the angles defined
by the 2D fits in the F625W 6 December 2006 observation. This ensures
that the same hotspots are found in all observations. Ellipses are
then fitted to the hotspots in all images. The best-fit estimates of
the position of \sna{} from the 33 images agree within 25~mas and are
shown in Figure~\ref{fig:pos} (right). The arithmetic mean of the 33
best-fit coordinates are
$\alpha = 05^\mathrm{h}\,35^\mathrm{m}\,27^\mathrm{s}\!.9875(11),
\delta = -69^\circ\,16'\,11\farcs{}107(4)$ (ICRF J2015.0), where the
1\nobreakdash{-}$\sigma{}$ uncertainties are estimated by
bootstrapping the hotspot locations. Unless otherwise stated, all
presented coordinates are at epoch J2015.0 and in the reference frame
of \textit{Gaia} DR1, which is effectively equivalent to ICRF (the
realization of ICRS) for the current level of precision. This will
henceforth be adopted as the position of \sna{} in this work.

We emphasize that the reported confidence interval only represents the
statistical uncertainty. Fitting to the ER continuum in the ALMA
observation results in a position offset of approximately 60~mas to
the east, but the ER is more diffuse at millimeter wavelengths and we
choose to use the optical observations. This is likely to be the best
approach because the hotspots are well-defined point sources, whereas
the millimeter emission originates from a larger volume above and
below the optically emitting ring.

The systematic error introduced by assuming the hotspots to be located
in an ellipse centered on the supernova position is checked by fitting
an elliptic annulus with a Gaussian radial profile to the entire inner
ring. This elliptical band is also allowed to rotate in the sky plane
and has a sinusoidal intensity along the azimuth. This describes the
inner ring as a continuum rather than as a collection of point sources
and serves as a relatively independent estimate of the position. The
center obtained using an elliptical annulus is
$\alpha = 05^\mathrm{h}\,35^\mathrm{m}\,27^\mathrm{s}\!.9866, \delta =
-69^\circ\,16'\,11\farcs{}108$ (ICRF J2015.0), shown in
Figure~\ref{fig:pos} (right, green pentagon). This position is offset
by 5~mas from the favored position of \sna{}.

The hotspot coordinates can be compared to the best estimate of the
location of the progenitor star \sleak{};
$\alpha = 05^\mathrm{h}\,35^\mathrm{m}\,27^\mathrm{s}\!.968(9), \delta
= -69^\circ\,16'\,11\farcs{}09(5)$~\citep[ICRF
J1991.5,][]{reynolds95}. In addition, \citet{reynolds95} used
observations by the Australia Telescope Compact Array from 21 October
1992 and 4--5 January 1993 at 8.8~GHz and reported a radio centroid
position;
$\alpha = 05^\mathrm{h}\,35^\mathrm{m}\,27^\mathrm{s}\!.994(12),
\delta = -69^\circ\,16'\,11\farcs{}08(5)$ (ICRF J1991.9). The
aforementioned coordinates are those reported by
\citet{reynolds95}. To compare with our estimates in
Figure~\ref{fig:pos}, the positions of \citet{reynolds95} are
corrected for the displacement between the epochs of observation. The
proper motion of the position of \sna{} in the Large Magellanic Cloud
(LMC) is 46~mas east and 13~mas north~\citep{kallivayalil13,
  van_der_marel14, van_der_marel16}. This assumes that the proper
motion of \sleak{}, and consequently also \sna{}, conforms to the
expected motion of its location within the LMC.

\subsection{Search Region}\label{sec:sea_reg}
An extended region is searched because the compact object created by
\sna{} is expected to have a kick velocity caused by the asymmetric
explosion. Typical 3D kick velocities of pulsars are
${\sim}$400\kmps{}~\citep{hobbs05,faucher-giguere06}. However, extreme
cases of velocities up to 1600\kmps{} have been
observed~\citep{cordes93,chatterjee02,chatterjee04,hobbs05}. When
searching for a point source in \sna{}, we assume a sky-plane
projected kick velocity of 800\kmps{}, which corresponds to a search
radius of ${\sim}100$~mas at the epoch of our observations. This is a
trade-off between having to search an excessively large region and the
risk of not including the true source position.

The effects of different choices of kick velocity are relatively
small. For a kick of 1600\kmps{}, the average correction factor to the
six \textit{HST} limits from 2009 (Section~\ref{sec:img_lim}) is
1.07. For a more typical speed of 400\kmps{} the corresponding factor
is 0.86. The reason for the small difference is that the brightness is
relatively uniform and the search-algorithm is dependent on both
surrounding morphology and brightness. This means that limits are not
directly proportional to the local brightness.

\subsection{Reddening}\label{sec:reddening}
The effect of interstellar reddening at UVOIR wavelengths is corrected
for using the model of \citet{cardelli89} with $R_V=3.1$ and
$E(\bv)=0.19$. The parameters are chosen based on the work by
\citet{france11}, which takes several studies of extinction to \sna{}
and the LMC into consideration~\citep{walker90, fitzpatrick90,
  scuderi96, michael03, gordon03}. The uncertainty in the de-reddening
is approximately 20\,\% below 3000~\AA{}, less than 10\,\% in optical,
and less than ${\sim}2$\,\% in NIR.

Large amounts of dust have been detected in the ejecta of
\sna{}~\citep{matsuura11, indebetouw14, wesson15, matsuura15, dwek15},
but we do not attempt to correct for it using the same method as for
interstellar reddening. The primary reason for this is that the
spatial distribution of the ejecta dust is poorly
constrained~\citep[e.g.][]{wesson15, dwek15}. Our treatment of ejecta
dust is explained in Section~\ref{sec:uvoir_dust}.

\subsection{Image Limits}
Image limits are obtained from ALMA, VLT/NACO, and
\textit{HST}/WFC3. The same method is used for all observations, apart
from two small differences for ALMA. The differences are how the PSF
of the instruments are determined and how the spatial positions are
chosen. For the interferometric ALMA images, the PSFs are the
well-defined reconstruction beams (Section~\ref{sec:obs_alm}). PSFs
for the UVOIR images are created using the
IRAF/DAOPHOT~\citep{stetson87} package following the guidelines for
fitting PSFs in \citet{davis94}. An empirical PSF is generated for
each observation by fitting to ${\sim}$10 bright, well-isolated
stars. A Gaussian is selected as the analytical component with a
linear look-up table. The quality of the PSFs is checked by estimating
the residuals when subtracting best fits from the original images. The
residuals are less than 5\,\% of the counts for most stars. In
addition, each individual PSF model is visually inspected for defects.

The ALMA limits are not required to be from the same point because
they are treated as three independent limits in the analysis. In
contrast, UVOIR image limits from the same epoch are at the same
spatial position for the different bands. The points are chosen such
that the highest total flux allowed by the limits in all bands is
maximized and allows us to combine the limits to constrain spectra.

The remainder of the process is identical for the ALMA and UVOIR
images. Limits are obtained by introducing artificial sources with
known fluxes that are recovered using a finding algorithm, which is
described in Appendix~\ref{app:find}. 

\subsection{Spectral Limits}\label{sec:met_spe}
The spectra are extracted from regions corresponding to the 800\kmps{}
kick region and wavelengths are corrected for the systematic
heliocentric velocity of \sna{} of 287\kmps{} away from
Earth~\citep{groningsson08, groningsson08b}. Limits are then
constructed from spectral regions that are relatively free of line
emission. These regions are assumed to contain contributions from weak
lines, gas continuum emission, and emission from the compact
object. Therefore, the determined limits are conservative limits on
the contribution from the compact object. We fit functions to the
observed flux within the regions that are free of line emission
(Section~\ref{sec:met_nir_spe} and~\ref{sec:met_opt_spe}). The
magnitude of these functions are then increased until the $\chi^2$
values have increased by 7.740. These values are then taken to be the
one-sided 3-$\sigma$ upper limits. We verified that the reduced
$\chi^2$ values are reasonably close to unity, which is required for
this method to be applicable.

\subsubsection{NIR Spectral Limits}\label{sec:met_nir_spe}
\begin{figure}
  \centering \includegraphics[width=\columnwidth]{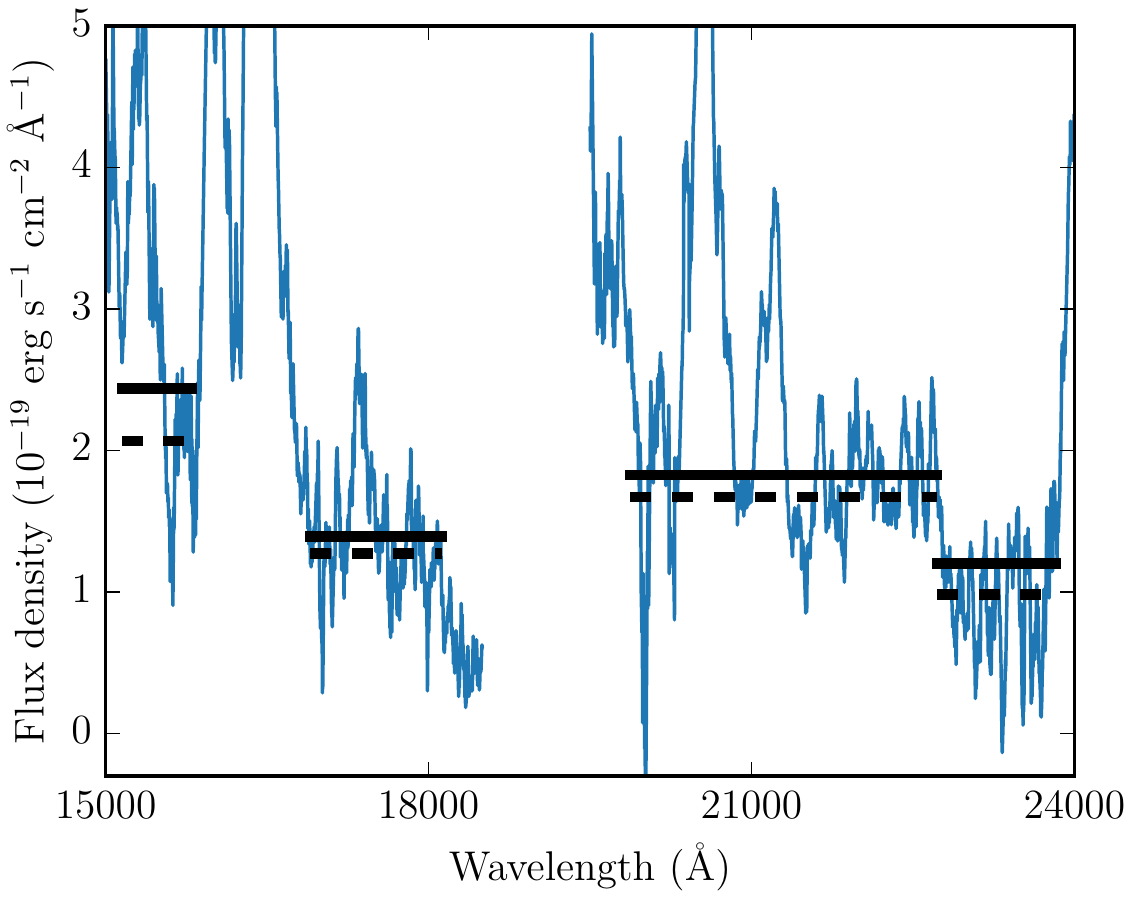}
  \caption{The 2014 SINFONI \textit{H} and \textit{K} spectra of the
    central region of \sna{}. Both spectra are extracted from regions
    that correspond to the 800\kmps{} kick region
    (Section~\ref{sec:sea_reg}, Figure~\ref{fig:pos}). The solid black
    lines are the upper limits and the dashed black lines are the
    average flux in the selected intervals
    (Section~\ref{sec:met_nir_spe}). The spectra have been corrected
    for spread light from the ER and background
    subtracted~(Section~\ref{sec:obs_snf}), and corrected for
    interstellar reddening~(Section~\ref{sec:reddening}). The observed
    spectra (blue) have been binned by a factor of 21 for visual
    clarity.\label{fig:nir_spc_lim}}
\end{figure}

The upper limits from the SINFONI spectra are set using the flux
density in regions that are free of strong line emission. The
resulting spectra are shown in Figure~\ref{fig:nir_spc_lim}. The
regions are selected to avoid intervals of emission lines identified
in \sna{} provided in Table~3 and~4 of \citet{kjaer07} and regions of
H$_2$ emission~\citep{fransson16}. Moreover, wavelength intervals
close to residual atmospheric lines are also excluded. These are
clearly seen in observations as narrow lines. In total, 16\,\% of all
data pass the aforementioned selection criteria in the \textit{H} band
and 28\,\% in the \textit{K} band. We combine many very narrow
intervals into four groups and define constant functions within the
groups (Figure~\ref{fig:nir_spc_lim}). We choose four regions because
the specific flux is relatively constant within the regions. The
constant values are the averages within each group, which are then
increased to a 3-$\sigma$ upper-limit level.

\subsubsection{Optical Spectral Limit}\label{sec:met_opt_spe}
\begin{figure}
  \centering \includegraphics[width=\columnwidth]{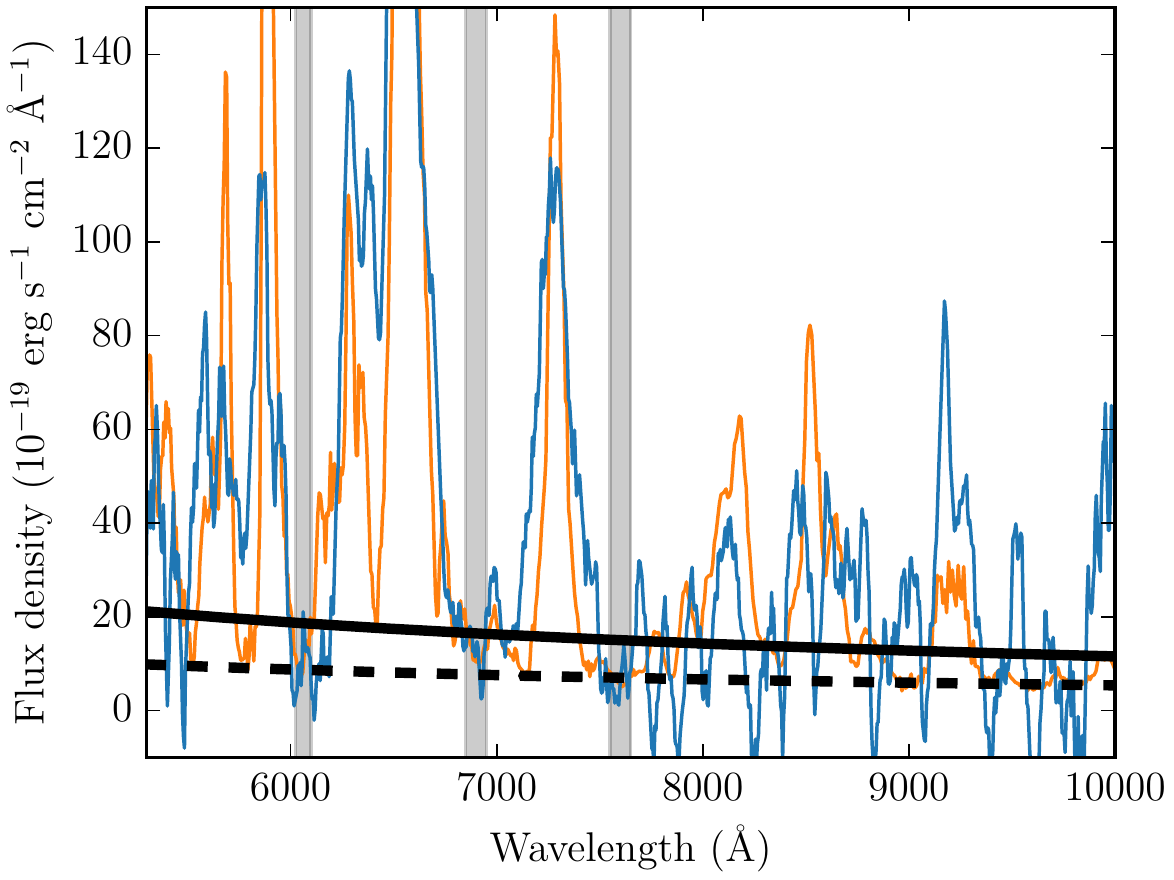}
  \caption{The 2014 STIS spectrum of the central region of \sna{}. The
    extraction region is a rectangle that approximately contains the
    800\kmps{} kick region (Section~\ref{sec:sea_reg},
    Figure~\ref{fig:pos}). The solid black line is the power law that
    represents the 3-$\sigma$ upper limit to continuum emission from a
    compact object. The dashed black line is the best-fit power law,
    which is a factor of $2.1$ lower than the limit. Gray regions
    indicate the wavelength intervals in which the power law is
    fitted. The spectra have been corrected for spread light from the
    ER and background subtracted~(Section~\ref{sec:obs_sts}), and
    corrected for interstellar
    reddening~(Section~\ref{sec:reddening}). The observed spectrum
    (blue) is smoothed by a factor of eleven for visual clarity. The
    orange line is a model spectrum of \sna{} taken from
    \citet{jerkstrand11} normalized to H$\alpha$. This model is only
    used for identifying regions that are free of line
    emission.\label{fig:opt_spc_lim}}
\end{figure}

To determine the compact object limit using the STIS observation, a
spectrum is extracted from the search region. Because of the
resolution of the instrument, the spectrum is from a region with a
width of two 100-mas slits and height of five 50-mas pixels, which is
approximately equivalent to a rectangle that just contains the
800\kmps{} extraction region (Figure~\ref{fig:pos}).  A power law is
fitted to regions that are free of strong lines (gray regions in
Figure~\ref{fig:opt_spc_lim}). The power law describes the
quasicontinuum well within the STIS wavelength range
5300--10\,000~\AA{}, which is why we do not use the same method as for
SINFONI (Section~\ref{sec:met_nir_spe}). No significant improvement in
the fit is seen for other simple functional forms. The selected
regions that are relatively free from line emission are 6025--6100,
6850--6950, and 7550--7650~\AA{}. The average flux in the middle
region is slightly higher, but excluding it results in a
${\sim}10$\,\% less constraining limit because of the reduced
statistics. The regions are found by visual inspection of the observed
spectrum and by searching the model spectrum of \citet{jerkstrand11}
for regions free of strong lines. The model computes an observed
spectrum by simulating the radiation transfer through the SN ejecta,
which is assumed to be powered by the radioactive decay of
$^{44}$Ti. We note that the predicted spectrum was summed over the
entire ejecta and worse agreement is expected for the smaller search
region. The model is only used to identify line-free regions and it is
included in Figure~\ref{fig:opt_spc_lim} for reference.

\subsection{X-ray Limits}\label{sec:cha_lim}
To determine the X-ray flux limit on the compact object, the ER
emission needs to be modeled. This is important because the observed
flux in the inner region is dominated by spread light from the ER. We
use spread light to refer to the result of angular blurring of
telescopes, also referred to as wings, leaked light, scattered light,
or glare. The spatial model allows for computation of the amount of
spread light in the inner region. Finally, the limits are set by
spectral fitting to the spectrum from the inner region with a model
that includes the spread light from the ER.

We also briefly inspected the 61\nobreakdash{-}ks ACIS observation
from 7 December 2000 (Obs.~ID 1967) and find that the X-ray limits
allow for approximately a factor of 2 higher luminosities because of
the higher ejecta absorption (Appendix~\ref{app:xray_abs}). It is
unlikely that more stringent limits can be placed using the High
Resolution Camera onboard \textit{Chandra} because of the very poor
energy resolution, which prevents separation of emission from the ER.

\subsubsection{Spatial Modeling}\label{sec:spa_mod}
\begin{figure}
  \centering
  \includegraphics[width=0.49\columnwidth]{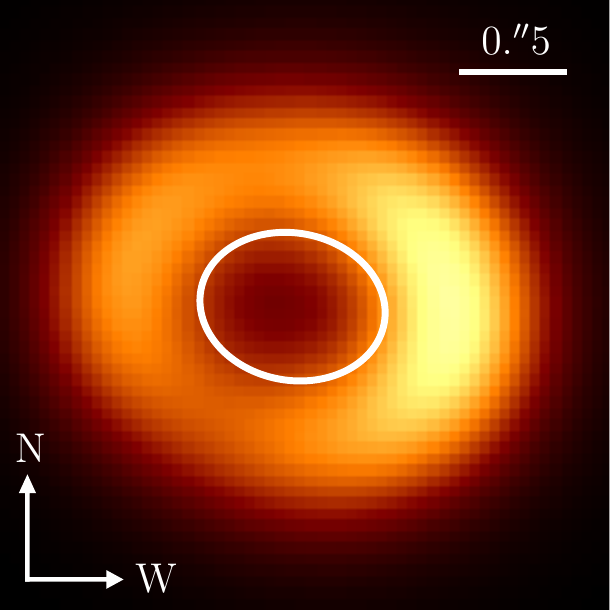}
  \hfill
  \includegraphics[width=0.49\columnwidth]{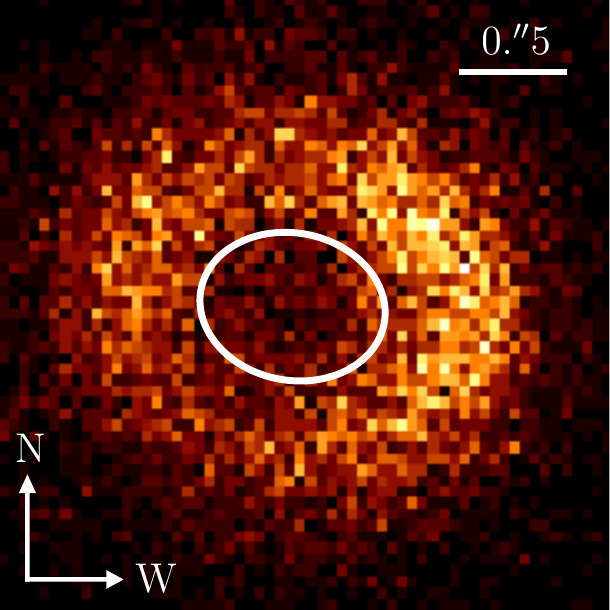}
  \caption{Folded model (left, see text) and the 17 September 2015
    \textit{Chandra}/ACIS observation binned to 50$^2$~mas$^2$ pixels
    in the energy range 0.3--10~keV (right). The ellipse shows the
    inner region where the ejecta spectrum is
    extracted.\label{fig:cha_obs}}
\end{figure}
The remnant is fitted with an ellipse of sinusoidal intensity along
the azimuth and Gaussian radial profile, which is then folded through
the modeled instrumental PSF. This is just a model used to describe
the observed morphology of the ER. A description of the simulation of
the \textit{Chandra} PSF is provided in Appendix~\ref{app:psf}. The
inner region covering the central ejecta is excluded from the fit to
reduce the effects of any contribution from a central
source. Observations show finer structure than allowed by this simple
model. Therefore, the pixels of the unfolded model are given some
freedom by assigning new values that are distributed as normal
distributions with the mean set to the original value and standard
deviation to one-third of the original value. Any negative values are
set to zero. The random reassignment of pixels is performed 10\,000
times and the folded model that gives the maximum likelihood for the
observed data is chosen, shown in Figure~\ref{fig:cha_obs}. The
goodness of fit is determined by simulating 100\,000 observations from
the folded best-fit model. A total of 53\,039 simulations resulted in
a higher statistical likelihood than the real observation, showing
that the fit is acceptable.

An inner and an outer region are defined using the best-fit model.
These are the regions from which spectra are extracted.  The inner
region (Figure~\ref{fig:cha_obs}) is defined as an inclined ellipse
with parameters given by the fitted model; a position angle of
83\degr{} to the semi-major axis (defined counter-clockwise from
north) and a ratio of semi-major to semi-minor axis of 1.37. The
magnitude of the semi-major axis of the inner region is set to
450~mas, which is chosen to maximize the signal-to-noise ratio (S/N)
of a central point source. This can be compared to the 870~mas
semi-major axis and 620~mas semi-minor axis of the best-fit
ellipse. The outer region is defined as an inclined elliptical annulus
with the inner region as the inner boundary.  The semi-major axis of
the outer boundary of the outer region is 10\arcsec{}.

An ER spectrum is extracted from the outer region and a central
spectrum from the inner region, henceforth referred to as the ejecta
spectrum. The ejecta spectrum has a total photon count of 624. The
option to correct for the encircled energy (EE) of the CIAO task
\texttt{specextract} is disabled. Instead, the correction factor for
the inner region is computed to be 0.46 using the simulated MARX
PSF. The value of 0.46 is computed for a point source at the center
and will be used to correct for the missing flux throughout this
analysis. No EE correction is applied to the outer region because the
physical flux of the ER is not of interest. This method is employed
because it allows for fitting of the fraction of spread light, which
can then be directly compared to the modeled fraction of spread light.

\subsubsection{Modeled Spread Light Fraction}
Light from the bright ER contaminates the central region of \sna{} in
the \textit{Chandra} observations. We estimate the fraction of spread
light ($f_\mathrm{s}$) using the spatial model and compare it to the
observed value, which is obtained by spectral fitting.  The fraction
of spread light is defined as the ratio of spread light flux in the
inner region to the flux in the outer region. The value
$f_\mathrm{s} = 0.073^{+0.011}_{-0.005}$ is computed using the spatial
model with no central source. The model is forward folded using the
PSF to simulate the angular response of the detector and the
uncertainties are obtained by simulations using the fitted model. We
assume that the energy dependence of the EE is relatively weak,
implying that the spectrum of the spread light in the inner region is
the same as the spectrum of the ER itself. This assumption is
partially motivated by the fact that the energy dependence is modest
over the range covered by the bulk of the photons. Additionally, the
small angular separations of $\lesssim 1\arcsec{}$ further reduce the
magnitude of this effect, see Section~3.2 of \citet{park10} for a more
detailed treatment of spread light.

\subsubsection{Observed Spread Light Fraction}\label{sec:obs_sca}
The observed fraction of spread light can be determined by
extracting spectra from both the ER and the inner region, and then
fitting the model that describes the ER to the inner spectrum, with
only the normalization left free to vary. This needs to be done
because there might be a significant contribution from the compact
object in the observation, which is assumed to have a spectrum
different from that of the ER.

The model we use for the ER spectrum consists of an ISM absorption
component and three source components; an ionization equilibrium
collisional plasma (\texttt{vequil}) at a temperature of 0.3~keV, a
constant-temperature plane-parallel shock plasma model
(\texttt{vpshock}) at a temperature of 2.1~keV, and a power-law
component. ISM absorption of all three components is modeled using the
\texttt{tbgrain}~\citep{wilms00} photoabsorption model with a frozen
hydrogen column density of
$N_\text{ISM} = 0.409\times 10^{22}$~cm$^{-2}$, of which
$0.144\times 10^{22}$~cm$^{-2}$ is molecular. These values are taken
from
\citet{willingale13}\footnote{\url{http://www.swift.ac.uk/analysis/nhtot/}}
and are approximately a factor of 2 higher than used by many previous
X-ray studies of \sna{}, which have neglected the molecular
component. We note that this only makes a difference of ${\sim}$3\,\%
for our results because of the high ejecta absorption. For this
reason, we also ignore any CSM or LMC absorption. The power law
reduced the fit statistic by $\Delta \chi^2 = -29$ for 2 additional
degrees of freedom (d.o.f.) and clearly improved the fit at energies
above 5~keV. The power law does not have a clear physical
interpretation, but the purpose of the model is only to represent the
spread light from the ER into the central region. The fit statistic
for the ER spectrum is $\chi^2=205$ for 202~d.o.f.

We then fit the ER model to the ejecta spectrum with all parameters
frozen apart from a constant factor. This represents an upper limit on
the spread light from the ER into the central region because it is
implicitly assumed that the contribution from the compact object is
negligible. The fit statistic is $\chi^2 = 29.8$ for 27~d.o.f. and the
fraction of spread light to ER flux is $0.062\pm 0.003$, which can be
compared to the value of $0.073^{+0.011}_{-0.005}$ predicted by the
model. The goodness of fit implies that the energy dependence of the
EE is small enough to be neglected in this case. The observed value is
marginally lower than predicted and implies that practically all flux
observed in the central region can be interpreted as spread light from
the ER. The purpose of comparing the modeled and observed
$f_\mathrm{s}$ is that an observed value that is significantly higher
than predicted would indicate an additional contribution in the
central region.

\subsubsection{Calculating X-ray Limits}
We construct X-ray limits on the compact object by adding components
to the ER spectrum model and re-fitting the model to the ejecta
spectrum. Then, we find the limiting value for a parameter of interest
of the additional component while fitting $f_\mathrm{s}$ and freezing
all other parameters. Leaving additional parameters free is not
possible because the low number of counts in the ejecta region is
insufficient to meaningfully constrain additional parameters. The
additional component represents the contribution from the compact
object and the SN ejecta absorption (Appendix~\ref{app:xray_abs}) is
only applied to this component using the
\texttt{tbvarabs}~\citep{wilms00} photoabsorption model. The
presented results are obtained using the Levenberg-Marquardt fitting
algorithm and the $\chi^2$ statistic. We also attempted an unbinned
analysis using the Cash-statistic~\citep{cash79}, but found
differences that are much smaller than other uncertainties.

Blackbody and power-law models with different
amounts of ejecta absorption are tested, as described in
Section~\ref{sec:res_xray}. The value $f_\mathrm{s}$ remained
$\geq 0.06$ for all models with absorption
(Section~\ref{sec:res_xray}), which means that the spread light is not
degenerate with the additional component. The upper limits are
obtained by requiring an increase of the $\chi^{2}$ statistic of
7.740. This corresponds to a one-sided 3-$\sigma$ limit, analogously
to the SINFONI and STIS spectral limits (Section~\ref{sec:met_spe}).

\section{Results}\label{sec:results}
We describe our model of the ejecta dust absorption in
Section~\ref{sec:uvoir_dust}, which is relevant for the results from
the UVOIR observations, but not the other wavelengths. The direct
millimeter, UVOIR, and X-ray limits are then presented in
Sections~\ref{sec:img_lim}--\ref{sec:res_xray}. In
Section~\ref{sec:bol_lim}, we present the bolometric limits, which are
partly dependent on the UVOIR observations. Therefore, we provide
bolometric limits for both dust-free and dust-obscured lines-of-sight.

\subsection{UVOIR Dust Absorption}\label{sec:uvoir_dust}
Effects of absorption by ejecta dust are important in UVOIR. The dust
is assumed to have a negligible impact on the millimeter observations
and have the same absorption properties as gas in the X-ray
regime~\citep{morrison83, draine03, alp18b}. Large amounts of dust
have been observed in \sna{}~\citep{matsuura11, lakicevic11,
  lakicevic12b, indebetouw14, wesson15, matsuura15, dwek15, bevan18}
and there is evidence that the dust resides in clumpy
structures~\citep[][]{lucy89, lucy91, fassia02, jerkstrand11}. The
latter means that the dust is modeled as a covering factor and not an
average optical depth. The diameter of the clumps of molecules have
been observed to be ${\sim}100$~mas~\citep[1000\kmps{} or
$8\times 10^{16}$~cm,][]{abellan17}, but it is possible that the size
of the dust clumps is different from the clumps of molecules. The
covering factor has been estimated to be 50--70\,\% by observing
asymmetries of emission lines and spectral modeling~\citep{lucy89,
  lucy91, wooden93, wang96, fassia02, jerkstrand11}.

The information about the ejecta dust in \sna{} is insufficient for
detailed corrections. In particular, the 3D SN explosion models used
for the X-ray absorption estimate (Appendix~\ref{app:xray_abs}) cannot
be used for the dust because dust formation, composition, and geometry
depends on additional unconstrained
parameters~\citep[e.g.][]{bevan18}. Instead, we assume that our
line-of-sight to the compact object in \sna{} is free of ejecta dust
clumps when placing limits in UVOIR. If there is a dust clump along
our line-of-sight to a compact object, then essentially all UVOIR
emission from the compact object would be absorbed and the presented
direct UVOIR limits would not apply.

However, we consider the reprocessing of the absorbed UVOIR emission
in the dust-obscured case in Section~\ref{sec:bol_lim}, where we
obtain bolometric limits based on the energy budget of the ejecta. We
assume that the dust is optically thick throughout the UVOIR part of
the spectrum because no indications of energy- or time-dependent
attenuation have been observed with VLT/SINFONI (since 2005) or
\textit{HST} \citep[since 1994,][]{larsson13, larsson16}.

\subsection{Millimeter \& UVOIR Image Limits}\label{sec:img_lim}
\begin{figure}
  \centering \includegraphics[width=\columnwidth]{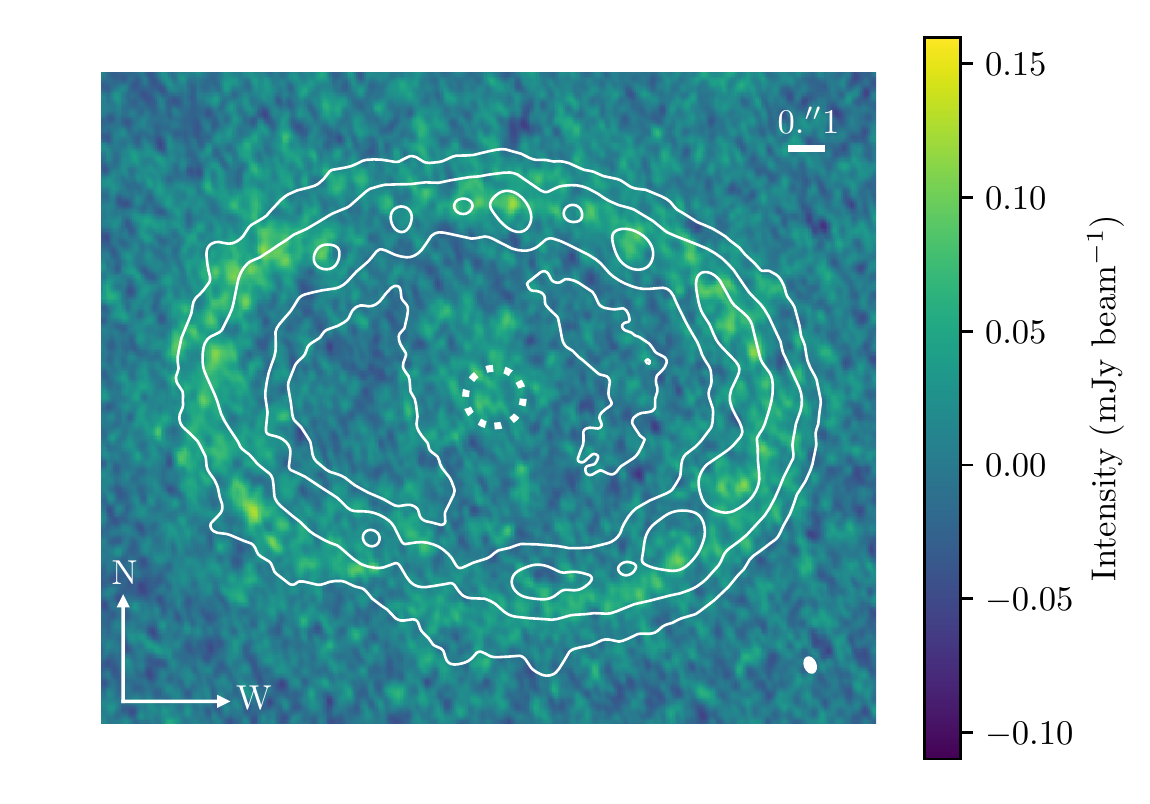}
  \caption{ALMA observation at 213~GHz (colormap) and the
    \textit{HST}/WFC3 observation from 15 June 2014 in the \textit{B}
    band (contours). The dotted white circle is the search region
    (Section~\ref{sec:sea_reg}) and the $57\times40$~mas$^{2}$ beam is
    shown in the lower right corner. The pixel size is $6^2$~mas$^{2}$
    and the off-source RMS noise is
    ${\sim}0.02$~mJy~beam$^{-1}$.\label{fig:alm}}
\end{figure}
\begin{deluxetable}{ccccccccc}
  \tablecaption{Millimeter Limits\label{tab:mm_lim}}
  \tablewidth{0pt}
  \tablehead{\colhead{Frequency} & \colhead{Flux Density} \\
    \colhead{(GHz)} & \colhead{(mJy)}} \startdata
  213 & 0.11 \\
  233 & 0.20 \\
  247 & 0.12 \\
  \enddata
\end{deluxetable}
We compute upper limits on the compact object in \sna{} at millimeter
wavelengths using the ALMA images. An image of the observation at
213~GHz is shown in Figure~\ref{fig:alm} and the limits are provided
in Table~\ref{tab:mm_lim}.  The ER and central ejecta structure are
clearly resolved and no obvious point source is seen in any of the
ALMA images. The level of spatially extended emission in the central
region is comparable to the noise level. The high noise level is a
consequence of constructing images of narrow frequency spans of
1--2~GHz (Section~\ref{sec:obs_alm}).

\begin{figure}
  \centering \includegraphics[width=\columnwidth]{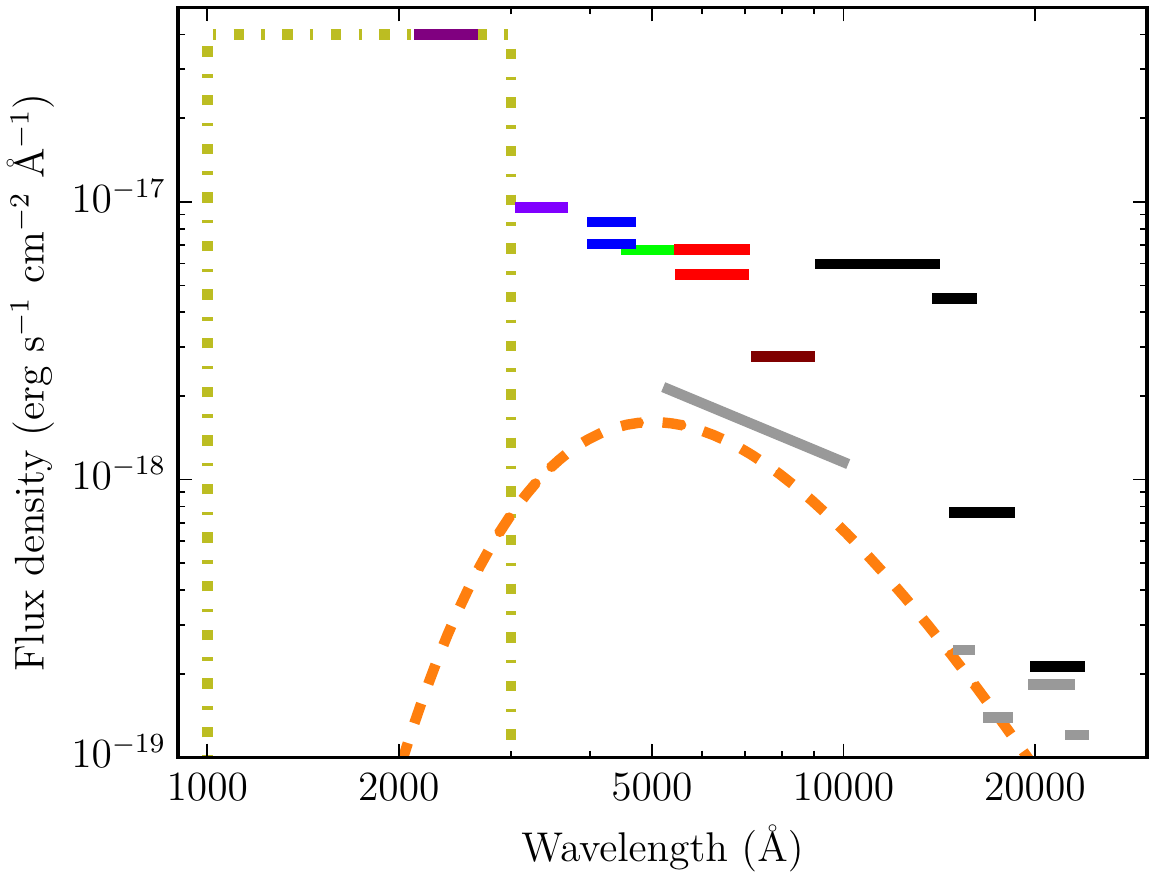}
  \caption{Limits in UVOIR on a point source in \sna{} inside a kick
    region corresponding to 800\kmps{}. Gray lines are spectral
    limits. Image limits from left to right: F225W, F336W, F438W,
    F555W, F625W, F814W, F110W, F160W, NACO \textit{H}, and NACO
    \ks{}. The less strict limits in F438W and F625W are from 2009 and
    are simultaneous with the other optical image limits, whereas the
    stricter F438W and F625W limits are from 2015. Limits from the
    same epoch are from the same spatial position. The dashed orange
    line is a blackbody spectrum for the temperature and radius of the
    Sun scaled to the distance of \sna{}. This can be taken as a limit
    on a surviving main-sequence companion if our line-of-sight is
    free of dust~(Section~\ref{sec:bin_com}). The dash-dotted yellow
    line is the assumed conservative spectrum corresponding to a
    luminosity of 6.6\Lsun{} used for the bolometric limit
    (Section~\ref{sec:bol_lim}).\label{fig:ubvri_lim}}
\end{figure}
\begin{deluxetable*}{lccccccccc}
  \tablecaption{UVOIR Limits\label{tab:res}}
  \tablewidth{0pt}
  \tablehead{\colhead{Instrument} & Filter/Grating & \colhead{Method} & \colhead{Epoch\tablenotemark{a}} &
    \colhead{Wavelength} & \colhead{Flux Density} & \colhead{Luminosity} & \colhead{Luminosity} \\
    \colhead{} & \colhead{} & \colhead{} & \colhead{(YYYY-mm-dd)} & \colhead{(\AA{})} &
    \colhead{(erg~s$^{-1}$~cm$^{-2}$~\AA$^{-1}$)} &
    \colhead{(erg~s$^{-1}$)} &
    \colhead{(L$_\Sun$)}}  \startdata
  WFC3    &      F225W & Spatial                   & 2009-12-13 & \phantom{0}2150--2615\phantom{0} & $4.0\times10^{-17}$ & $5.8\times10^{33}$ & 1.5\phantom{000} \\
  WFC3    &      F336W & Spatial                   & 2009-12-13 & \phantom{0}3102--3617\phantom{0} & $9.5\times10^{-18}$ & $1.5\times10^{33}$ & 0.40\phantom{00} \\
  WFC3    &      F438W & Spatial                   & 2009-12-12 & \phantom{0}4026--4638\phantom{0} & $8.4\times10^{-18}$ & $1.6\times10^{33}$ & 0.42\phantom{00} \\
  WFC3    &      F555W & Spatial                   & 2009-12-13 & \phantom{0}4556--6112\phantom{0} & $6.7\times10^{-18}$ & $3.3\times10^{33}$ & 0.86\phantom{00} \\
  WFC3    &      F625W & Spatial                   & 2009-12-12 & \phantom{0}5525--6991\phantom{0} & $6.7\times10^{-18}$ & $3.1\times10^{33}$ & 0.81\phantom{00} \\
  WFC3    &      F841W & Spatial                   & 2009-12-13 & \phantom{0}7284--8864\phantom{0} & $2.8\times10^{-18}$ & $1.4\times10^{33}$ & 0.36\phantom{00} \\
  WFC3    &      F110W & Spatial                   & 2011-01-05 & \phantom{0}9203--13901           & $6.0\times10^{-18}$ & $8.8\times10^{33}$ & 2.3\phantom{000} \\
  WFC3    &      F160W & Spatial                   & 2011-01-05 &           14027--15925           & $4.5\times10^{-18}$ & $2.7\times10^{33}$ & 0.70\phantom{00} \\
  WFC3    &      F438W & Spatial                   & 2015-05-24 & \phantom{0}4022--4639\phantom{0} & $7.1\times10^{-18}$ & $1.4\times10^{33}$ & 0.36\phantom{00} \\
  WFC3    &      F625W & Spatial                   & 2015-05-24 & \phantom{0}5529--6982\phantom{0} & $5.5\times10^{-18}$ & $2.5\times10^{33}$ & 0.65\phantom{00} \\
  NACO    & \textit{H} & Spatial                   & 2010-10-26 &           14950--18250           & $7.6\times10^{-19}$ & $7.9\times10^{32}$ & 0.21\phantom{00} \\
  NACO    &      \ks{} & Spatial                   & 2012-12-14 &           20050--23550           & $2.1\times10^{-19}$ & $2.3\times10^{32}$ & 0.061\phantom{0} \\
  STIS    &      G750L & Spectral & 2014-08-16 & \phantom{0}5300--10000           & $1.5\times10^{-18}$\tablenotemark{b} & $2.3\times10^{33}$ & 0.59\phantom{00} \\
  SINFONI &  \nodata{} & Spectral                  & 2014-10-10 &           15150--15800           & $2.4\times10^{-19}$ & $5.0\times10^{31}$ & 0.013\phantom{0} \\
  SINFONI &  \nodata{} & Spectral                  & 2014-10-10 &           16900--18125           & $1.4\times10^{-19}$ & $5.3\times10^{31}$ & 0.014\phantom{0} \\
  SINFONI & \textit{H} & Spectral & 2014-10-10 &           15000--18500\tablenotemark{c}           & $1.8\times10^{-19}$ & $2.0\times10^{32}$ & 0.051\phantom{0} \\
  SINFONI &  \nodata{} & Spectral                  & 2014-10-12 &           19875--22725           & $1.8\times10^{-19}$ & $1.6\times10^{32}$ & 0.043\phantom{0} \\
  SINFONI &  \nodata{} & Spectral                  & 2014-10-12 &           22725--23825           & $1.2\times10^{-19}$ & $4.2\times10^{31}$ & 0.011\phantom{0} \\
  SINFONI & \textit{K} & Spectral & 2014-10-12 &           19500--24000\tablenotemark{c}           & $1.7\times10^{-19}$ & $2.3\times10^{32}$ & 0.061\phantom{0} \\
  \enddata
  \tablenotetext{a}{Start of first exposure if multi-day observation.}
  \tablenotetext{b}{Average of the limiting power law.}
  \tablenotetext{c}{Extrapolated slightly outside of, and interpolated
    between fitted intervals.}
\end{deluxetable*}
The upper limits in UVOIR are shown in Figure~\ref{fig:ubvri_lim} and
listed in Table~\ref{tab:res}. There are limits in six filters from
2009; F225W, F336W, F438W, F555W, F625W, and F841W. The more
constraining limits in the \textit{B} and \textit{R} bands are the
2015 observations.  The four NIR limits are: WFC3 F110W and F160W from
2011, and NACO \textit{H} from 2010 and \ks{} from 2012. All limits
from the same epoch are at the same spatial position. In contrast to
the ALMA images, the limits at UVOIR are dominated by the ejecta
emission.

\subsection{UVOIR Spectral Limits}\label{sec:opt_spc_lim}
The SINFONI \textit{H} and \textit{K} spectra from the central region
and limits on the compact object are shown in
Figures~\ref{fig:nir_spc_lim} and~\ref{fig:ubvri_lim}. The limits for
the individual wavelength intervals are listed in
Table~\ref{tab:res}. We note that the four intervals are for
presentation only and are groups consisting of many narrow ranges,
which are used to compute the limits
(Section~\ref{sec:met_nir_spe}). Images of the selected wavelength
intervals that are relatively free of line emission are also studied
and no clear point source is seen in the resolved image of the
ejecta. The spatial distribution of the emission in the central
regions of the ejecta in the continuum images is essentially uniform
at the resolution of SINFONI in both the \textit{H} and \textit{K}
band.

The STIS spectrum for the central region of \sna{} is shown in
Figure~\ref{fig:opt_spc_lim}. The limiting power law is given by 
\begin{equation}
  \label{eq:1}
  F_\lambda = 1.2\times 10^{-18}\,
  \left(\frac{\lambda}{10^4\text{~\AA{}}}\right)^{-0.95}\text{~erg~s$^{-1}$~cm$^{-2}$~\AA$^{-1}$,}
\end{equation}
where $F_\lambda$ is the spectral flux density and $\lambda$ is the
wavelength. The power-law limit is shown in
Figures~\ref{fig:opt_spc_lim} and~\ref{fig:ubvri_lim} and included in
Table~\ref{tab:res}. We note that the limits from spectra are more
constraining than image limits at corresponding frequencies.

\subsection{X-ray limits}\label{sec:res_xray}
\begin{deluxetable*}{lccccccccc}
  \tablecaption{X-ray Limits\label{tab:xray_lim}}
  \tablewidth{0pt}
  \tablehead{
    \colhead{Model} & \colhead{Absorption} & \colhead{$\chi^2$} & \colhead{d.o.f.} & \colhead{$\chi^2$/d.o.f.} & \colhead{$f_s$\tablenotemark{a}} & \colhead{$T$\tablenotemark{b}} & \colhead{$\Gamma$\tablenotemark{c}} & \colhead{$K_\infty{}$\tablenotemark{b}} & \colhead{$L_\infty{}$\tablenotemark{d}} \\
    \colhead{}      & \colhead{} & \colhead{}         & \colhead{}    & \colhead{}             & \colhead{($10^{-2}$)} & \colhead{(MK)} & \colhead{} & \colhead{($10^{-4}$~keV$^{\Gamma{}-1}$~s$^{-1}$~cm$^{-2}$)} & \colhead{(erg~s$^{-1}$)}} \startdata
  Blackbody   &                  No SN abs. &  28.6 &  26 &   1.10 & $5.0\pm0.3$ &       4.4 & \nodata{} & \nodata{} & $1.6\times{}10^{35}$ \\
  Blackbody   & 10$^\mathrm{th}$ percentile &  29.8 &  26 &   1.15 & $5.9\pm0.3$ &       8.2 & \nodata{} & \nodata{} & $1.9\times{}10^{36}$ \\
  Blackbody   &                     Average &  29.4 &  26 &   1.13 & $5.9\pm0.3$ &       8.9 & \nodata{} & \nodata{} & $2.6\times{}10^{36}$ \\
  Blackbody   & 90$^\mathrm{th}$ percentile &  29.1 &  26 &   1.12 & $6.0\pm0.3$ &       9.6 & \nodata{} & \nodata{} & $3.6\times{}10^{36}$ \\
  Carbon atm. &                     Average &  29.0 &  26 &   1.12 & $6.0\pm0.3$ &       7.7 & \nodata{} & \nodata{} & $1.5\times{}10^{36}$ \\
  Power law   &                  No SN abs. &  29.7 &  26 &   1.14 & $5.6\pm0.3$ & \nodata{} &      1.63 &       2.9 & $4.1\times{}10^{34}$ \\  
  Power law   & 10$^\mathrm{th}$ percentile &  29.1 &  26 &   1.12 & $6.0\pm0.3$ & \nodata{} &      1.63 &       2.1 & $3.0\times{}10^{35}$ \\
  Power law   &                     Average &  28.9 &  26 &   1.11 & $6.0\pm0.3$ & \nodata{} &      1.63 &       3.5 & $4.9\times{}10^{35}$ \\
  Power law   & 90$^\mathrm{th}$ percentile &  29.1 &  26 &   1.12 & $6.0\pm0.3$ & \nodata{} &      1.63 &       6.1 & $8.8\times{}10^{35}$ \\
  Power law   &                     Average &  28.9 &  26 &   1.11 & $6.0\pm0.3$ & \nodata{} &     2.108 &       7.7 & $5.3\times{}10^{35}$ \\
  \enddata
  \tablenotetext{a}{The fraction of spread light from the ER at
    the limiting values of the parameter of interest (either $T$ or
    $K_\infty{}$). This can be compared to the predicted value of
    $0.073^{+0.011}_{-0.005}$
    (Section~\ref{sec:obs_sca}). Uncertainties are $1\sigma{}$.}
  \tablenotetext{b}{3-$\sigma$ upper limits.}
  \tablenotetext{c}{Frozen during fits.}
  \tablenotetext{d}{Observed bolometric luminosity for the thermal
    components. Luminosity in the observed 2--10~keV range for the
    power laws.}
\end{deluxetable*}
X-ray limits on the compact object are set using the 17 September 2015
\textit{Chandra}/ACIS observation and the ejecta absorption estimate
from the SN model B15 (Appendix~\ref{app:xray_abs}). We investigate
the standard blackbody model, the XSPEC thermal model
\texttt{nsmaxg}~\citep{mori07, ho08b}, and two power laws with photon
indices ($\Gamma{}$) of 1.63, corresponding to the Crab Pulsar, and
2.108, corresponding to the Crab Nebula~\citep{willingale01}. The
\texttt{nsmaxg} model describes thermal emission from a NS for
different atmospheric compositions and magnetic field strengths. We
only explore the case of a NS with a carbon atmosphere and a magnetic
field strength of $10^{12}$~G. A carbon-atmosphere model was reported
to fit the NS in Cas~A~\citep{ho09, posselt13}. See Figure~2 of
\citet{ho09} for a comparison of different atmospheric
compositions. Apart from the spread light fraction, only the
temperature is allowed to vary for the two thermal components and the
normalization for the two power laws. The remaining parameters are
frozen under the assumptions (discussed in
Section~\ref{sec:thermal_emission}) of a gravitational mass of
1.4\Msun{}, local (unredshifted) NS radius of 10~km, and uniform
emission from the entire surface. The assumed mass and radius give a
gravitational redshift factor of 0.766.

All X-ray limits are listed in Table~\ref{tab:xray_lim}. The presented
effective surface temperatures ($T$) are given in the local
(unredshifted) frame. The parameter ($K_\infty{}$) is the XSPEC power
law normalization and is given in the observed frame (``at
infinity''). The luminosities ($L_\infty{}$) are also given in the
observed frame to facilitate comparisons with other observational
studies. The luminosities for the thermal components are bolometric
whereas the power law luminosities are given for the observed
2--10~keV range.

The limits for the standard blackbody and the $\Gamma = 1.63$ power
law are given for three different levels of SN ejecta absorption. The
selected amounts are the average, 10$^\mathrm{th}$ percentile, and
90$^\mathrm{th}$ percentile of the optical depth. The general trend
for the two thermal components and the two power laws are the same;
the difference between the 10$^\mathrm{th}$ and 90$^\mathrm{th}$
percentiles is a factor of ${\sim}$2 in luminosity for the thermal
components and ${\sim}$3 for the power laws. Absorption by the ISM is
always included, but we also provide limits for no SN ejecta
absorption. This represents the minimum amount of absorption, in case
the SN explosion model describes \sna{} poorly and we happen to have a
very favorable line-of-sight.

The reason for the extremely high limiting temperatures and
luminosities of the thermal models is that the high optical depths at
energies below ${\sim}$4~keV effectively absorb all emission. We point
out that an important factor to the thermal components crossing the
detection threshold is the shift of the emission toward higher
energies where the optical depth is lower.  The power-law models are
more constrained in the sense that the limiting luminosity in the
2--10~keV range is much lower because the power-law components extend
to higher energies.

\subsection{Bolometric Limit}\label{sec:bol_lim}
The bolometric luminosity of the compact object can be constrained by
the total energy budget of \sna{}. The energy inputs are radioactive
decay of $^{44}$Ti and the unknown contribution from the compact
object. The energy outputs are far-infrared (FIR) dust emission and
UVOIR de-excitation and recombination emission lines. Detailed models
of \sna{} predict that much of the emission powered by $^{44}$Ti would
emerge as fine-structure lines in MIR~\citep{jerkstrand11}, but
observations severely constrain these lines~\citep{lundqvist01,
  bouchet06}. This implies that the MIR emission is reprocessed and
escapes as thermal dust emission in the sub-mm to
FIR~\citep{matsuura11, indebetouw14, matsuura15, dwek15}.

The lifetime of $^{44}$Ti is $\tau = 85.0$ years~\citep[half-life of
58.9 years,][]{ahmad06} and decays into $^{44}$Sc, which emits
596\nobreakdash{-}keV positrons when promptly decaying to stable
$^{44}$Ca. All positrons deposit their energy locally under the
assumption of the presence of a weak magnetic
field~\citep{ruiz_lapuente98} and a fraction $f_\mathrm{h}$ of the
energy goes into heating and the rest goes into excitation and
ionization~\citep{kozma92, jerkstrand11}. We assume that the
ionization fraction is slightly higher at ${\sim}10\,000$~days
(current epochs) than at 2875~days, which is what was modeled by
\citet{jerkstrand11}. A higher ionization fraction results in a higher
heating fraction~\citep[Figure~5 of][]{kozma92}.

The emission processes relevant for the compact object are thermal
surface emission, accretion or pulsar wind activity. We assume that
the emission is dominated by X-ray emission below 10~keV, which is
absorbed locally because of the high optical depth~\citep{alp18b} and
escapes as thermal dust emission or UVOIR lines. Both surface emission
and accretion would be observed as unresolved point
sources. \citet{chevalier92} investigated the early evolution of young
pulsars and their effect on the surrounding ejecta and found that the
bubble expansion velocity is ${\sim}500$--800\kmps{} for a pulsar
luminosity of $10^{39}$\ergps{} above 13.6~eV. The current limits on
the compact object constrains the luminosity, and consequently the
expansion velocity, to be orders of magnitude lower. For expansion
velocities less than ${\sim}$100\kmps{}, it is reasonable to treat a
possible pulsar wind nebula (PWN) as a point source. Assuming the
compact object to be point-like allows us to separate the cases where
the line-of-sight is free of dust and dust-obscured, and use the
point-source image limits in the dust-free case.

In the case where our line-of-sight is free of dust, we assume that
70\,\% of the input from the compact object goes into heating and the
remaining 30\,\% escapes as UVOIR emission lines. These fractions are
distinct from those for the positron input, but we assume them to be
the same~\citep{kozma92, jerkstrand11}. We do not consider further
reprocessing of the energy that goes into heating, which most likely
escapes as thermal dust emission~\citep{bouchet06, jerkstrand11}. The
compact object is situated in the central regions where the
photoabsorption is dominated by iron~\citep[Figure~2 of][]{alp18b},
which implies that the line spectrum of the reprocessed emission from
the compact object could be different from that of the
$^{44}$Ti-powered ejecta. Therefore, we choose a conservative limit on
the reprocessed UVOIR emission from the compact object to be
$4\times 10^{-17}$\ergps{} in the range 1000--3000~\AA{}, which is an
extrapolation of the 3\nobreakdash{-}$\sigma{}$ UV (F225W)
\textit{HST} point-source limit (Section~\ref{sec:img_lim} and
Figure~\ref{fig:ubvri_lim}). This spectral shape was chosen because it
results in the least constraining limit. The wavelength range covers
the region where many of the metal lines are expected to
escape~\citep[Figure~3--5 of][]{jerkstrand11} and longer wavelengths
are strongly constrained by the limits (Figure~\ref{fig:ubvri_lim} and
Table~\ref{tab:res}). The flux limit corresponds to an allowed UVOIR
luminosity of 6.6\Lsun{}, which for the assumed heating fraction of
70\,\% results in a bolometric limit of 22\Lsun{}. The epoch of this
limit is December 2009, which is set by the \textit{HST} UV
observation.

The situation is different in the case where our line-of-sight to the
compact object is obscured by dust. In this case, the contribution
from the compact object is added to the contribution to dust heating
from $^{44}$Ti. Out of the fraction $1-f_\mathrm{h}$ of the positron
input that goes into excitation and ionization, a fraction
$f_\mathrm{d}$ is absorbed by dust. The case is simpler for the
electromagnetic input from the compact object, all of which goes into
dust heating in the dust-obscured case. This implicitly assumes
spherical symmetry and means that the total dust luminosity is
expected to be
\begin{equation}
  \label{eq:Ldust}
  L_\mathrm d = \left[f_\mathrm{h} + f_\mathrm{d}\,
    (1-f_\mathrm{h})\right]\,L_\mathrm{Ti} + L_\bullet{},
\end{equation}
where $L_\mathrm{Ti}$ is the $^{44}$Ti positron decay luminosity and
$L_\bullet{}$ is the contribution from the compact object.

A limit on $L_\bullet{}$ can now be determined. The values of the
other parameters are taken to be
$L_\mathrm d = 295 \pm 17$\Lsun{}~\citep[][weighted values from 2010
and 2012, and scaled to 51.2~kpc]{matsuura15, dwek15},
$L_\mathrm{Ti} = 298 \pm 36$\Lsun{}~\citep[initial $^{44}$Ti mass of
$1.6\times 10^{-4}$\Msun{} for 51.2~kpc,][]{jerkstrand11, boggs15},
$f_\mathrm h = 0.55$--0.85~\citep{kozma92, jerkstrand11}, and
$f_\mathrm d = 0.5$--0.7~\citep[][see also
Section~\ref{sec:uvoir_dust}]{lucy89, lucy91, wooden93, wang96,
  fassia02, jerkstrand11}. The values are scaled to 9090~days after
explosion (January 2012), which is the time of the dedicated
\textit{Herschel} observations of the dust
luminosity~\citep{matsuura15}. The distributions are assumed to be
Gaussian and confidence intervals are \mbox{1-$\sigma$} except for the
fractions, which are assumed to be uniformly distributed within the
intervals. This is clearly a very primitive model but the uncertainty
in the Ti-mass estimate is the largest source of uncertainty and an
improvement in the determination of $L_\mathrm{Ti}$ in the near future
is unlikely. Therefore, a more detailed model of the energy budget
would not improve the limit on $L_\bullet{}$ by much. Following the
above reasoning, the estimate of the compact object luminosity is
$L_\bullet{} = 33^{+37}_{-38}$\Lsun{}, which shows that an additional
contribution from the compact object is not statistically significant.
The 3-$\sigma$ upper limit is $L_\bullet{} < 138$\Lsun{}.

Some simplifications have implicitly been made. It is possible that
some fraction of the energy emitted by the compact object escapes the
remnant before being reprocessed into observable wavebands, for
example in MIR~\citep[Section~\ref{sec:the_sur_emi},][]{bouchet06,
  bouchet14} or as high-energy gamma rays
(Section~\ref{sec:fav_exp}). The X-ray emission from the ring provides
an additional energy source for the ejecta~\citep{larsson11}, but this
primarily affects the outer H and He envelope~\citep{fransson13,
  larsson13}.

\section{Discussion}\label{sec:discussion}
All limits on the compact object in \sna{} presented in
Section~\ref{sec:results} apply to both NSs and BHs. However, the
expected emission characteristics for the two classes of objects are
very different.  This discussion primarily focuses on NSs because most
studies favor the creation of a NS in \sna{} (Section~\ref{sec:intro})
and NSs power a wider diversity of physical processes. In contrast,
BHs primarily reveal themselves through accretion, which is explored
in Section~\ref{sec:accretion} and more comprehensively in
\citet{graves05}.

\begin{deluxetable*}{lccccccccc}
  \tablecaption{Model-Dependent Constraints on Physical Parameters\label{tab:lim_par}}
  \tablewidth{0pt}
  \tablehead{
     \colhead{Model} & \colhead{Method\tablenotemark{a}} & \colhead{Observation\tablenotemark{b}} & \colhead{Dust\tablenotemark{c}} &
     \colhead{Spectrum} & \colhead{Lum.} & \colhead{Limit
       on physical parameter} \\
          \colhead{} & \colhead{} & \colhead{} & \colhead{}       &         \colhead{} &  \colhead{(L$_{\sun}$)} &     \colhead{}} \startdata
        NS Surface &               DO & \textit{Chandra}      & \nodata{} &           Thermal & 450 &   $T < 8$~MK   ($T_\infty < 6$~MK) \\
        NS Surface &               EB & \textit{HST} F225W    &         N &           Thermal &  22 & $T < 3.8$~MK ($T_\infty < 2.9$~MK) \\
        NS Surface &               EB & \textit{Herschel}     &         Y &           Thermal & 138 & $T < 5.9$~MK ($T_\infty < 4.5$~MK) \\
        Accretion  &               DO & \textit{Chandra}      & \nodata{} &   X-ray dominated & 300 & $\dot M < 2.0\times10^{-11}\,\eta^{-1}\text{\Msun{}~yr}^{-1}$ \\
        Accretion  &               EB & \textit{HST} F225W    &         N &   X-ray dominated &  22 & $\dot M < 1.5\times10^{-12}\,\eta^{-1}\text{\Msun{}~yr}^{-1}$ \\
        Accretion  &               EB & \textit{Herschel}     &         Y &   X-ray dominated & 138 & $\dot M < 9.2\times10^{-12}\,\eta^{-1}\text{\Msun{}~yr}^{-1}$ \\
        PWN        &               DO & SINFONI 1.7~\micron{} &         N &       Crab Nebula &   3 & $B < 1.8\times 10^{13}\,P^2$~G~s$^{-2}$ \\
        PWN        &               DO & ALMA 213~GHz          & \nodata{} &       Crab Nebula &  10 & $B < 3.2\times 10^{13}\,P^2$~G~s$^{-2}$ \\
        PWN        &               DO & SINFONI 1.7~\micron{} &         N &       Crab Pulsar & 528 & $B < 2.3\times 10^{14}\,P^2$~G~s$^{-2}$ \\
        PWN        &               DO & \textit{Chandra}      & \nodata{} &       Crab Pulsar & 830 & $B < 2.9\times 10^{14}\,P^2$~G~s$^{-2}$ \\
        PWN        &               EB & \textit{HST} F225W    &         N &   X-ray dominated &  22 & $B < 4.7\times 10^{13}\,P^2$~G~s$^{-2}$ \\
        PWN        &               EB & \textit{Herschel}     &         Y &   X-ray dominated & 138 & $B < 1.2\times 10^{14}\,P^2$~G~s$^{-2}$ \\
  \enddata
  \tablenotetext{a}{Either constrained by direct observations (DO) or by
   the energy budget (EB).}
  \tablenotetext{b}{Which observation that is
    constraining. \textit{Herschel} observations~\citep{matsuura15,
      dwek15} are used for the bolometric limit (Section~\ref{sec:bol_lim}).}
  \tablenotetext{c}{Ejecta dust along the line-of-sight; Yes, No, or blank if limit is insensitive to dust.}
\end{deluxetable*}
Table~\ref{tab:lim_par} summarizes all limits on physical parameters.
The combination of all available information favors that the compact
object is a dust-obscured thermally-emitting neutron star
(Section~\ref{sec:rem_pos}). The discussion is organized as
follows. We compile literature limits for a comprehensive overview of
\sna{} observations across the entire electromagnetic spectrum in
Section~\ref{sec:glo_lim} and compare our limits with previous works
in Section~\ref{sec:comp}. Implications of the limits based on direct
observations for thermal surface emission are discussed in
Section~\ref{sec:thermal_emission}, accretion in
Section~\ref{sec:accretion}, and pulsar activity in
Section~\ref{sec:BP}. We relate the bolometric limit to physical
parameters in Section~\ref{sec:energy_budget} and extrapolate the
limits to other epochs using simple models in Section~\ref{sec:birth}.
Lastly, we explore constraints on a possible binary main-sequence
companion in Section~\ref{sec:bin_com}, remaining possibilities for
the compact object in Section~\ref{sec:rem_pos}, and briefly look into
future prospects in Section~\ref{sec:future}.

\subsection{Global Limits}\label{sec:glo_lim}
\begin{figure*}
  \centering \includegraphics[width=\textwidth]{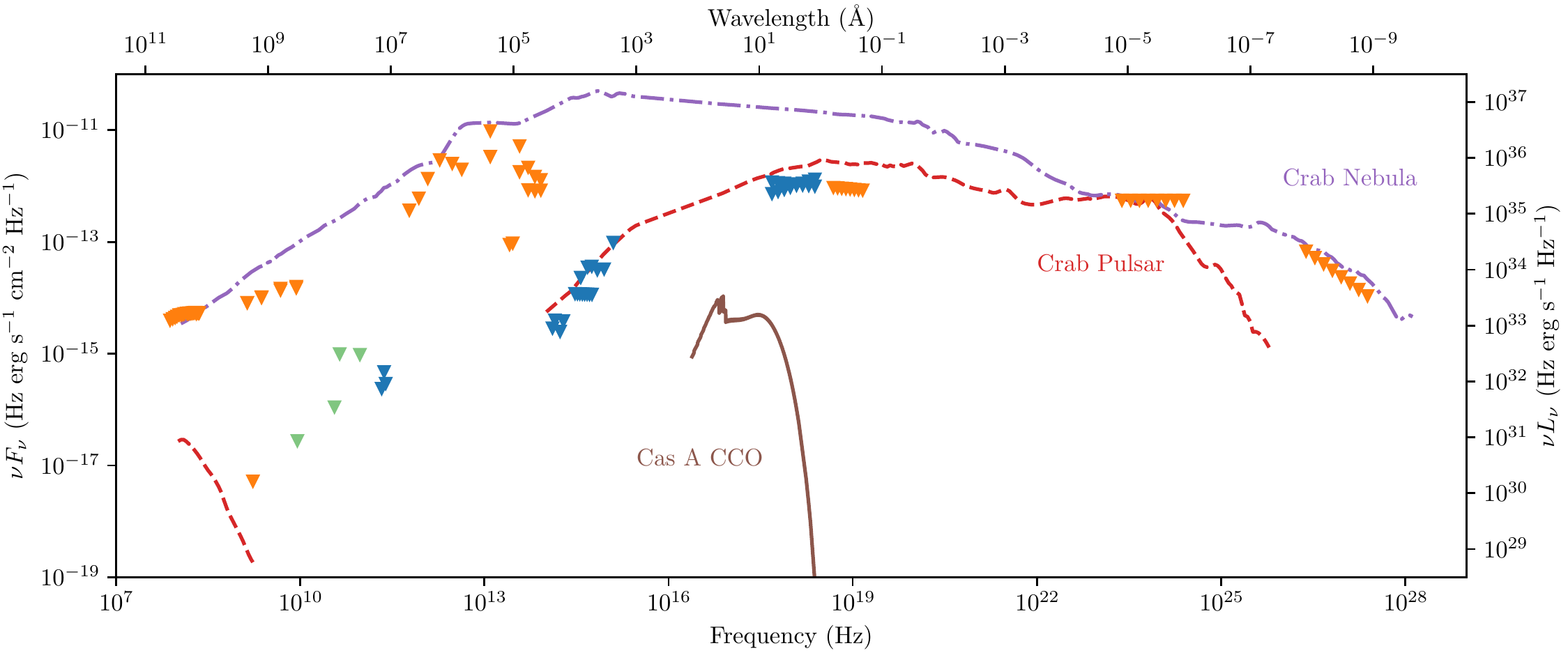}
  \caption{Limits (triangles) on the compact object in \sna{}, shown together with
    the spectra of the Crab Nebula~\citep[dash-dotted purple,][and
    references therein]{buhler14}, the Crab Pulsar~\citep[dashed
    red,][and references therein]{buhler14}, and the CCO in Cas A
    modeled as a NS with a carbon atmosphere~\citep[solid
    brown,][]{posselt13}. The spectra are scaled to the distance of
    \sna{}. Blue limits are presented in this work, and orange and
    green limits are literature limits. The green limits are
    super-resolved using deconvolution algorithms and all orange
    limits are unresolved except for the VLBI observation at 1.7~GHz
    and the Gemini/T-ReCS observations at
    10~\micron{}.\label{fig:glo_lim}}
\end{figure*}
\begin{longrotatetable}
\begin{deluxetable*}{rrrcccl}
  \tablecaption{Literature Limits on \sna{}\label{tab:lit_lim}}
  \tablewidth{0pt}
  \tablehead{\colhead{Frequency} &
    \colhead{Flux Density} &
    \colhead{Epoch} &
    \colhead{Instr.\tablenotemark{a}} &
    \colhead{Resolution\tablenotemark{b}} &
    \colhead{Conf. level\tablenotemark{c}} &
    \colhead{Reference} \\
    \colhead{(Hz)} &
    \colhead{(erg~s$^{-1}$~cm$^{-2}$~Hz$^{-1}$)} &
    \colhead{([YYYY--]YYYY)} &
    \colhead{} &
    \colhead{} &
    \colhead{} &
    \colhead{}} \startdata
  0.076--8.642$\,\times 10^{9}$ & 5.1--$0.17\times10^{-23}$ & 2013--2014                  &         MWA, ATCA & U &  \nodata{} & \citet{callingham16} \\
           1.7$\,\times 10^{9}$ &       $0.3\times10^{-26}$ &       2008                  &              VLBI & R & 3-$\sigma$ & \citet{ng11} \\
             9$\,\times 10^{9}$ &       $0.3\times10^{-26}$ &       1996\tablenotemark{d} &              ATCA & S & 3-$\sigma$ & \citet{ng08} \\
          36.2$\,\times 10^{9}$ & $0.3\pm0.2\times10^{-26}$ &       2008                  &              ATCA & S &          E & \citet{potter09} \\
            44$\,\times 10^{9}$ &       $2.2\times10^{-26}$ &       2011                  &              ATCA & S &          E & \citet{zanardo13} \\
            94$\,\times 10^{9}$ &         $1\times10^{-26}$ &       2011                  &              ATCA & S & 2-$\sigma$ & \citet{lakicevic12} \\
     0.6--4.3$\,\times 10^{12}$ &   50--$150\times10^{-26}$ &       2012                  &       SPIRE, PACS & U &  \nodata{} & \citet{matsuura15} \\
       12--83$\,\times 10^{12}$ &   1.0--$76\times10^{-26}$ & 2003--2015                  &        MIPS, IRAC & U &  \nodata{} & \citet{arendt16} \\
           26$\,\times 10^{12}$ &      $0.34\times10^{-26}$ &       2005                  &            T-ReCS & R & 3-$\sigma$ & \citet{bouchet06} \\
           29$\,\times 10^{12}$ &      $0.32\times10^{-26}$ &       2003                  &            T-ReCS & R &          E\tablenotemark{e} & \citet{bouchet04} \\
     0.5--1.5$\,\times 10^{19}$ &  1.9--$0.6\times10^{-31}$ & 2010--2011                  &              IBIS & U &  \nodata{} & \citet{grebenev12} \\
    2.4--24.2$\,\times 10^{23}$ & 2.2--$0.22\times10^{-36}$ & 2008--2014                  &               LAT & U &  \nodata{} & \citet{ackermann16} \\ 
    2.4--24.2$\,\times 10^{26}$ & 2.8--$0.04\times10^{-40}$ & 2003--2012                  &              HESS & U &  \nodata{} & \citet{hess15} \\ 
  \enddata
  \tablenotetext{a}{The abbreviations and acronyms are: Murchison
    Widefield Array (MWA); Australia Telescope Compact Array (ATCA);
    Very Long Baseline Interferometry (VLBI) using ATCA, Parkes,
    Mopra, and the NASA DSS 43 antenna at Tidbinbilla; Spectral and
    Photometric Imaging Receiver (SPIRE) and Photodetector Array
    Camera and Spectrometer (PACS) onboard \textit{Herschel};
    Multiband Imaging Photometer (MIPS) and Infrared Array Camera
    (IRAC) onboard \textit{Spitzer}; Thermal Region Camera and
    Spectrograph (T-ReCS) attached to the Gemini South 8~m Telescope;
    Imager onboard the INTEGRAL Satellite (IBIS) onboard
    \textit{INTEGRAL}; The Large Area Telescope (LAT) onboard
    \textit{Fermi}; and the High Energy Stereoscopic System (HESS).}
  
  \tablenotetext{b}{Either unresolved (U) meaning that the flux
    densities are given for the ER and ejecta combined, 
    super-resolved (S) meaning that images are restored using a
    deconvolution algorithm, or resolved (R) meaning that the ER and
    central ejecta structure are spatially resolved.}

  \tablenotetext{c}{Flux densities presented as estimates of a point source
    rather than upper limits are denoted by ``E''. This is left blank
    for
    values presented as measured fluxes of the ejecta and ER combined.}
  
  \tablenotetext{d}{Limits from other epochs are very similar.}

  \tablenotetext{e}{This was reported as potential dust emission.}
\end{deluxetable*}
\end{longrotatetable}
Limits on a point source collected from the literature are included to
give a complete coverage over the entire electromagnetic spectrum. We
only include literature limits at frequencies not covered by this
work. Limits covering the same bands as our limits are instead
discussed in Section~\ref{sec:comp}. An overview of limits at all
frequencies is shown in Figure~\ref{fig:glo_lim}. More details on the
literature limits are provided in Table~\ref{tab:lit_lim}. We
categorize the limits based on the methods employed. Unresolved
imaging in this context implies that the ER and ejecta are not
spatially resolved. These limits are just the total flux of the ER and
ejecta combined, resulting in very conservative limits. The unresolved
radio limit is dominated by ejecta-ER
interactions~\citep{callingham16}, unresolved IR limits by thermal
dust emission~\citep{matsuura15, arendt16}, unresolved X-ray limit
most likely by ejecta-ER interactions~\citep{grebenev12}, and
gamma-ray limits most likely by spread light from the nearby objects
N~157B and 30~Dor~C~\citep{hess15, ackermann16}.  Super-resolved
implies that images are restored using a deconvolution algorithm,
which introduces additional assumptions and is model-dependent in some
cases. This is especially unreliable because the compact object is
surrounded by bright ejecta~\citep[e.g.][]{white94}. Resolved images
clearly distinguish the ER from the central ejecta and are the most
robust measurements.  The VLT/SINFONI, \textit{HST}/STIS, and
\textit{Chandra}/ACIS limits from this work are based on spectra,
whereas all other limits are determined using images.

The radio limits of \citet{potter09} and \citet{zanardo13} in
Table~\ref{tab:lit_lim} are referred to as estimates. These are excess
sources inside the ER that were interpreted as possible indications of
a pulsar. However, the evidence remains inconclusive and we are not
able to compare our limits with their observations.

\subsection{Comparison with Previous Limits}\label{sec:comp}
Earlier studies have presented limits on the compact object in \sna{}
in (sub-)mm, optical, UV, and X-rays. \citet{zanardo14} discussed the
possibility of a PWN with a flux of 3~mJy in the range 102~GHz to
672~GHz. This is not directly comparable to our limits of
${\sim}$0.1~mJy at 213--247~GHz because our limits apply to point
sources and a PWN might be spatially extended (cf.\
Figure~\ref{fig:alm}).

\citet{graves05} placed limits in optical and UV using data from
\textit{HST}. The image limits from \citet{graves05} are lower than
ours by a factor of ${\sim}2$, but the STIS limit presented in this
work is ${\sim}30\,\%$ more constraining than any previous limit in
the same wavelength range. There are numerous factors that contribute
to the differences. Their STIS spectrum is from December 1999 and
images are from November 2003 taken by the Advanced Camera for Surveys
(ACS), which has a higher angular resolution than WFC3. Our
observations are from later epochs, which implies that the ejecta have
expanded significantly, the shock interactions with the ER have
evolved, the ER X-ray illumination has increased~\citep{larsson11},
and our search region needs to be larger. Additionally, slightly
different values for the reddening and equivalent widths of the
filters are used, as well as a different search algorithm
(Appendix~\ref{app:find}). We verified that the combined effect of all
factors explains the differences between our limits and those of
\citet{graves05}.

Many authors have presented upper limits on the X-ray luminosity of
the compact object in \sna{} using observations from \textit{Chandra}
and \textit{XMM-Newton}~\citep{burrows00, park02, park04,
  shtykovskiy05, haberl06, ng09, orlando15, frank16,
  esposito18}. Reported luminosity limits are in the range
0.3--$60\times 10^{34}$\ergps{} for different instruments, methods,
assumed spectra, and energy ranges (often 2--10~keV).  This should be
compared to our limit of $4\times 10^{34}$\ergps{} for the
$\Gamma = 1.63$ power-law model without ejecta absorption
(Table~\ref{tab:xray_lim}). However, most of our X-ray limits are
approximately an order of magnitude less constraining than previous
X-ray limits because we use a more realistic model of the soft X-ray
photoabsorption of the SN ejecta based on 3D neutrino-driven SN
explosion models~\citep{alp18b}. We also employ a method
(Section~\ref{sec:cha_lim}) that uses the angular resolution of
\textit{Chandra}/ACIS in conjunction with its spectral resolution.

\subsection{Model Comparisons}
\subsubsection{Thermal Emission}\label{sec:thermal_emission}
The direct limits do not strongly constrain the surface temperature of
a NS (Tables~\ref{tab:xray_lim} and~\ref{tab:lim_par}). The remainder
of this section provides the information needed to draw this
conclusion. We note that more constraining limits are obtained from
the bolometric limits (Sections~\ref{sec:energy_budget}
and~\ref{sec:rem_pos}).
  
To relate the surface temperature of a NS to observed luminosity, it
is necessary to adopt a mass and radius. We assume a gravitational
mass of 1.4\Msun{} and a local (unredshifted) radius of 10~km for a NS
in \sna{}. Recent best estimates based on Bayesian analyses of
low-mass X-ray binaries~\citep{steiner13}, nuclear physics and
observational constraints on the neutron-star equation of
state~\citep{hebeler13}, and the binary NS merger
GW170817~\citep{bauswein17} favor radii in the range 11--13~km, see
also Figure~10 of \citet{ozel16}. The primary reason for choosing a
radius of 10~km is that the limits are more conservative because all
reported temperature limits decrease for an increasing NS radius and
fixed mass. The decrease in limiting temperature for a radius of 12~km
is approximately a factor of 0.89 for $T$ and 0.94 for $T_\infty{}$,
because of the different dependencies on the gravitational redshift
factor.

Typical surface temperatures of young NSs are of the order of a few
million Kelvin and correspond to spectral peaks at soft X-ray energies
of ${\sim}$1~keV. The characteristic temperature for a given NS age
depends on the relatively unknown cooling properties of
NSs~\citep{yakovlev04}. For a NS at 30 years that has not undergone
thermal relaxation, a typical temperature is
$T\approx{}3.3$~MK~\citep[$T_\infty{}\approx{}2.5$~MK,][]{gnedin01,
  shternin08, page09, klochkov15}. A more extreme case is for a NS
with a carbon heat blanket. Carbon is more heat transparent and gives
$T\approx{}4.1$~MK~\citep[$T_\infty{}\approx{}3.2$~MK,][]{yakovlev11,
  klochkov15}. These values are at the high end of temperatures
predicted for a NS in \sna{}. If thermal relaxation has started, the
temperature would be decreasing quickly at current
epochs~\citep{gnedin01, yakovlev04, shternin08, page09}.

Limits on thermal emission from a NS in \sna{} based on the X-ray
observation are provided in Tables~\ref{tab:xray_lim}
and~\ref{tab:lim_par}. The X-ray limits constrain thermal spectra much
more strictly than the UVOIR limits. The limiting temperatures are
approximately 8~MK for all expected levels of ejecta absorption and NS
atmospheres. This is clearly above the predicted values of
$T\lesssim 4$~MK, implying that the direct observations do not exclude
any models. Thus, a scenario where \sna{} created a central compact
object~\citep[e.g.][]{posselt13, bogdanov14} is consistent with the
observational limits.

\subsubsection{Accretion}\label{sec:accretion}
It is possible that the compact object is accreting a significant
amount of matter. An extensive study of many different accretion
scenarios was made by \citet[their Sections~5 and~6]{graves05}, to
which the reader is referred for a comprehensive analysis of accretion
in \sna{}. We restrict our discussion of accretion to the simplest
model with the purpose of estimating the luminosity, and find that
most predictions for fallback are excluded unless the accretion
efficiency is less than $0.03$ (cf.\ Section~\ref{sec:birth}).

The simplest accretion model is to assume that a significant amount of
the gravitational binding energy of the infalling material is
converted into radiation. The accretion luminosity ($L_\mathrm{a}$) is
then given by
\begin{equation}
  \begin{split}
    \label{eq:La1}
    L_\mathrm{a} &= \eta \dot M c^2 \\
    & = 5.7\times 10^{46}\,\eta\left(\frac{\dot
        M}{\mathrm{M_{\sun{}}~yr}^{-1}}\right)\!\text{~erg~s}^{-1}.
  \end{split}
\end{equation}
where $\eta$ is the accretion efficiency, $\dot M$ the accretion rate,
and $c$ the speed of light in vacuum. A typical accretion efficiency
is $\eta \approx 0.1$ for a NS of mass 1.4\Msun{} and radius 10~km,
assuming the accreted gas radiatively cools
efficiently~\citep[e.g.][]{mccray79}. The efficiency of accretion onto
BHs is more model-dependent. Possible values of black-hole accretion
efficiencies range from $10^{-10}$ for spherically symmetric
accretion~\citep{shapiro73} to 0.4 for disk accretion~\citep{frank02}.
For reference, the Eddington luminosity ($L_\text{Edd}$) for an object
of mass 1.4\Msun{} is $1.8\times 10^{38}$\ergps{}, which corresponds
to an Eddington accretion rate ($\dot{\mathrm M}_\mathrm{Edd}$) of
$3.1\times 10^{-9}\,\eta^{-1}$\Msun{}~yr$^{-1}$. This relies on some
standard assumptions that are inapplicable in this case, but we choose
to use the Eddington luminosity as a unit for comparison with other
works.

The amount of fallback onto NSs after SN explosions has been estimated
to $\lesssim 0.1$\Msun{}, which mostly accrete on timescales of
$\lesssim 1$~year~\citep{chevalier89, houck91, brown94,
  chatterjee00}. \citet{brown94} estimated that a mass of
$10^{-4}$--$10^{-3}$\Msun{} remains bound to the NS in \sna{} after
${\sim}$3~years. Relevant timescales for accretion of this remaining
mass is $\gtrsim 1000$~years, and it is possible that most of the
remaining mass is expelled~\citep{houck91, chatterjee00}. As an
example, we assume a relatively conservative fallback mass of
$10^{-5}$\Msun{} that is uniformly accreted over $10^4$~yr.  This
results in an accretion rate of $10^{-9}$\Msun{}~yr$^{-1}$, which
corresponds to $L_\mathrm{a} = 6\times 10^{37}\,\eta$~erg~s$^{-1}$
($\approx 10^4\,\eta$\Lsun{}, Equation~\ref{eq:La1}).

The X-ray limits are 0.04--$3.6\times 10^{36}$\ergps{}
(Table~\ref{tab:xray_lim}) at 10433 days (September 2015). Given that
the spectrum is not known,
$\dot M < 2\times 10^{-11}\,\eta^{-1}$\Msun{}~yr$^{-1}$
($\approx
10^{36}$~erg~s$^{-1} \approx 6\times10^{-3}\text{~}\dot{\mathrm
  M}_\mathrm{Edd}$) can be taken as a limit on the current accretion
rate in \sna{} based on the X-ray observation. This is only consistent
with the prediction of $6\times 10^{37}\,\eta$~erg~s$^{-1}$ if $\eta$
is $<0.03$. The discrepancy between models and observations is even
clearer if the temporal evolution of the accretion rate is considered
(Section~\ref{sec:birth}).

\subsubsection{Magnetic Field and Rotation}\label{sec:BP}
We constrain the surface magnetic field strength ($B$) and rotational
period ($P$) of a NS in \sna{} using a simple model of a
rotation-powered PWN and assuming spectra in the form of the Crab
Nebula and Pulsar. An analogue of the Crab Nebula or the Crab Pulsar
in \sna{} is ruled out even if our line-of-sight is dust obscured
(Table~\ref{tab:lim_par}).

The total luminosity of a NS can be modeled by a rotating magnetic
dipole in vacuum~\citep[e.g.\ Equation~10.5.4 of][]{shapiro83}
\begin{equation}
  \begin{split}
    \label{eq:energy_rpp}
    L & = \frac{2^5\,\pi^4\,B^2\,R^6}{3\,c^3\,P^4} \\
    & = 3.9\times 10^{31}\left(\frac{B}{10^{12}\text{~G}}\right)^{2}\!\left(\frac{P}{\text{s}}\right)^{-4}\!\left(\frac{R}{\text{10~km}}\right)^{6}\text{~erg~s}^{-1}.
  \end{split}
\end{equation}
Several assumptions have implicitly been made; the rotation axis and
the magnetic dipole axis are orthogonal and the dominating source of
energy is the rotational energy of the NS. We follow the convention of
letting the ``surface'' magnetic field strength be the magnetic
equator field strength and using the vacuum formula~\citep[e.g.\ the
ATNF Pulsar Catalogue,][]{manchester05}. The magnetic field strength
at the magnetic poles is a factor of 2 higher. We also note that the
force-free magnetic dipole formula would imply 1.7 times lower
magnetic field strengths than the vacuum formula for orthogonal
rotation and magnetic axes~\citep{spitkovsky06}. In the simplest
picture, the energy emitted by the NS is deposited into the
surroundings where it is reprocessed. We do not attempt to model the
complex interactions that generate the observed spectrum of a PWN.
Instead, we assume that the deposited energy emerges with the spectrum
of the Crab Nebula or the Crab Pulsar, and study these two cases
separately. The spectrum of the pulsar is the pulsed component, which
corresponds to radiation originating predominantly from the immediate
surroundings of the pulsar and to a lesser extent from the pulsar
itself. The Crab is one of the most extreme sources in the sky and it
is not a typical PWN. For a high-energy comparison of the pulsed
emission, see Figure~28 of \citet{kuiper15}, and for a multiwavelength
comparison of the Crab and PSR~B0540\nobreakdash{-}69.3, see Figure~15
of \citet{serafimovich04}. However, the Crab is relatively young,
well-observed, and frequently used as a reference. The physical
scenario is that a NS created by \sna{} is a pulsar that drives a
younger and smaller analogue of the Crab in the SN remnant.

\begin{figure}
  \centering \includegraphics[width=\columnwidth]{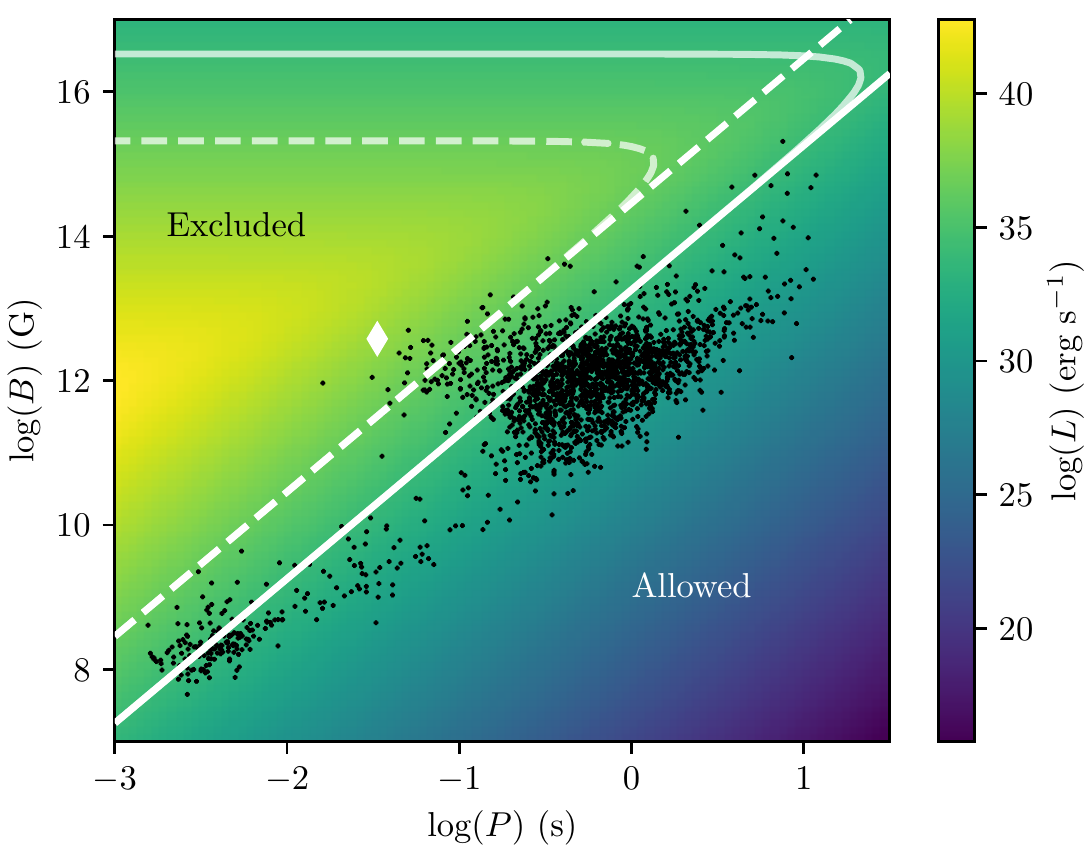}
  \caption{Limits on $B$ and $P$ of a NS in \sna{}, together with the
    pulsars (black points) in the ATNF Pulsar
    Catalogue~\citep{manchester05}, which are included for reference
    and do not represent the expected properties of the compact object
    in \sna{}. The large white diamond is the Crab Pulsar. The solid
    line is the limit from direct observations for a Crab Nebula
    spectrum in the dust-free case and the dotted line is for a Crab
    Pulsar spectrum obscured by dust. These limits are chosen because
    they are the two extremes, implying that all other $BP$-plane
    limits are covered within the range. The shaded extensions that
    are partially horizontal correspond to the respective lines but
    for constraints on birth values of $B$ and $P$ for an assumed
    temporal evolution (see Section~\ref{sec:birth}). The colormap
    represents the current luminosity of a rotating dipole in vacuum
    as modeled by Equation~\eqref{eq:energy_rpp} as a function of
    birth values of $B$ and $P$.\label{fig:BP}}
\end{figure}
With Equation~\eqref{eq:energy_rpp} and a spectral shape, we can
compute the region in the $BP$-plane that is allowed by the upper
limits. This is done by taking the Crab Nebula spectrum and the Crab
Pulsar spectrum, scaling to the distance of \sna{}, and then scaling
the spectra such that they are consistent with the limits (cf.\
Figure~\ref{fig:glo_lim}). If our line-of-sight is free of dust, both
spectra are constrained by the SINFONI data point at 1.7~\micron{},
else the nebula spectrum is constrained by the 213\nobreakdash{-}GHz
limit and the pulsar spectrum by the $\Gamma = 1.63$ X-ray power-law
limit. We only use the new limits presented in this work and we assume
that a PWN is point-like~\citep[Section~\ref{sec:bol_lim},
and][]{chevalier92}. The allowed luminosities vary from 3\Lsun{},
corresponding to $8.2\times 10^{-5}$ of the Crab Nebula, to 830\Lsun,
corresponding to 0.42 of the Crab Pulsar
(Table~\ref{tab:lim_par}). The luminosities can be translated to
limits in the $BP$-plane by rearranging
Equation~\eqref{eq:energy_rpp}. The local (unredshifted) radius is
taken to be 10~km. The limits span a range from
$B < 1.8\times 10^{13}\,P^2$~G~s$^{-2}$ to
$B<2.9\times 10^{14}\,P^2$~G~s$^{-2}$, and are included in
Table~\ref{tab:lim_par} and shown in Figure~\ref{fig:BP}. The limits
are all relatively low and any possible PWN activity is most likely
very weak.

\subsection{Energy Budget}\label{sec:energy_budget}
The bolometric limit on the compact object in \sna{} is 22\Lsun{} at
8329 days (December 2009) if our line-of-sight is free of dust and
138\Lsun{} at 9090 days (January 2012) in the dust-obscured case. The
limits rely on many assumptions~(Section~\ref{sec:bol_lim}) and the
direct observations are arguably less model-dependent. The
direct-observation limits and the bolometric limits can be viewed as
independent limits on the compact object in \sna{}. All constraints on
physical parameters are provided in Table~\ref{tab:lim_par}. Only the
bolometric limits on the effective temperature are substantially more
constraining than the corresponding direct-observation limits.

In the dust-free case, the bolometric limit constrains the blackbody
temperature to $T < 3.8$~MK ($T_\infty < 2.9$~MK) for a NS radius of
10~km and mass of 1.4\Msun{}. The corresponding value for the
dust-obscured case is $T < 5.9$~MK ($T_\infty < 4.5$~MK). This is much
stronger than the constraints based on direct X-ray observations,
which are approximately $T < 8.5$~MK ($T_\infty < 6.5$~MK,
Table~\ref{tab:xray_lim}). Additionally, the bolometric limit on the
effective blackbody temperature is independent of the composition of
the atmosphere of the NS. Interestingly, the dust-free limit of 3.8~MK
is close to some theoretical predictions
(Section~\ref{sec:thermal_emission}). Given that the limit is
conservative, it can be taken as an indication that the compact object
is obscured by dust if it is a NS.

The constraints on the accretion rate and pulsar activity from the
bolometric limits were obtained using Equations~\eqref{eq:La1}
and~\eqref{eq:energy_rpp}. The results are within a factor of 3 to
those of the direct limits, as summarized in Table~\ref{tab:lim_par}.

\subsection{Implications for Other Epochs}\label{sec:birth}
In the above discussion, we have focused on constraints on physical
properties at the times of observation. Here, we briefly explore
extrapolations of the limits to other epochs. In the case of thermal
emission from a NS, the surface temperature is expected to be
relatively constant from a year after explosion to current
epochs~\citep{shternin08}, which implies that the current limits apply
to earlier times as well. However, if thermal relaxation has occurred,
then it is possible that the surface temperature has been higher than
our current limits.

Accretion and pulsar properties are expected to have evolved, which
implies that the current limits need to be modified if extrapolated to
other epochs. These extrapolations are uncertain and rely on models of
how accretion and pulsar properties evolve over time. For accretion,
the basic picture is a period of rapid fallback followed by a
declining tail with a time dependence of $t^{-5/3}$. This dependence
can be derived from simple arguments for marginally bound
gas~\citep{rees88, phinney89, evans89}. This means that the limit at
current epochs of around $10^{-11}\,\eta^{-1}\text{\Msun{}~yr}^{-1}$
corresponds to $3\times{}10^{-9}\,\eta^{-1}\text{\Msun{}~yr}^{-1}$ one
year after explosion. The accretion rate was predicted to be about
$10^{-4}$\Msun{}~yr$^{-1}$ one year after
explosion~\citep{chevalier89, houck91}.  The discrepancy with
observations was clear a few years after
explosion~\citep[e.g.][]{suntzeff92}, and is stronger now. A feedback
is indicated, which could be the radiation pressure when the radiation
can first escape from the shocked region~\citep{houck91}.

The temporal evolution of a rotating dipole in vacuum
(Equation~\ref{eq:energy_rpp}) is given by
\begin{equation}
  \label{eq:evo_rpp}
  L = \frac{2^5\,\pi^4\,B^2\,R^6}{3\,c^3\,P_0^4\,(1+t/t')^2}
\end{equation}
where $P_0$ is the birth period, $t$ the time since birth, and
\begin{equation}
  \begin{split}
    \label{eq:tp}
    & t' \equiv P/2\dot{P} =\\
    &4.1\times{}10^{14}\left(\frac{M_\bullet{}}{\text{M$_{\sun}$}}\right)\!
    \left(\frac{B}{10^{12}\text{~G}}\right)^{-2}\!\left(\frac{P}{\text{s}}\right)^{2}\!\left(\frac{R}{\text{10~km}}\right)^{-4}\text{~s},
  \end{split}
\end{equation}
where $\dot{P}$ is the period derivative and $M_\bullet{}$ the NS
mass. This models the NS as a homogeneous sphere, and assumes constant
magnetic field and inclination angle. This allows us to translate the
current observational limits to constraints on birth properties
(Figure~\ref{fig:BP}). For all but very high magnetic field strengths,
the limits are essentially the same because the spin-down timescale is
long. However, for magnetic field strengths above ${\sim}10^{16}$~G,
the current period is practically independent of the birth period.

For the rotational energy of the NS to contribute a significant
fraction of the explosion energy, the rotational period has to be a
few milliseconds. Figure~\ref{fig:BP} excludes all initial periods
shorter than 10~ms unless the magnetic field is unusually weak or
unusually strong. If the field has not evolved and the pulsar formula
is applicable, the rotation of the NS is thus unlikely to have
contributed a significant fraction of the explosion energy, which
lends some support to the hypothesis that \sna{} was a
neutrino-powered event.

\subsection{Limits on a Binary Companion}\label{sec:bin_com}
The UVOIR limits can also be used to constrain a possible surviving
binary companion in \sna{}. The evolution of the progenitor \sleak{}
is still not fully understood and some theories involve binary
interaction as an explanation for the three circumstellar
rings~\citep{blondin93, morris07, morris09} and the peculiar
properties of \sna{}~\citep[e.g.][]{menon17, kochanek18}.

In Figure~\ref{fig:ubvri_lim} we show a blackbody spectrum
corresponding to the temperature and radius of the Sun scaled by the
distance to \sna{}. It happens to just fit below all UVOIR limits and
is constrained by the SINFONI data point at 1.7~\micron{}. Therefore,
the Sun can serve as the limit for a possible main-sequence companion
in \sna{}. We note that this limit only applies if our line-of-sight
to a companion is free of dust.

We can use the bolometric limit on the compact object of 138\Lsun{} to
constrain a main-sequence companion even if our line-of-sight is
obscured by dust. The mass-luminosity (M-L) relation for
  companion masses $2.4 < M_\mathrm{c} \leq 7$\Msun{}
  is~\citep{eker15}
\begin{equation}
  \label{eq:mas_lum}
  \left(\frac{L_\mathrm{c}}{\mathrm{L}_{\sun}}\right) =
  1.32\left(\frac{M_\mathrm{c}}{\mathrm{M}_{\sun}}\right)^{3.96},
\end{equation}
where $L_\mathrm{c}$ is the companion luminosity. By imposing that
$L_\mathrm{c} < 138$\Lsun{}, we find that $M_\mathrm{c} <
3.2$\Msun{}. We test the sensitivity of this result by comparing to
limits from other M-L relations. The M-L relation
$L_\mathrm{c}/\mathrm{L}_{\sun} =
(M_\mathrm{c}/\mathrm{M}_{\sun})^4$~\citep[][p.\ 20]{duric03} results
in a limit of $M_\mathrm{c} < 3.4$\Msun{} and
$M_\mathrm{c} < 3.2$--3.7\Msun{} (depending on angular momentum and
metallicity) from Figure~5.11 of \citet{salaris05}.

\citet{morris09} proposed that a 15-M$_{\sun}$ primary and a
5-M$_{\sun}$ companion merged to form \sleak{}. The constraints on a
binary companion show that such a companion did not survive as a
5-M$_{\sun}$ main-sequence star.

\subsection{Remaining Possibilities}\label{sec:rem_pos}
In this section, we combine all available information and explore the
remaining possibilities for the compact object, which results in much
stronger conclusions. Even though the limits in individual frequency
intervals are relatively weak, only a few possible options remain for
the compact object in \sna{}. This is because the direct limits are
corrected for absorption, but do not consider reprocessing of the
absorbed energy. This is a limitation with important consequences
because some of the limits (Table~\ref{tab:lim_par}) require more than
100\Lsun{} to be absorbed and it is not obvious how such large amounts
of energy can escape undetected. The bolometric limits address this
limitation by including reprocessing of the emission. However, the
bolometric limits only consider the cases when the UVOIR emission can
escape and when the UVOIR emission is absorbed by dust. This is
effectively equivalent to having a minimal 1D spherical geometry and
disregarding the spatial information of the observations. We stress
that the bolometric limits rely on additional assumptions and is much
more model-dependent.

Below, we first describe in detail the reprocessing of the thermal
emission that is expected in all models involving NSs in
Section~\ref{sec:the_sur_emi}. This is followed by our favored
explanation and reasons for rejecting additional components in
Section~\ref{sec:fav_exp}.

\subsubsection{Thermal Surface Emission}\label{sec:the_sur_emi}
If the compact object is a NS, then at least thermal surface emission
is expected. For a gravitational mass of 1.4\Msun{} and a local radius
of 10~km, 3.1~MK corresponds to a luminosity at infinity (redshifted,
observer's frame) of 10\Lsun{}
(Section~\ref{sec:thermal_emission}). The choice of 10~km is
conservative as a choice of 12~km would increase the luminosity by
60\,\% for a fixed temperature (increased emitting area and decreased
gravitational redshift).

The thermal emission peaks at soft X-ray energies, which is
photoabsorbed locally (on-the-spot) due to the high optical depth of
the ejecta for soft X-rays~\citep{alp18b}. The X-ray emission that is
absorbed by the ejecta is reprocessed into dust continuum emission,
and optical and UV emission lines. The fact that the very conservative
limit of 22\Lsun{} (Section~\ref{sec:bol_lim}) is close to the
expected luminosity of 10\Lsun{} indicates that the compact object is
dust obscured. Regardless of whether or not there are dust clumps
directly along the line-of-sight, it is likely that a significant
amount of the X-ray input would escape as thermal dust emission at
\mbox{(sub\nobreakdash{-})m}illimeter and FIR wavelengths. If the
reprocessing into dust heating occurs on-the-spot and thermal dust
emission escapes directly, a NS would appear as a point source in the
dust emission. To fully explore this scenario, we need to analyze
observations at frequencies where the dust emission peaks; investigate
the dust lifetime close to the NS; model the dust composition,
geometry, and temperature; model the distribution of $^{44}$Ti; and
compute the radiation propagation of UVOIR photons powered by X-ray
emission from a NS. This is beyond the scope of this paper and will be
the subject of future studies.

The most likely alternative scenario to on-the-spot dust heating is if
the mean free path of UVOIR photons in the ejecta is comparable to the
spatial extent of the ejecta. In this case, the emission would diffuse
on scales comparable to the size of the ejecta and be spatially mixed
with the emission powered by the decay of $^{44}$Ti. This is
effectively what was assumed for the dust-obscured bolometric limit of
138\Lsun{} (Section~\ref{sec:bol_lim}) because it did not consider the
spatial distribution of the escaping radiation. However, if the
UVOIR-photons have a long mean free path, the escaping reprocessed
UVOIR emission is expected to be directly observed. This means that
only a certain range of intermediate mean free paths allow a NS to be
hidden in the ejecta.

Future observations that spatially resolve the dust emission will
provide information about the mean free path of UVOIR photons. It is
unlikely that clumps of $^{44}$Ti would appear as point-sources if the
mean free path is long because observable overdensities require the
clumps of $^{44}$Ti to be well-obscured by dust. In addition, it is
not clear if the intrinsic distribution of $^{44}$Ti allows for
overdensities that can be confused with a NS~\citep{wongwathanarat13,
  wongwathanarat15}.

A less likely scenario is if the input from the NS surface emission
somehow escapes in MIR where observational limits are poor, possibly
as hot-dust emission~\citep[see][]{bouchet04, bouchet14}. This is
unlikely because it requires the primary emission to be reprocessed by
dust with a temperature tuned such that the emission escapes
detection. Furthermore, we know that the MIR fine-structure lines that
are predicted to be the primary cooling channels~\citep{jerkstrand11}
have not been observed~\citep{lundqvist01, bouchet06}, implying that
MIR emission cannot escape the ejecta.

\subsubsection{Favored Explanation \& Additional Components}\label{sec:fav_exp}
All things considered, we find the most likely scenario to be that the
compact object in \sna{} is a dust-obscured thermally-emitting NS. We
favor this scenario regardless of whether or not the dust absorbs the
UVOIR emission locally or if the mean free path for UVOIR photons is
comparable to the size of the ejecta.

The bolometric limit of 138\Lsun{} leaves little room for accretion
and pulsar wind activity, which would appear as additional
contributions to the expected thermal surface emission of
${\sim}$10\Lsun{} (Table~\ref{tab:lim_par}). The effects vary
depending on the spectrum. For accretion, the input is most likely in
the form of soft X-rays and can be treated analogously to the thermal
surface emission and simply be added to the thermal luminosity in the
current framework.

Pulsar wind activity is more complicated since it could extend over
the entire electromagnetic spectrum. However, the luminosity from
millimeter to soft X-rays is limited by the bolometric limits and the
spatial extent~\citep{chevalier92} is constrained to less than
${\sim}$100\kmps{} (Section~\ref{sec:bol_lim}). The only realistic
scenario for a PWN to contribute more than 138\Lsun{} would be if the
spectrum is heavily gamma-ray dominated~\citep[e.g.\ Vela and
Geminga,][]{danilenko11, abdo13, kuiper15}. High-energy gamma-rays
escape the ejecta and are not expected to be reprocessed into lower
frequencies~\citep{alp18b}. The \textit{Fermi}/LAT limit
(Table~\ref{tab:lit_lim}) is not stringent enough to rule out this
scenario.

If the compact object is a radio pulsar, it would emit narrow beams of
radio emission. The total radio power is ${\sim}10^{29}$\ergps{} for
typical radio pulsars~\citep{lorimer12, szary14}. Even if the ejecta
are free-free thick at radio wavelengths, the energy input is
insufficient to significantly contribute to the heating of the ejecta.
The only realistic avenue to distinguish a thermally-emitting NS from
a radio pulsar is if the free-free depth is low enough and the radio
beams sweep our line-of-sight, in which case pulsed radio emission
would be detected~\citep[for recent limits, see][]{zhang18}.

For completeness, the compact object in \sna{} could be a BH. However,
as discussed in Section~\ref{sec:intro}, most studies predict that a
NS formed in the explosion.

\subsection{Future Observations}\label{sec:future}
Below, we review the prospects for detecting the compact object in
\sna{} with future facilities. The best constraints in radio will come
from the Square Kilometre Array~\citep[SKA,][]{dewdney09,
  taylor13}. SKA-low will not be able to resolve the central ejecta
from the ER, but can perform timing observations to search for pulsed
emission. Because of the sidelobes of ${\sim}$1\,\%, the sensitivity
of SKA-mid is limited to ${\sim}$1\,\% of the ER, which will have
2~mJy spots at 0.1~arcsec resolution at 8~GHz~\citep{zanardo13}. A
point-limit of ${\sim}$0.02~mJy can therefore be expected, assuming
free-free absorption to be negligible.  The limit could possibly be
improved by an order of magnitude depending on the
$uv$\nobreakdash{-}coverage and the ability to self-calibrate.

As discussed in Section~\ref{sec:rem_pos}, the thermal surface
emission from a NS could be reprocessed into a point-like source in
the thermal ejecta dust emission. The dust emission peaks at
200~\micron{} (1500~GHz) and has been observed at low spatial
resolution by ALMA and \textit{Herschel}~\citep{indebetouw14,
  zanardo14, matsuura15}.  Higher-resolution observations may be able
to detect a region of NS-heated dust.

The \textit{James Webb Space
  Telescope}~\citep[\textit{JWST},][]{gardner06}, The Giant Magellan
Telescope~\citep[GMT,][]{johns12}, and The European Extremely Large
Telescope~\citep[E-ELT,][]{gilmozzi07} will allow for significantly
deeper searches using both imaging and spectral observations in IR and
optical. However, it remains uncertain if the compact object in \sna{}
is bright at IR or optical wavelengths. A point-like source of a few
L$_{\sun}$ is expected from reprocessing of thermal X-ray emission
from a NS surface into UVOIR, if not obscured by dust clouds
(Section~\ref{sec:rem_pos}).

The \textit{Advanced Telescope for High ENergy Astrophysics}
\citep[\textit{Athena},][]{barcons15, collon15, barcons17} will be
unable to spatially resolve \sna{} and any emission from the compact
object will be blended with the ER emission. In addition,
\citet{orlando15} predict that the X-ray emission from the ER will
fade, but the central parts will become brighter, primarily driven by
interaction with the reverse shock. This means that \sna{} will become
brighter in X-rays toward the center where the compact object is
expected to reside. However, these difficulties will be partly
counteracted by the decreasing optical depth. The optical depth in the
homologous expansion phase scales as $\tau \propto t^{-2}$, where $t$
is the time elapsed since the explosion. The optical depth at 2~keV is
expected to reach 3 by $2066\pm10$~\citep{alp18b}. The error bar
accounts for asymmetries of the explosion, but excludes any
uncertainty in the explosion model, variance introduced by the compact
object being kicked by the explosion, and CSM structure. At higher
energies, this will occur much earlier, as is relevant for a PWN.

\section{Summary \& Conclusions}\label{sec:conclusions}
We have placed upper limits on the compact object in \sna{} using
observations at millimeter wavelengths from ALMA; NIR from VLT;
optical and UV from \textit{HST}; and X-rays from \textit{Chandra}. We
assume that the compact object would appear as a point source in
images and that it only contributes to the continuum component in
observed spectra. We also place constraints on the bolometric
luminosity of the compact object by investigating the total energy
budget of \sna{}. Our main conclusions are the following.

\begin{itemize}
\item The only model-independent results are the direct flux
  limits. They are corrected for absorption, but do not include
  information about the reprocessing of the absorbed emission nor the
  geometry of the system. The most constraining limit in the
  millimeter range of ALMA is 0.11~mJy at 213~GHz. The deepest UVOIR
  limits are from the spectra taken by VLT/SINFONI in NIR and
  \textit{HST}/STIS in optical. The allowed luminosity of the compact
  object in the UVOIR band is approximately 1\Lsun{}. The X-ray limits
  allow luminosities less than ${\sim}10^{36}$\ergps{}, but are very
  sensitive to the assumed spectrum.
  
\item The total energy budget of \sna{} places a bolometric limit of
  22\Lsun{} on the compact object if our line-of-sight is free of
  dust, or 138\Lsun{} if dust-obscured. This is based on assumptions
  and models of the emission reprocessing, but relies on a minimal 1D
  spherical model of the geometry.
  
\item The limits can be used to constrain the effective local
  (unredshifted) blackbody temperature of a NS. Only the limit of
  $3.8$~MK from the dust-free bolometric limit is close to
  constraining any theoretical predictions, which typically are in the
  range 3--4~MK. This can be taken as a marginal indication that the
  compact object is obscured by dust if it is a NS.
  
\item The current accretion rate is limited to less than about
  $10^{-11}\,\eta^{-1}$\Msun{}~yr$^{-1}$ for the simplest model of
  accretion. This excludes most predictions for fallback in
  \sna{}~\citep{chevalier89, houck91, suntzeff92} and indicates some
  kind of feedback~\citep{houck91}.
  
\item The limits constrain PWN activity to 3--830\Lsun{}, depending on
  assumptions about dust and spectral shape. The luminosities can be
  related to the magnetic field strength and spin period by modeling
  the NS as a rotating dipole in vacuum. The limits constrain $B$ to
  be less than 1.8--$29\times10^{13}\,P^2$~G~s$^{-2}$. However,
  because of the rapid spin-down, we cannot exclude birth magnetic
  field strengths higher than $10^{16}$~G.

\item By combining all available information about radiation
  reprocessing and geometry, the most likely remaining scenario is
  that the compact object is a dust-obscured thermally-emitting NS. In
  this case, the thermal surface emission from the NS would be
  reprocessed into thermal dust emission. For realistic assumptions
  about the dust properties and geometry, only a small parameter space
  remains open for additional accretion and pulsar-wind components. We
  stress that this result is model-dependent. The most promising
  avenues for detecting reprocessed surface emission from a NS is
  provided by \textit{JWST}, GMT, E-ELT, and ALMA.
\end{itemize}
  
\acknowledgments{This research was funded by the Knut \& Alice
  Wallenberg Foundation and the Swedish Research Council. SEW at UCSC
  was supported by the NASA Theory Program (NNX14AH34G). At Garching,
  support by the Deutsche Forschungsgemeinschaft through Excellence
  Cluster Universe (EXC~153) and Sonderforschungsbereich SFB~1258
  ``Neutrinos and Dark Matter in Astro- and Particle Physics'', and by
  the European Research Council through grant ERC-AdG No.\
  341157-COCO2CASA is acknowledged. The research of JCW is supported
  in part by the Samuel T. and Fern Yanagisawa Regents
  Professorship. This paper makes use of the following ALMA data:
  ADS/JAO.ALMA\#2013.1.00280.S and ADS/JAO.ALMA\#2015.1.00631.S. ALMA
  is a partnership of ESO (representing its member states), NSF (USA)
  and NINS (Japan), together with NRC (Canada) and NSC and ASIAA
  (Taiwan) and KASI (Republic of Korea), in cooperation with the
  Republic of Chile. The Joint ALMA Observatory is operated by ESO,
  AUI/NRAO and NAOJ. The National Radio Astronomy Observatory is a
  facility of the National Science Foundation operated under
  cooperative agreement by Associated Universities, Inc. Based on
  observations collected at the European Organisation for Astronomical
  Research in the Southern Hemisphere under ESO programme
  294.D-5033(A). This research has made use of data obtained through
  the High Energy Astrophysics Science Archive Research Center Online
  Service, provided by the NASA/Goddard Space Flight Center. Support
  for Programs GO-13181, GO-13405, GO-13810, GO-14333, GO-13401,
  GO-14753, and GO-15256 was provided by NASA through a grant from the
  Space Telescope Science Institute, which is operated by the
  Association of Universities for Research in Astronomy, Incorporated,
  under NASA contract NAS5-26555. STSDAS and PyRAF are products of the
  Space Telescope Science Institute, which is operated by AURA for
  NASA. The scientific results reported in this article are based to a
  significant degree on data obtained from the \textit{Chandra} Data
  Archive. This research has made use of software provided by the
  \textit{Chandra} X-ray Center (CXC) in the application package
  CIAO. This research has made use of NASA's Astrophysics Data
  System.}

\vspace{5mm} \facilities{ALMA, \textit{HST}(UVIS, STIS), VLT(FORS2,
  NACO, SINFONI), \textit{CXO}(ACIS)}
\software{\texttt{astropy}~\citep{astropy13}, CASA~\citep{mcmullin07},
  CIAO/CALDB~\citep{fruscione06}, DAOPHOT~\citep{stetson87},
  DrizzlePac~\citep{gonzaga12}, IRAF/PyRAF,
  \texttt{matplotlib}~\citep{hunter07}, MARX~\citep{davis12},
  \texttt{numpy}~\citep{jones01, van_der_walt11},
  \texttt{scipy}~\citep{jones01}, STSDAS, Synphot~\citep{bushouse94},
  XSPEC~\citep{arnaud96}}

\appendix
\section{Circular Polarimetry of \sna{}}\label{app:pol}
\subsection{Observations}
All observations were acquired with FORS2 mounted at the Cassegrain
focus of the UT1 VLT. The observations were obtained in imaging
polarimetric mode (IPOL), through the V$\_$HIGH FORS2 standard filter
($\lambda_0$ = 555~nm, FWHM = 123.2~nm) and with two different
quarter-wave retarder plate angles of $\theta=\pm45^\circ$ per epoch,
during four epochs: 15, 16, 18, and 23 February 2015. We obtain four
exposures per angle, each of 350~s. In the night of February~16, the
instrument has been rotated by 90\arcdeg{}. In IPOL mode, the image is
split by the Wollaston prism into two orthogonal polarised outgoing
beams, ordinary (o) and extra-ordinary (e), and the MOS Slitlets strip
mask is inserted to avoid overlapping of the beams. One of the
observations is shown in Figure~\ref{fig:4x4}.

\subsection{Methods}
\begin{figure}
  \begin{center}
    \includegraphics[width=\textwidth]{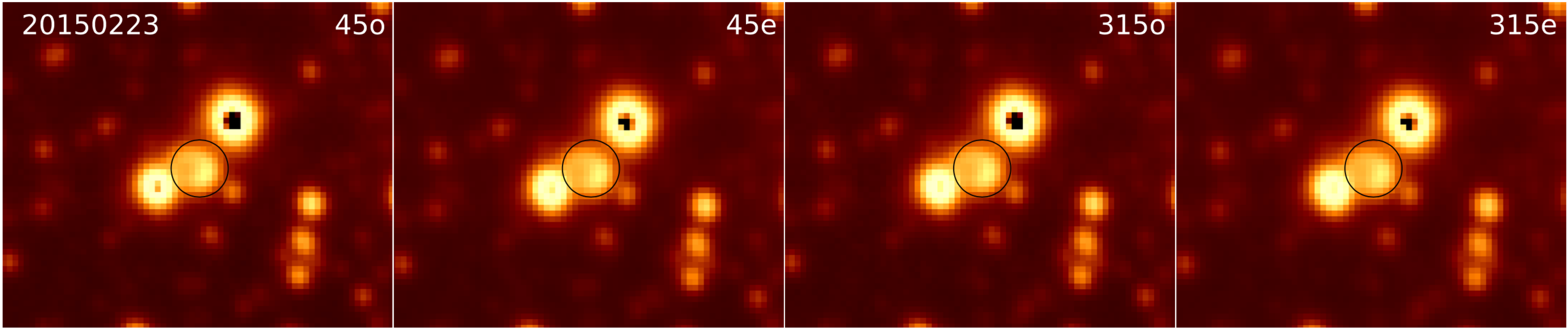}
    \caption{VLT/FORS2 circular-polarization observation of \sna{}
      from 23 February 2015. The images show ordinary (o) and
      extra-ordinary (e) beams at two quarter-wave retarder plate
      angles of $\theta=\pm45^\circ$. The circle centered at the
      position of SN 1987A ($\alpha=83.866246^\circ$,
      $\delta=-69.269722^\circ$) marks an aperture of 5-pixel
      (1250-mas) radius within which the flux is measured. Images at
      other epochs show no significant variability.\label{fig:4x4}}
  \end{center}
\end{figure}

All frames are bias subtracted using the corresponding calibration
bias frames. A flat-field correction is not performed because the
flat-field effect~\citep{patat06, obrien15}, the additional
polarization caused by the color dependent offset to the nominal
retarder plate position, and the effect of the incomplete retardation
of the quarter wave plate~\citep{obrien15} gets canceled out when
calculating the circular polarization using two angles. For each
epoch, we group the science frames according to the quarter-wave
retarder plate angle, split the ordinary and extra-ordinary beams, and
create separate science frames, align them, and calculate the median
of the four exposures.  Finally, we investigate the circular
polarization of SN 1987A by performing aperture photometry with a set
of different aperture radii, centered at the position of SN 1987A
(Figure~\ref{fig:4x4}) in ordinary and extra-ordinary beams using the
DAOPHOT.PHOT package, and calculating the circular polarization from
the determined fluxes.  We ignore the observations of February~18,
because of variable weather conditions, which makes aperture
photometry difficult. We determine the amount of circular polarization
by the equation below, as described in \citet{obrien15}:
\begin{equation}
  P_V =\frac{1}{2} \left[ \left(\frac{f^\mathrm{o} - f^\mathrm{e}}{f^\mathrm{o} + f^\mathrm{e}}\right)_{\theta=45^{\circ}} - \left(\frac{f^\mathrm{o} - f^\mathrm{e}}{f^\mathrm{o} + f^\mathrm{e}}\right)_{\theta=-45^{\circ}}  \right]
\end{equation}
where $f^\mathrm{o}$ and $f^\mathrm{e}$ is the measured flux in the
ordinary and extra-ordinary beam, respectively. The error is
calculated by propagating the photometry uncertainties.

\subsection{Results}
\begin{figure}
  \begin{center}
    \includegraphics[trim=0mm 0mm 0mm 0mm, width=9cm, clip=true]{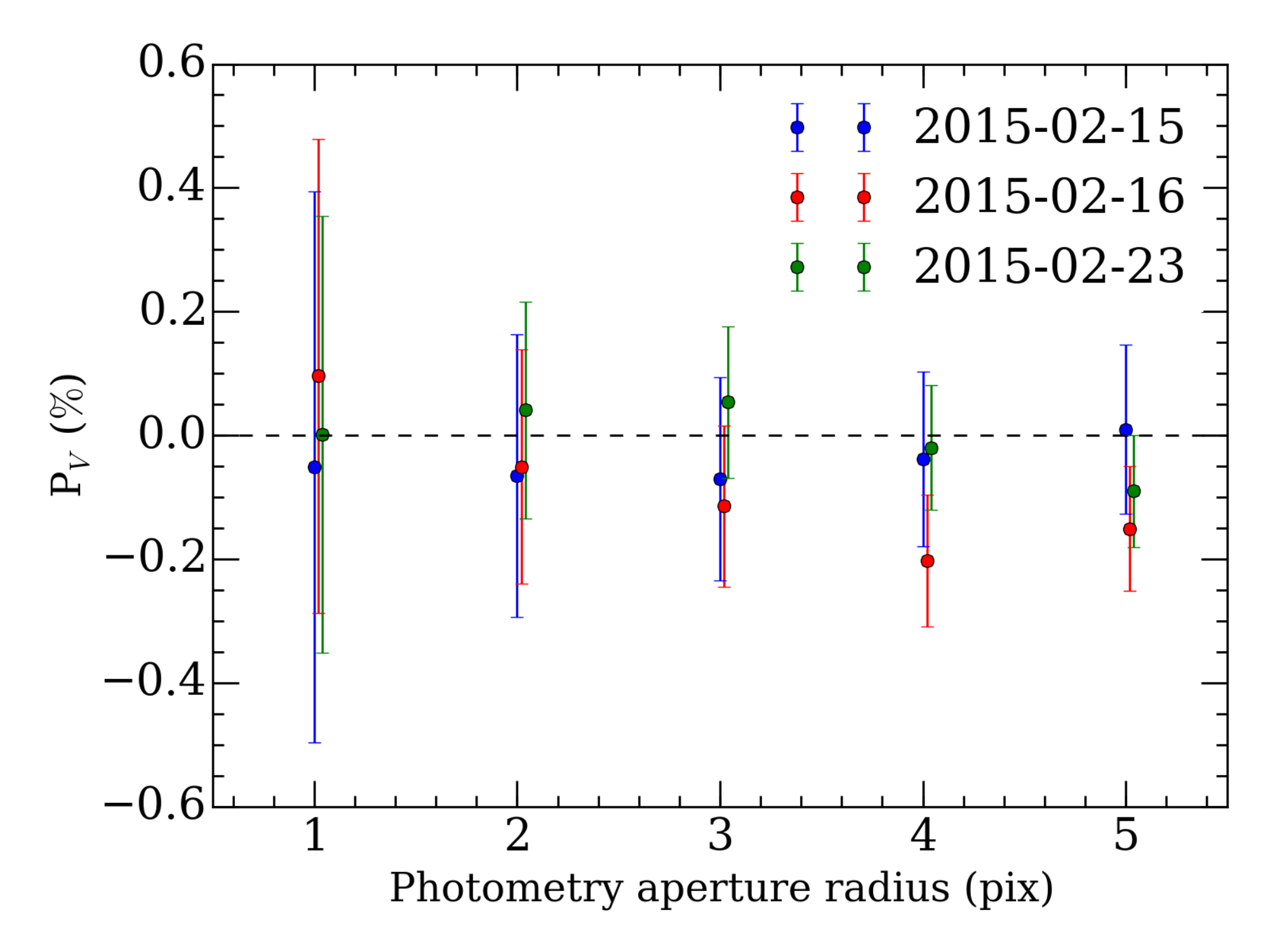}
    \caption{Circular polarization of SN 1987A measured using
      different aperture radii of 1, 2, 3, 4 and 5~pixels
      (250~mas~pixel$^{-1}$). The circular polarization is consistent
      with 0\,\%.\label{fig:pv}}
  \end{center}
\end{figure}

We calculated the circular polarization of SN 1987A from fluxes
determined by performing aperture photometry using different aperture
radii of 1, 2, 3, 4 and 5~pixels (250~mas~pixel$^{-1}$), centered at
the position of SN 1987A. We found that the circular polarization is
consistent with zero. Figure~\ref{fig:pv} shows the circular
polarization, Stokes V, values for the different aperture radii. We
are unable to determine an upper limit on the circular polarization of
the central ejecta because the angular resolution is insufficient for
resolving the structure of \sna{}. This leaves the circular
polarization from the compact object unconstrained.

The Crab was also observed using the same setup but for a shorter
duration. We are unable to detect any circular polarization in the
Crab Nebula, which possibly implies that the method is relatively
insensitive. A possible explanation for this is that the
phase-averaged polarization is essentially zero.

\section{Finding Algorithm}\label{app:find}
Searches for point sources in the ejecta are made using the DAOPHOT
task \texttt{daofind}. The most important input parameter for daofind
is the local noise level (\texttt{sigma}). For ALMA images, this is
set to the off-source noise in the images. We verified that this
method gives essentially the same noise estimates as measurements of
the noise in the visibility amplitudes. The parameter \texttt{sigma}
for the UVOIR images is chosen to account for both the sky background
and the Poisson noise of the ejecta. We note that both contributions
are of comparable magnitude. The sky background is determined by
setting a 3-$\sigma$ threshold such that essentially no noise peaks
pass while as many real sources as possible are detected. This
transition is clearly seen as a break in the detection-threshold
relation~\citep{davis94}. The ejecta Poisson noise is set to the
square root of the maximum photon count in the original detector
pixels within the search region. The parameter \texttt{sigma} is then
computed as the square root of sky background and ejecta noise added
in quadrature. It is verified that the sample deviation of small
segments of the central ejecta is comparable to the computed
\texttt{sigma}. The effects of super-resolving when drizzling and
instrument gain on the Poisson noise properties are accounted for by
including a multiplicative correction factor for the geometric mapping
and another factor for the gain.

Limits are determined by inserting artificial PSFs and finding the PSF
flux such that it just crosses the 3-$\sigma$ threshold of
\texttt{daofind}. The procedure is then repeated inside the search
region at 12.5~mas intervals, which is chosen to be the half-pixel
size of the WFC3 and NACO images. At a few points, the detection
threshold is crossed without added artificial sources, implying that
point sources are detected. These are all just slightly above the
detection threshold of 3-$\sigma$ and are interpreted as structure in
the ejecta. The limit in these points are set to the maximum flux of a
PSF that is consistent with the observation. None of the threshold
crossing events are spatially coincident in several adjacent filters
and they are not significantly increasing the upper limits.

\section{X-ray Ejecta Absorption}\label{app:xray_abs}
\begin{deluxetable}{lccccccccc}
  \tablecaption{X-ray Absorption Parameters\label{tab:col_den}}
  \tablewidth{0pt}
  \tablehead{
    \colhead{Element} & \colhead{$N_\mathrm{SN}$\tablenotemark{a}} &
    \colhead{$A_\mathrm{SN}$\tablenotemark{b}} & \colhead{$A_\mathrm{SN}/A_\mathrm{ISM}$\tablenotemark{c}} \\
    \colhead{}      & \colhead{(cm$^{-2}$)}     & \colhead{}     & \colhead{}} \startdata
  H                          & $1.4\times{}10^{22}$ & $                 1$ & $      1$ \\
  He                         & $1.1\times{}10^{22}$ & $7.8\times{}10^{-1}$ & $      8$ \\
  C                          & $1.4\times{}10^{20}$ & $1.0\times{}10^{-2}$ & $     43$ \\
  O                          & $2.0\times{}10^{20}$ & $1.4\times{}10^{-2}$ & $     29$ \\
  Ne                         & $4.1\times{}10^{19}$ & $3.0\times{}10^{-3}$ & $     34$ \\
  Mg                         & $4.5\times{}10^{18}$ & $3.2\times{}10^{-4}$ & $     13$ \\
  Si                         & $5.3\times{}10^{19}$ & $3.9\times{}10^{-3}$ & $    208$ \\
  S                          & $6.6\times{}10^{18}$ & $4.8\times{}10^{-4}$ & $     39$ \\
  Ar                         & $2.6\times{}10^{18}$ & $1.9\times{}10^{-4}$ & $     74$ \\
  Ca                         & $1.3\times{}10^{19}$ & $9.4\times{}10^{-4}$ & $    591$ \\
  Fe                         & $5.5\times{}10^{19}$ & $4.0\times{}10^{-3}$ & $    148$ \\
  H$_{0.1}$\tablenotemark{d} & $1.0\times{}10^{22}$ & $              0.73$ & \nodata{} \\
  H$_{0.9}$\tablenotemark{d} & $1.8\times{}10^{22}$ & $              1.31$ & \nodata{} \\
  \enddata
  \tablenotetext{a}{Direction-averaged SN column number density.}
  \tablenotetext{b}{$A_\mathrm{SN}(X)\equiv N_\mathrm{SN}(X)/
    N_\mathrm{SN}(\mathrm H)$, where $X$ is a chemical element.}
  \tablenotetext{c}{$A_\mathrm{ISM}(X)$ is the abundance of $X$ in the
    ISM from \citet{wilms00}.}
  \tablenotetext{d}{The quantity $\mathrm{H}_{0.1}$ is the hydrogen
    column density scaled by the ratio of the direction-averaged
    optical depth to the 10$^\mathrm{th}$ percentile of the optical
    depth at 2~keV, and $\mathrm{H}_{0.9}$ is the 90$^\mathrm{th}$
    analogue. The column densities of other elements are assumed to be
    scaled by the same fraction~\citep{alp18b}.}
\end{deluxetable}
X-ray absorption by SN ejecta is explored in detail using 3D
neutrino-driven SN explosion models~\citep{wongwathanarat13,
  wongwathanarat15} in an accompanying paper~\citep{alp18b}. One of
the main conclusions is that the optical depth of the SN ejecta for
X-rays below 10~keV is very high at the age of \sna{}. For a
discussion of the transport of the absorbed energy, see
Section~\ref{sec:bol_lim}. Here, we use the absorption estimates based
on the B15 explosion model~\citep{woosley88} for our X-ray
analysis. The B15 model is a single-star model that was evolved to
core-collapse in one dimension without mass loss. It explodes as a
blue supergiant with a mass of 15.4\Msun{} and is designed to
represent \sna{}.

From the models, we compute the column number densities
($N_\mathrm{SN}$) of H, He, C, O, Ne, Mg, Si, S, Ar, Ca, and Fe using
the explosion model. A single estimate cannot be made because of the
asymmetries of the SN explosion. Instead, we focus on the
direction-averaged column number densities and the hydrogen number
densities corresponding to the 10$^\mathrm{th}$ and 90$^\mathrm{th}$
percentiles of the optical depth, shown in
Table~\ref{tab:col_den}. The percentiles are used to represent the
variance introduced by SN explosion asymmetries. Even though the
hydrogen column number density $N_\mathrm{SN}(\mathrm{H})$ is
relatively low, the high metallicity of the ejecta results in an
optical depth of ${\sim}25$ at 2~keV at current epochs.

\section{Spatial Alignment}\label{app:spa_ali}
All observations need to be accurately registered for us to use the
position determined in Section~\ref{sec:pos}, which is needed to
define the search regions for the images and the extraction regions
for the spectra. Only ALMA has good enough absolute astrometry. The
other observations are aligned with the \textit{HST} observations
using either nearby stars or the ER, as described below.

ALMA has an absolute astrometric accuracy of less than $\sim
10$~mas. The accuracy is determined by measuring the phase RMS and
using it to estimate the phase transfer error, which likely results in
a quite conservative estimate. Applying a self-calibration gain table
to a point source and measuring the offset, and considering the
accuracy in measuring baselines both yield uncertainty estimates that
are smaller than 10~mas.

Both NACO images are mapped onto the \textit{HST}/WFC3 images using
the IRAF tasks \texttt{geomap} and \texttt{geotran}. We choose ten
bright stars in the common FOV, use polynomial fitting functions, and
a general geometry, which consists of shifts, scale factors, a
rotation and a skew. This aligns the images and resizes them to a
common pixel size of $25^2$~mas$^2$. The magnification increases the
pixel size from the original detector scale of $13.27^2$~mas$^2$, but
this does not affect the measurements because the FWHM of the point
spread function (PSF) is ${\sim}100$~mas. Comparisons of NACO and
\textit{HST} images show that the spatial alignments are better than
${\sim}25$~mas at the position of \sna{}, and the rotations and skews
have a negligible impact on the region relevant to this work.

The SINFONI images are aligned by fitting an elliptical band with
Gaussian radial profile to the ER. The center of the ellipse is then
matched with the \textit{HST} image position presented in
Section~\ref{sec:pos}. The accuracy is better than 20~mas, which is
estimated using the diagonal elements of the covariance
matrix. Alignment using nearby stars is not possible because the small
FOV of SINFONI does not extend much beyond the ER. The hotspots are
also poorly resolved and faint in the SINFONI images, which is why an
elliptical band is used to fit the ER as an extended source.

The position of \sna{} in the STIS observation is determined by
mapping onto the \textit{HST} images. The right ascension is
determined with respect to a nearby, isolated reference star that was
used for the spacecraft pointing when performing the five individual
slit observations. Spectral lines from the ER are used to match the
declination with the \textit{HST} images. This is done by fitting
Gaussians to the north and south 1D profiles of the ER in the
slits. The position of \sna{} is known relative to the ER from
Section~\ref{sec:pos}. The five slits are first matched individually,
allowing the sample variance to serve as an estimate of the
statistical uncertainty. The average position of the five alignments
is then used to align all slits. The 1-$\sigma$ uncertainty in the
final declination of all slits is 0.14 pixels or 7.0~mas. 

The ER in the \textit{Chandra} observation is modeled by fitting an
ellipse of sinusoidal intensity along the azimuth and Gaussian radial
profile (Section~\ref{sec:spa_mod}), which is also used for
alignment. There are no point sources visible in the \textit{Chandra}
FOV that can be used for alignment. The error in position using this
method is ${\sim}$40~mas, which is small compared to the PSF FWHM of
${\sim}$700~mas of \textit{Chandra}/ACIS. The uncertainty of the
position is determined by simulating samples from the model and then
applying the same fitting method to resample the position.

\section{Observations used for Astrometric Registration of \sna{}}\label{app:pos_obs}
\begin{deluxetable}{ccccccccc}
  \tablecaption{\textit{HST} Observations used for Astrometric Registration of \sna{}\label{tab:pos_obs}}
  \tablewidth{0pt}
  \tablehead{
    \colhead{Epoch}        & \colhead{Instrument} & \colhead{Band} & \colhead{Filter} & \colhead{Exposure} & \colhead{Band} & \colhead{Filter} & \colhead{Exposure} \\
    \colhead{(YYYY-mm-dd)} & \colhead{}           &     \colhead{} &       \colhead{} &      \colhead{(s)} &     \colhead{} &       \colhead{} &      \colhead{(s)}} \startdata
    2003-01-05             &                ACS   &     \textit{R} &            F625W &   \hphantom{00}800 &     \textit{B} &            F435W &   \hphantom{0}1200 \\
    2003-08-12             &                ACS   &     \textit{R} &            F625W &   \hphantom{00}480 &     \textit{B} &            F435W &   \hphantom{00}800 \\
    2003-11-28             &                ACS   &     \textit{R} &            F625W &   \hphantom{00}800 &     \textit{B} &            F435W &   \hphantom{0}1600 \\
    2004-12-15             &                ACS   &      \nodata{} &        \nodata{} &          \nodata{} &     \textit{B} &            F435W &   \hphantom{0}1600 \\
    2005-04-02             &                ACS   &      \nodata{} &        \nodata{} &          \nodata{} &     \textit{B} &            F435W &   \hphantom{0}1200 \\
    2005-09-26             &                ACS   &     \textit{R} &            F625W &              12000 &      \nodata{} &        \nodata{} &          \nodata{} \\
    2005-09-28             &                ACS   &     \textit{R} &            F625W &   \hphantom{00}720 &      \nodata{} &        \nodata{} &          \nodata{} \\ 
    2006-04-15             &                ACS   &     \textit{R} &            F625W &   \hphantom{0}1200 &     \textit{B} &            F435W &   \hphantom{0}1200 \\
    2006-04-29             &                ACS   &     \textit{R} &            F625W &   \hphantom{00}720 &      \nodata{} &        \nodata{} &          \nodata{} \\
    2006-12-06             &                ACS   &     \textit{R} &            F625W &   \hphantom{0}1200 &     \textit{B} &            F435W &   \hphantom{0}1800 \\
    2007-05-12             &                WFPC2 &     \textit{R} &            F675W &   \hphantom{0}2700 &     \textit{B} &            F439W &   \hphantom{0}3000 \\
    2008-02-19             &                WFPC2 &     \textit{R} &            F675W &   \hphantom{0}1600 &     \textit{B} &            F439W &   \hphantom{0}2400 \\
    2009-04-29             &                WFPC2 &     \textit{R} &            F675W &   \hphantom{0}1600 &     \textit{B} &            F439W &   \hphantom{0}2000 \\
    2009-12-12             &                WFC3  &     \textit{R} &            F625W &   \hphantom{0}3000 &     \textit{B} &            F438W &   \hphantom{00}800 \\
    2011-01-05             &                WFC3  &     \textit{R} &            F625W &   \hphantom{0}1000 &     \textit{B} &            F438W &   \hphantom{0}1400 \\
    2013-02-06             &                WFC3  &     \textit{R} &            F625W &   \hphantom{0}1200 &     \textit{B} &            F438W &   \hphantom{0}1200 \\
    2014-06-15             &                WFC3  &     \textit{R} &            F625W &   \hphantom{0}1200 &     \textit{B} &            F438W &   \hphantom{0}1200 \\
    2015-05-24             &                WFC3  &     \textit{R} &            F625W &   \hphantom{0}1200 &     \textit{B} &            F438W &   \hphantom{0}1200 \\
    2016-06-08             &                WFC3  &     \textit{R} &            F625W &   \hphantom{00}600 &     \textit{B} &            F438W &   \hphantom{00}600 \\
    \enddata
\end{deluxetable}
The observations that are used to determine the position of \sna{} are
listed in Table~\ref{tab:pos_obs}. The WFC3 observations are reduced
as described in Section~\ref{sec:obs_wfc3}. The previous observations
are described in \citet{larsson11}.

\section{\textit{Chandra} PSF}\label{app:psf}
The \textit{Chandra} PSF is created using MARX 5.3.2~\citep{davis12},
which is called from the CIAO task \texttt{simulate\_psf}. The PSF is
created for the position of \sna{} on the CCD chip, 22 pixels off-axis
(approximately the distance between the default aimpoint and optical
axis), using an input spectrum extracted from the source region. The
quantum efficiency of the detector is included by disabling the
\texttt{ideal} option. Simulation of the readout streak and pileup are
both performed. The \texttt{extended} option is disabled because the
PSF is used as a convolution kernel. The PSF is simulated onto a pixel
size of 50$^2$~mas$^2$ and statistical fluctuations of the simulation
are reduced by making 200 iterations. The default value of 70~mas for
the parameter \texttt{AspectBlur}, which is the measured uncertainty
of the aspect solution, is used because it better matches the
observation according to the statistical likelihood. We note that this
is smaller than the ``merely suggested'' value of
${\sim}$280~mas\footnote{\url{http://cxc.harvard.edu/ciao/why/aspectblur.html}},
which is based on a limited number of observations~\citep{primini11}.
The source of this additional blurring for at least some observations
is currently not fully understood.

\bibliography{references}

\begin{thebibliography}{}
\expandafter\ifx\csname natexlab\endcsname\relax\def\natexlab#1{#1}\fi
\providecommand{\url}[1]{\href{#1}{#1}}
\providecommand{\dodoi}[1]{doi:~\href{http://doi.org/#1}{\nolinkurl{#1}}}
\providecommand{\doeprint}[1]{\href{http://ascl.net/#1}{\nolinkurl{http://ascl.net/#1}}}
\providecommand{\doarXiv}[1]{\href{https://arxiv.org/abs/#1}{\nolinkurl{https://arxiv.org/abs/#1}}}

\bibitem[{{Abdo} {et~al.}(2013){Abdo}, {Ajello}, {Allafort}, {Baldini},
  {Ballet}, {Barbiellini}, {Baring}, {Bastieri}, {Belfiore}, {Bellazzini}, \&
  et~al.}]{abdo13}
{Abdo}, A.~A., {Ajello}, M., {Allafort}, A., {et~al.} 2013, \apjs, 208, 17,
  \dodoi{10.1088/0067-0049/208/2/17}

\bibitem[{{Abell{\'a}n} {et~al.}(2017){Abell{\'a}n}, {Indebetouw}, {Marcaide},
  {Gabler}, {Fransson}, {Spyromilio}, {Burrows}, {Chevalier}, {Cigan},
  {Gaensler}, {Gomez}, {Janka}, {Kirshner}, {Larsson}, {Lundqvist}, {Matsuura},
  {McCray}, {Ng}, {Park}, {Roche}, {Staveley-Smith}, {van Loon}, {Wheeler}, \&
  {Woosley}}]{abellan17}
{Abell{\'a}n}, F.~J., {Indebetouw}, R., {Marcaide}, J.~M., {et~al.} 2017,
  \apjl, 842, L24, \dodoi{10.3847/2041-8213/aa784c}

\bibitem[{{Ackermann} {et~al.}(2016){Ackermann}, {Albert}, {Atwood}, {Baldini},
  {Ballet}, {Barbiellini}, {Bastieri}, {Bellazzini}, {Bissaldi}, {Bloom},
  {Bonino}, {Brandt}, {Bregeon}, {Bruel}, {Buehler}, {Caliandro}, {Cameron},
  {Caragiulo}, {Caraveo}, {Cavazzuti}, {Cecchi}, {Charles}, {Chekhtman},
  {Chiang}, {Chiaro}, {Ciprini}, {Cohen-Tanugi}, {Cutini}, {D'Ammando}, {de
  Angelis}, {de Palma}, {Desiante}, {Digel}, {Drell}, {Favuzzi}, {Ferrara},
  {Focke}, {Franckowiak}, {Fusco}, {Gargano}, {Gasparrini}, {Giglietto},
  {Giordano}, {Godfrey}, {Grenier}, {Grondin}, {Guillemot}, {Guiriec},
  {Harding}, {Hill}, {Horan}, {J{\'o}hannesson}, {Kn{\"o}dlseder}, {Kuss},
  {Larsson}, {Latronico}, {Li}, {Li}, {Longo}, {Loparco}, {Lubrano}, {Maldera},
  {Martin}, {Mayer}, {Mazziotta}, {Michelson}, {Mizuno}, {Monzani}, {Morselli},
  {Murgia}, {Nuss}, {Ohsugi}, {Orienti}, {Orlando}, {Ormes}, {Paneque},
  {Pesce-Rollins}, {Piron}, {Pivato}, {Porter}, {Rain{\`o}}, {Rando},
  {Razzano}, {Reimer}, {Reimer}, {Romani}, {S{\'a}nchez-Conde}, {Schulz},
  {Sgr{\`o}}, {Siskind}, {Smith}, {Spada}, {Spandre}, {Spinelli}, {Suson},
  {Takahashi}, {Thayer}, {Tibaldo}, {Torres}, {Tosti}, {Troja}, {Vianello},
  {Wood}, \& {Zimmer}}]{ackermann16}
{Ackermann}, M., {Albert}, A., {Atwood}, W.~B., {et~al.} 2016, \aap, 586, A71,
  \dodoi{10.1051/0004-6361/201526920}

\bibitem[{{Ahmad} {et~al.}(2006){Ahmad}, {Greene}, {Moore}, {Ghelberg}, {Ofan},
  {Paul}, \& {Kutschera}}]{ahmad06}
{Ahmad}, I., {Greene}, J.~P., {Moore}, E.~F., {et~al.} 2006, \prc, 74, 065803,
  \dodoi{10.1103/PhysRevC.74.065803}

\bibitem[{{Ahola} {et~al.}(2018)}]{ahola18}
{Ahola}, A., {et~al.} 2018, in preparation

\bibitem[{{Alekseev} {et~al.}(1987){Alekseev}, {Alekseeva}, {Volchenko}, \&
  {Krivosheina}}]{alekseev87}
{Alekseev}, E.~N., {Alekseeva}, L.~N., {Volchenko}, V.~I., \& {Krivosheina},
  I.~V. 1987, Soviet Journal of Experimental and Theoretical Physics Letters,
  45, 589

\bibitem[{{Alexeyev} {et~al.}(1988){Alexeyev}, {Alexeyeva}, {Krivosheina}, \&
  {Volchenko}}]{alexeyev88}
{Alexeyev}, E.~N., {Alexeyeva}, L.~N., {Krivosheina}, I.~V., \& {Volchenko},
  V.~I. 1988, Physics Letters B, 205, 209, \dodoi{10.1016/0370-2693(88)91651-6}

\bibitem[{{Alp} {et~al.}(2018){Alp}, {Larsson}, {Fransson}, {Gabler},
  {Wongwathanarat}, \& {Janka}}]{alp18b}
{Alp}, D., {Larsson}, J., {Fransson}, C., {et~al.} 2018, ArXiv e-prints.
\newblock \doarXiv{1805.04528}

\bibitem[{{Arendt} {et~al.}(2016){Arendt}, {Dwek}, {Bouchet}, {Danziger},
  {Frank}, {Gehrz}, {Park}, \& {Woodward}}]{arendt16}
{Arendt}, R.~G., {Dwek}, E., {Bouchet}, P., {et~al.} 2016, \aj, 151, 62,
  \dodoi{10.3847/0004-6256/151/3/62}

\bibitem[{{Arnaud}(1996)}]{arnaud96}
{Arnaud}, K.~A. 1996, in Astronomical Society of the Pacific Conference Series,
  Vol. 101, Astronomical Data Analysis Software and Systems V, ed. G.~H.
  {Jacoby} \& J.~{Barnes}, 17

\bibitem[{{Arnett} {et~al.}(1989){Arnett}, {Bahcall}, {Kirshner}, \&
  {Woosley}}]{arnett89}
{Arnett}, W.~D., {Bahcall}, J.~N., {Kirshner}, R.~P., \& {Woosley}, S.~E. 1989,
  \araa, 27, 629, \dodoi{10.1146/annurev.aa.27.090189.003213}

\bibitem[{{Asaki} {et~al.}(2014){Asaki}, {Matsushita}, {Kawabe}, {Fomalont},
  {Barkats}, \& {Corder}}]{asaki14}
{Asaki}, Y., {Matsushita}, S., {Kawabe}, R., {et~al.} 2014, in \procspie, Vol.
  9145, Ground-based and Airborne Telescopes V, 91454K

\bibitem[{{Astropy Collaboration} {et~al.}(2013){Astropy Collaboration},
  {Robitaille}, {Tollerud}, {Greenfield}, {Droettboom}, {Bray}, {Aldcroft},
  {Davis}, {Ginsburg}, {Price-Whelan}, {Kerzendorf}, {Conley}, {Crighton},
  {Barbary}, {Muna}, {Ferguson}, {Grollier}, {Parikh}, {Nair}, {Unther},
  {Deil}, {Woillez}, {Conseil}, {Kramer}, {Turner}, {Singer}, {Fox}, {Weaver},
  {Zabalza}, {Edwards}, {Azalee Bostroem}, {Burke}, {Casey}, {Crawford},
  {Dencheva}, {Ely}, {Jenness}, {Labrie}, {Lim}, {Pierfederici}, {Pontzen},
  {Ptak}, {Refsdal}, {Servillat}, \& {Streicher}}]{astropy13}
{Astropy Collaboration}, {Robitaille}, T.~P., {Tollerud}, E.~J., {et~al.} 2013,
  \aap, 558, A33, \dodoi{10.1051/0004-6361/201322068}

\bibitem[{{Barcons} {et~al.}(2015){Barcons}, {Nandra}, {Barret}, {den Herder},
  {Fabian}, {Piro}, {Watson}, \& {the Athena Team}}]{barcons15}
{Barcons}, X., {Nandra}, K., {Barret}, D., {et~al.} 2015, in Journal of Physics
  Conference Series, Vol. 610, Journal of Physics Conference Series, 012008

\bibitem[{{Barcons} {et~al.}(2017){Barcons}, {Barret}, {Decourchelle}, {den
  Herder}, {Fabian}, {Matsumoto}, {Lumb}, {Nandra}, {Piro}, {Smith}, \&
  {Willingale}}]{barcons17}
{Barcons}, X., {Barret}, D., {Decourchelle}, A., {et~al.} 2017, Astronomische
  Nachrichten, 338, 153, \dodoi{10.1002/asna.201713323}

\bibitem[{{Bauswein} {et~al.}(2017){Bauswein}, {Just}, {Janka}, \&
  {Stergioulas}}]{bauswein17}
{Bauswein}, A., {Just}, O., {Janka}, H.-T., \& {Stergioulas}, N. 2017, \apjl,
  850, L34, \dodoi{10.3847/2041-8213/aa9994}

\bibitem[{{Bevan}(2018)}]{bevan18}
{Bevan}, A. 2018, ArXiv e-prints.
\newblock \doarXiv{1803.10241}

\bibitem[{{Bionta} {et~al.}(1987){Bionta}, {Blewitt}, {Bratton}, {Casper}, \&
  {Ciocio}}]{bionta87}
{Bionta}, R.~M., {Blewitt}, G., {Bratton}, C.~B., {Casper}, D., \& {Ciocio}, A.
  1987, Physical Review Letters, 58, 1494, \dodoi{10.1103/PhysRevLett.58.1494}

\bibitem[{{Blondin} \& {Lundqvist}(1993)}]{blondin93}
{Blondin}, J.~M., \& {Lundqvist}, P. 1993, \apj, 405, 337,
  \dodoi{10.1086/172366}

\bibitem[{{Blum} \& {Kushnir}(2016)}]{blum16}
{Blum}, K., \& {Kushnir}, D. 2016, \apj, 828, 31,
  \dodoi{10.3847/0004-637X/828/1/31}

\bibitem[{{Bogdanov}(2014)}]{bogdanov14}
{Bogdanov}, S. 2014, \apj, 790, 94, \dodoi{10.1088/0004-637X/790/2/94}

\bibitem[{{Boggs} {et~al.}(2015){Boggs}, {Harrison}, {Miyasaka},
  {Grefenstette}, {Zoglauer}, {Fryer}, {Reynolds}, {Alexander}, {An}, {Barret},
  {Christensen}, {Craig}, {Forster}, {Giommi}, {Hailey}, {Hornstrup},
  {Kitaguchi}, {Koglin}, {Madsen}, {Mao}, {Mori}, {Perri}, {Pivovaroff},
  {Puccetti}, {Rana}, {Stern}, {Westergaard}, \& {Zhang}}]{boggs15}
{Boggs}, S.~E., {Harrison}, F.~A., {Miyasaka}, H., {et~al.} 2015, Science, 348,
  670, \dodoi{10.1126/science.aaa2259}

\bibitem[{{Bonnet} {et~al.}(2004){Bonnet}, {Abuter}, {Baker}, {Bornemann},
  {Brown}, {Castillo}, {Conzelmann}, {Damster}, {Davies}, {Delabre},
  {Donaldson}, {Dumas}, {Eisenhauer}, {Elswijk}, {Fedrigo}, {Finger},
  {Gemperlein}, {Genzel}, {Gilbert}, {Gillet}, {Goldbrunner}, {Horrobin}, {Ter
  Horst}, {Huber}, {Hubin}, {Iserlohe}, {Kaufer}, {Kissler-Patig}, {Kragt},
  {Kroes}, {Lehnert}, {Lieb}, {Liske}, {Lizon}, {Lutz}, {Modigliani}, {Monnet},
  {Nesvadba}, {Patig}, {Pragt}, {Reunanen}, {R{\"o}hrle}, {Rossi}, {Schmutzer},
  {Schoenmaker}, {Schreiber}, {Stroebele}, {Szeifert}, {Tacconi}, {Tecza},
  {Thatte}, {Tordo}, {van der Werf}, \& {Weisz}}]{bonnet04}
{Bonnet}, H., {Abuter}, R., {Baker}, A., {et~al.} 2004, The Messenger, 117, 17

\bibitem[{{Bouchet} \& {Danziger}(2014)}]{bouchet14}
{Bouchet}, P., \& {Danziger}, J. 2014, in IAU Symposium, Vol. 296, Supernova
  Environmental Impacts, ed. A.~{Ray} \& R.~A. {McCray}, 9--14

\bibitem[{{Bouchet} {et~al.}(2004){Bouchet}, {De Buizer}, {Suntzeff},
  {Danziger}, {Hayward}, {Telesco}, \& {Packham}}]{bouchet04}
{Bouchet}, P., {De Buizer}, J.~M., {Suntzeff}, N.~B., {et~al.} 2004, \apj, 611,
  394, \dodoi{10.1086/421936}

\bibitem[{{Bouchet} {et~al.}(2006){Bouchet}, {Dwek}, {Danziger}, {Arendt}, {De
  Buizer}, {Park}, {Suntzeff}, {Kirshner}, \& {Challis}}]{bouchet06}
{Bouchet}, P., {Dwek}, E., {Danziger}, J., {et~al.} 2006, \apj, 650, 212,
  \dodoi{10.1086/505929}

\bibitem[{{Bratton} {et~al.}(1988){Bratton}, {Casper}, {Ciocio}, {Claus},
  {Crouch}, {Dye}, {Errede}, {Gajewski}, {Goldhaber}, {Haines}, {Jones},
  {Kielczewska}, {Kropp}, {Learned}, {Losecco}, {Matthews}, {Miller}, {Mudan},
  {Price}, {Reines}, {Schultz}, {Seidel}, {Sinclair}, {Sobel}, {Stone},
  {Sulak}, {Svoboda}, {Thornton}, \& {van der Velde}}]{bratton88}
{Bratton}, C.~B., {Casper}, D., {Ciocio}, A., {et~al.} 1988, \prd, 37, 3361,
  \dodoi{10.1103/PhysRevD.37.3361}

\bibitem[{{Brown} {et~al.}(1992){Brown}, {Bruenn}, \& {Wheeler}}]{brown92}
{Brown}, G.~E., {Bruenn}, S.~W., \& {Wheeler}, J.~C. 1992, Comments on
  Astrophysics, 16, 153

\bibitem[{{Brown} \& {Weingartner}(1994)}]{brown94}
{Brown}, G.~E., \& {Weingartner}, J.~C. 1994, \apj, 436, 843,
  \dodoi{10.1086/174961}

\bibitem[{{B{\"u}hler} \& {Blandford}(2014)}]{buhler14}
{B{\"u}hler}, R., \& {Blandford}, R. 2014, Reports on Progress in Physics, 77,
  066901, \dodoi{10.1088/0034-4885/77/6/066901}

\bibitem[{{Burrows}(1988)}]{burrows88}
{Burrows}, A. 1988, \apj, 334, 891, \dodoi{10.1086/166885}

\bibitem[{{Burrows} {et~al.}(2000){Burrows}, {Michael}, {Hwang}, {McCray},
  {Chevalier}, {Petre}, {Garmire}, {Holt}, \& {Nousek}}]{burrows00}
{Burrows}, D.~N., {Michael}, E., {Hwang}, U., {et~al.} 2000, \apjl, 543, L149,
  \dodoi{10.1086/317271}

\bibitem[{{Bushouse} \& {Simon}(1994)}]{bushouse94}
{Bushouse}, H., \& {Simon}, B. 1994, in Astronomical Society of the Pacific
  Conference Series, Vol.~61, Astronomical Data Analysis Software and Systems
  III, ed. D.~R. {Crabtree}, R.~J. {Hanisch}, \& J.~{Barnes}, 339

\bibitem[{{Callingham} {et~al.}(2016){Callingham}, {Gaensler}, {Zanardo},
  {Staveley-Smith}, {Hancock}, {Hurley-Walker}, {Bell}, {Dwarakanath},
  {Franzen}, {Hindson}, {Johnston-Hollitt}, {Kapi{\'n}ska}, {For}, {Lenc},
  {McKinley}, {Morgan}, {Offringa}, {Procopio}, {Wayth}, {Wu}, \&
  {Zheng}}]{callingham16}
{Callingham}, J.~R., {Gaensler}, B.~M., {Zanardo}, G., {et~al.} 2016, \mnras,
  462, 290, \dodoi{10.1093/mnras/stw1489}

\bibitem[{{Canizares} {et~al.}(2005){Canizares}, {Davis}, {Dewey}, {Flanagan},
  {Galton}, {Huenemoerder}, {Ishibashi}, {Markert}, {Marshall}, {McGuirk},
  {Schattenburg}, {Schulz}, {Smith}, \& {Wise}}]{canizares05}
{Canizares}, C.~R., {Davis}, J.~E., {Dewey}, D., {et~al.} 2005, \pasp, 117,
  1144, \dodoi{10.1086/432898}

\bibitem[{{Cardelli} {et~al.}(1989){Cardelli}, {Clayton}, \&
  {Mathis}}]{cardelli89}
{Cardelli}, J.~A., {Clayton}, G.~C., \& {Mathis}, J.~S. 1989, \apj, 345, 245,
  \dodoi{10.1086/167900}

\bibitem[{{Cash}(1979)}]{cash79}
{Cash}, W. 1979, \apj, 228, 939, \dodoi{10.1086/156922}

\bibitem[{{Chatterjee} {et~al.}(2000){Chatterjee}, {Hernquist}, \&
  {Narayan}}]{chatterjee00}
{Chatterjee}, P., {Hernquist}, L., \& {Narayan}, R. 2000, \apj, 534, 373,
  \dodoi{10.1086/308748}

\bibitem[{{Chatterjee} \& {Cordes}(2002)}]{chatterjee02}
{Chatterjee}, S., \& {Cordes}, J.~M. 2002, \apj, 575, 407,
  \dodoi{10.1086/341139}

\bibitem[{{Chatterjee} \& {Cordes}(2004)}]{chatterjee04}
---. 2004, \apjl, 600, L51, \dodoi{10.1086/381498}

\bibitem[{{Chevalier}(1989)}]{chevalier89}
{Chevalier}, R.~A. 1989, \apj, 346, 847, \dodoi{10.1086/168066}

\bibitem[{{Chevalier} \& {Fransson}(1992)}]{chevalier92}
{Chevalier}, R.~A., \& {Fransson}, C. 1992, \apj, 395, 540,
  \dodoi{10.1086/171674}

\bibitem[{{Chita} {et~al.}(2008){Chita}, {Langer}, {van Marle},
  {Garc{\'{\i}}a-Segura}, \& {Heger}}]{chita08}
{Chita}, S.~M., {Langer}, N., {van Marle}, A.~J., {Garc{\'{\i}}a-Segura}, G.,
  \& {Heger}, A. 2008, \aap, 488, L37, \dodoi{10.1051/0004-6361:200810087}

\bibitem[{{Cohen} {et~al.}(2003){Cohen}, {Wheaton}, \& {Megeath}}]{cohen03}
{Cohen}, M., {Wheaton}, W.~A., \& {Megeath}, S.~T. 2003, \aj, 126, 1090,
  \dodoi{10.1086/376474}

\bibitem[{{Collon} {et~al.}(2015){Collon}, {Vacanti}, {G{\"u}nther}, {Yanson},
  {Barri{\`e}re}, {Landgraf}, {Vervest}, {Chatbi}, {Beijersbergen}, {Bavdaz},
  {Wille}, {Haneveld}, {Koelewijn}, {Leenstra}, {Wijnperle}, {van Baren},
  {M{\"u}ller}, {Krumrey}, {Burwitz}, {Pareschi}, {Conconi}, \&
  {Christensen}}]{collon15}
{Collon}, M.~J., {Vacanti}, G., {G{\"u}nther}, R., {et~al.} 2015, in \procspie,
  Vol. 9603, Society of Photo-Optical Instrumentation Engineers (SPIE)
  Conference Series, 96030K

\bibitem[{{Cordes} {et~al.}(1993){Cordes}, {Romani}, \& {Lundgren}}]{cordes93}
{Cordes}, J.~M., {Romani}, R.~W., \& {Lundgren}, S.~C. 1993, \nat, 362, 133,
  \dodoi{10.1038/362133a0}

\bibitem[{{Danilenko} {et~al.}(2011){Danilenko}, {Zyuzin}, {Shibanov}, \&
  {Zharikov}}]{danilenko11}
{Danilenko}, A.~A., {Zyuzin}, D.~A., {Shibanov}, Y.~A., \& {Zharikov}, S.~V.
  2011, \mnras, 415, 867, \dodoi{10.1111/j.1365-2966.2011.18753.x}

\bibitem[{{Davies}(2007)}]{davies07}
{Davies}, R.~I. 2007, \mnras, 375, 1099,
  \dodoi{10.1111/j.1365-2966.2006.11383.x}

\bibitem[{{Davis} {et~al.}(2012){Davis}, {Bautz}, {Dewey}, {Heilmann}, {Houck},
  {Huenemoerder}, {Marshall}, {Nowak}, {Schattenburg}, {Schulz}, \&
  {Smith}}]{davis12}
{Davis}, J.~E., {Bautz}, M.~W., {Dewey}, D., {et~al.} 2012, in \procspie, Vol.
  8443, Space Telescopes and Instrumentation 2012: Ultraviolet to Gamma Ray,
  84431A

\bibitem[{Davis(1994)}]{davis94}
Davis, L.~E. 1994, IRAF Programming Group, NOAO, Tucson

\bibitem[{{Dewdney} {et~al.}(2009){Dewdney}, {Hall}, {Schilizzi}, \&
  {Lazio}}]{dewdney09}
{Dewdney}, P.~E., {Hall}, P.~J., {Schilizzi}, R.~T., \& {Lazio}, T.~J.~L.~W.
  2009, IEEE Proceedings, 97, 1482, \dodoi{10.1109/JPROC.2009.2021005}

\bibitem[{{Draine}(2003)}]{draine03}
{Draine}, B.~T. 2003, \apj, 598, 1026, \dodoi{10.1086/379123}

\bibitem[{Duric(2003)}]{duric03}
Duric, N. 2003, Advanced Astrophysics (Cambridge University Press),
  \dodoi{10.1017/CBO9780511800177}

\bibitem[{{Dwek} \& {Arendt}(2015)}]{dwek15}
{Dwek}, E., \& {Arendt}, R.~G. 2015, \apj, 810, 75,
  \dodoi{10.1088/0004-637X/810/1/75}

\bibitem[{{Eisenhauer} {et~al.}(2003){Eisenhauer}, {Abuter}, {Bickert},
  {Biancat-Marchet}, {Bonnet}, {Brynnel}, {Conzelmann}, {Delabre}, {Donaldson},
  {Farinato}, {Fedrigo}, {Genzel}, {Hubin}, {Iserlohe}, {Kasper},
  {Kissler-Patig}, {Monnet}, {Roehrle}, {Schreiber}, {Stroebele}, {Tecza},
  {Thatte}, \& {Weisz}}]{eisenhauer03}
{Eisenhauer}, F., {Abuter}, R., {Bickert}, K., {et~al.} 2003, in \procspie,
  Vol. 4841, Instrument Design and Performance for Optical/Infrared
  Ground-based Telescopes, ed. M.~{Iye} \& A.~F.~M. {Moorwood}, 1548--1561

\bibitem[{{Eker} {et~al.}(2015){Eker}, {Soydugan}, {Soydugan}, {Bilir}, {Yaz
  G{\"o}k{\c c}e}, {Steer}, {T{\"u}ys{\"u}z}, {{\c S}eny{\"u}z}, \&
  {Demircan}}]{eker15}
{Eker}, Z., {Soydugan}, F., {Soydugan}, E., {et~al.} 2015, \aj, 149, 131,
  \dodoi{10.1088/0004-6256/149/4/131}

\bibitem[{{Ertl} {et~al.}(2016){Ertl}, {Janka}, {Woosley}, {Sukhbold}, \&
  {Ugliano}}]{ertl16}
{Ertl}, T., {Janka}, H.-T., {Woosley}, S.~E., {Sukhbold}, T., \& {Ugliano}, M.
  2016, \apj, 818, 124, \dodoi{10.3847/0004-637X/818/2/124}

\bibitem[{{Esposito} {et~al.}(2018){Esposito}, {Rea}, {Lazzati}, {Matsuura},
  {Perna}, \& {Pons}}]{esposito18}
{Esposito}, P., {Rea}, N., {Lazzati}, D., {et~al.} 2018, \apj, 857, 58,
  \dodoi{10.3847/1538-4357/aab6b6}

\bibitem[{{Evans} \& {Kochanek}(1989)}]{evans89}
{Evans}, C.~R., \& {Kochanek}, C.~S. 1989, \apjl, 346, L13,
  \dodoi{10.1086/185567}

\bibitem[{{Fassia} {et~al.}(2002){Fassia}, {Meikle}, \&
  {Spyromilio}}]{fassia02}
{Fassia}, A., {Meikle}, W.~P.~S., \& {Spyromilio}, J. 2002, \mnras, 332, 296,
  \dodoi{10.1046/j.1365-8711.2002.05293.x}

\bibitem[{{Faucher-Gigu{\`e}re} \& {Kaspi}(2006)}]{faucher-giguere06}
{Faucher-Gigu{\`e}re}, C.-A., \& {Kaspi}, V.~M. 2006, \apj, 643, 332,
  \dodoi{10.1086/501516}

\bibitem[{{Fitzpatrick} \& {Walborn}(1990)}]{fitzpatrick90}
{Fitzpatrick}, E.~L., \& {Walborn}, N.~R. 1990, \aj, 99, 1483,
  \dodoi{10.1086/115432}

\bibitem[{{France} {et~al.}(2011){France}, {McCray}, {Penton}, {Kirshner},
  {Challis}, {Laming}, {Bouchet}, {Chevalier}, {Garnavich}, {Fransson}, {Heng},
  {Larsson}, {Lawrence}, {Lundqvist}, {Panagia}, {Pun}, {Smith}, {Sollerman},
  {Sonneborn}, {Sugerman}, \& {Wheeler}}]{france11}
{France}, K., {McCray}, R., {Penton}, S.~V., {et~al.} 2011, \apj, 743, 186,
  \dodoi{10.1088/0004-637X/743/2/186}

\bibitem[{{Frank} {et~al.}(2002){Frank}, {King}, \& {Raine}}]{frank02}
{Frank}, J., {King}, A., \& {Raine}, D.~J. 2002, {Accretion Power in
  Astrophysics: Third Edition} (Cambridge University Press), 398

\bibitem[{{Frank} {et~al.}(2016){Frank}, {Zhekov}, {Park}, {McCray}, {Dwek}, \&
  {Burrows}}]{frank16}
{Frank}, K.~A., {Zhekov}, S.~A., {Park}, S., {et~al.} 2016, \apj, 829, 40,
  \dodoi{10.3847/0004-637X/829/1/40}

\bibitem[{{Fransson} \& {Kozma}(2002)}]{fransson02}
{Fransson}, C., \& {Kozma}, C. 2002, \nar, 46, 487,
  \dodoi{10.1016/S1387-6473(02)00188-4}

\bibitem[{{Fransson} {et~al.}(2016){Fransson}, {Larsson}, {Spyromilio},
  {Leibundgut}, {McCray}, \& {Jerkstrand}}]{fransson16}
{Fransson}, C., {Larsson}, J., {Spyromilio}, J., {et~al.} 2016, \apjl, 821, L5,
  \dodoi{10.3847/2041-8205/821/1/L5}

\bibitem[{{Fransson} {et~al.}(2013){Fransson}, {Larsson}, {Spyromilio},
  {Chevalier}, {Gr{\"o}ningsson}, {Jerkstrand}, {Leibundgut}, {McCray},
  {Challis}, {Kirshner}, {Kjaer}, {Lundqvist}, \& {Sollerman}}]{fransson13}
---. 2013, \apj, 768, 88, \dodoi{10.1088/0004-637X/768/1/88}

\bibitem[{{Fransson} {et~al.}(2015){Fransson}, {Larsson}, {Migotto}, {Pesce},
  {Challis}, {Chevalier}, {France}, {Kirshner}, {Leibundgut}, {Lundqvist},
  {McCray}, {Spyromilio}, {Taddia}, {Jerkstrand}, {Mattila}, {Smith},
  {Sollerman}, {Wheeler}, {Crotts}, {Garnavich}, {Heng}, {Lawrence}, {Panagia},
  {Pun}, {Sonneborn}, \& {Sugerman}}]{fransson15}
{Fransson}, C., {Larsson}, J., {Migotto}, K., {et~al.} 2015, \apjl, 806, L19,
  \dodoi{10.1088/2041-8205/806/1/L19}

\bibitem[{{Fruchter} \& {Hook}(2002)}]{fruchter02}
{Fruchter}, A.~S., \& {Hook}, R.~N. 2002, \pasp, 114, 144,
  \dodoi{10.1086/338393}

\bibitem[{{Fruscione} {et~al.}(2006){Fruscione}, {McDowell}, {Allen},
  {Brickhouse}, {Burke}, {Davis}, {Durham}, {Elvis}, {Galle}, {Harris},
  {Huenemoerder}, {Houck}, {Ishibashi}, {Karovska}, {Nicastro}, {Noble},
  {Nowak}, {Primini}, {Siemiginowska}, {Smith}, \& {Wise}}]{fruscione06}
{Fruscione}, A., {McDowell}, J.~C., {Allen}, G.~E., {et~al.} 2006, in
  \procspie, Vol. 6270, Society of Photo-Optical Instrumentation Engineers
  (SPIE) Conference Series, 62701V

\bibitem[{{Fryer}(1999)}]{fryer99}
{Fryer}, C.~L. 1999, \apj, 522, 413, \dodoi{10.1086/307647}

\bibitem[{{Gaia Collaboration} {et~al.}(2016{\natexlab{a}}){Gaia
  Collaboration}, {Prusti}, {de Bruijne}, {Brown}, {Vallenari}, {Babusiaux},
  {Bailer-Jones}, {Bastian}, {Biermann}, {Evans}, \& et~al.}]{gaia16b}
{Gaia Collaboration}, {Prusti}, T., {de Bruijne}, J.~H.~J., {et~al.}
  2016{\natexlab{a}}, \aap, 595, A1, \dodoi{10.1051/0004-6361/201629272}

\bibitem[{{Gaia Collaboration} {et~al.}(2016{\natexlab{b}}){Gaia
  Collaboration}, {Brown}, {Vallenari}, {Prusti}, {de Bruijne}, {Mignard},
  {Drimmel}, {Babusiaux}, {Bailer-Jones}, {Bastian}, \& et~al.}]{gaia16}
{Gaia Collaboration}, {Brown}, A.~G.~A., {Vallenari}, A., {et~al.}
  2016{\natexlab{b}}, \aap, 595, A2, \dodoi{10.1051/0004-6361/201629512}

\bibitem[{{Gardner} {et~al.}(2006){Gardner}, {Mather}, {Clampin}, {Doyon},
  {Greenhouse}, {Hammel}, {Hutchings}, {Jakobsen}, {Lilly}, {Long}, {Lunine},
  {McCaughrean}, {Mountain}, {Nella}, {Rieke}, {Rieke}, {Rix}, {Smith},
  {Sonneborn}, {Stiavelli}, {Stockman}, {Windhorst}, \& {Wright}}]{gardner06}
{Gardner}, J.~P., {Mather}, J.~C., {Clampin}, M., {et~al.} 2006, \ssr, 123,
  485, \dodoi{10.1007/s11214-006-8315-7}

\bibitem[{{Garmire} {et~al.}(2003){Garmire}, {Bautz}, {Ford}, {Nousek}, \&
  {Ricker}}]{garmire03}
{Garmire}, G.~P., {Bautz}, M.~W., {Ford}, P.~G., {Nousek}, J.~A., \& {Ricker},
  Jr., G.~R. 2003, in \procspie, Vol. 4851, X-Ray and Gamma-Ray Telescopes and
  Instruments for Astronomy., ed. J.~E. {Truemper} \& H.~D. {Tananbaum}, 28--44

\bibitem[{{Gilmozzi} \& {Spyromilio}(2007)}]{gilmozzi07}
{Gilmozzi}, R., \& {Spyromilio}, J. 2007, The Messenger, 127

\bibitem[{{Gnedin} {et~al.}(2001){Gnedin}, {Yakovlev}, \&
  {Potekhin}}]{gnedin01}
{Gnedin}, O.~Y., {Yakovlev}, D.~G., \& {Potekhin}, A.~Y. 2001, \mnras, 324,
  725, \dodoi{10.1046/j.1365-8711.2001.04359.x}

\bibitem[{Gonzaga {et~al.}(2012)Gonzaga, Hack, Fruchter, \& Mack}]{gonzaga12}
Gonzaga, S., Hack, W., Fruchter, A., \& Mack, J. 2012, {The DrizzlePac
  Handbook} (Baltimore, STScI operated by AURA for NASA)

\bibitem[{{Gordon} {et~al.}(2003){Gordon}, {Clayton}, {Misselt}, {Landolt}, \&
  {Wolff}}]{gordon03}
{Gordon}, K.~D., {Clayton}, G.~C., {Misselt}, K.~A., {Landolt}, A.~U., \&
  {Wolff}, M.~J. 2003, \apj, 594, 279, \dodoi{10.1086/376774}

\bibitem[{{Gould} \& {Uza}(1998)}]{gould98}
{Gould}, A., \& {Uza}, O. 1998, \apj, 494, 118, \dodoi{10.1086/305193}

\bibitem[{{Graves} {et~al.}(2005){Graves}, {Challis}, {Chevalier}, {Crotts},
  {Filippenko}, {Fransson}, {Garnavich}, {Kirshner}, {Li}, {Lundqvist},
  {McCray}, {Panagia}, {Phillips}, {Pun}, {Schmidt}, {Sonneborn}, {Suntzeff},
  {Wang}, \& {Wheeler}}]{graves05}
{Graves}, G.~J.~M., {Challis}, P.~M., {Chevalier}, R.~A., {et~al.} 2005, \apj,
  629, 944, \dodoi{10.1086/431422}

\bibitem[{{Grebenev} {et~al.}(2012){Grebenev}, {Lutovinov}, {Tsygankov}, \&
  {Winkler}}]{grebenev12}
{Grebenev}, S.~A., {Lutovinov}, A.~A., {Tsygankov}, S.~S., \& {Winkler}, C.
  2012, \nat, 490, 373, \dodoi{10.1038/nature11473}

\bibitem[{{Gr{\"o}ningsson} {et~al.}(2008{\natexlab{a}}){Gr{\"o}ningsson},
  {Fransson}, {Leibundgut}, {Lundqvist}, {Challis}, {Chevalier}, \&
  {Spyromilio}}]{groningsson08b}
{Gr{\"o}ningsson}, P., {Fransson}, C., {Leibundgut}, B., {et~al.}
  2008{\natexlab{a}}, \aap, 492, 481, \dodoi{10.1051/0004-6361:200810551}

\bibitem[{{Gr{\"o}ningsson} {et~al.}(2008{\natexlab{b}}){Gr{\"o}ningsson},
  {Fransson}, {Lundqvist}, {Lundqvist}, {Leibundgut}, {Spyromilio},
  {Chevalier}, {Gilmozzi}, {Kj{\ae}r}, {Mattila}, \&
  {Sollerman}}]{groningsson08}
{Gr{\"o}ningsson}, P., {Fransson}, C., {Lundqvist}, P., {et~al.}
  2008{\natexlab{b}}, \aap, 479, 761, \dodoi{10.1051/0004-6361:20077604}

\bibitem[{{Haberl} {et~al.}(2006){Haberl}, {Geppert}, {Aschenbach}, \&
  {Hasinger}}]{haberl06}
{Haberl}, F., {Geppert}, U., {Aschenbach}, B., \& {Hasinger}, G. 2006, \aap,
  460, 811, \dodoi{10.1051/0004-6361:20066198}

\bibitem[{{Hebeler} {et~al.}(2013){Hebeler}, {Lattimer}, {Pethick}, \&
  {Schwenk}}]{hebeler13}
{Hebeler}, K., {Lattimer}, J.~M., {Pethick}, C.~J., \& {Schwenk}, A. 2013,
  \apj, 773, 11, \dodoi{10.1088/0004-637X/773/1/11}

\bibitem[{{Helder} {et~al.}(2013){Helder}, {Broos}, {Dewey}, {Dwek}, {McCray},
  {Park}, {Racusin}, {Zhekov}, \& {Burrows}}]{helder13}
{Helder}, E.~A., {Broos}, P.~S., {Dewey}, D., {et~al.} 2013, \apj, 764, 11,
  \dodoi{10.1088/0004-637X/764/1/11}

\bibitem[{{H.E.S.S.~Collaboration} {et~al.}(2015){H.E.S.S.~Collaboration},
  {Abramowski}, {Aharonian}, {Ait Benkhali}, {Akhperjanian}, {Ang{\"u}ner},
  {Backes}, {Balenderan}, {Balzer}, {Barnacka}, \& et~al.}]{hess15}
{H.E.S.S.~Collaboration}, {Abramowski}, A., {Aharonian}, F., {et~al.} 2015,
  Science, 347, 406, \dodoi{10.1126/science.1261313}

\bibitem[{{Hirata} {et~al.}(1987){Hirata}, {Kajita}, {Koshiba}, {Nakahata}, \&
  {Oyama}}]{hirata87}
{Hirata}, K., {Kajita}, T., {Koshiba}, M., {Nakahata}, M., \& {Oyama}, Y. 1987,
  Physical Review Letters, 58, 1490, \dodoi{10.1103/PhysRevLett.58.1490}

\bibitem[{{Hirata} {et~al.}(1988){Hirata}, {Kajita}, {Koshiba}, {Nakahata},
  {Oyama}, {Sato}, {Suzuki}, {Takita}, {Totsuka}, {Kifune}, {Suda},
  {Takahashi}, {Tanimori}, {Miyano}, {Yamada}, {Beier}, {Feldscher}, {Frati},
  {Kim}, {Mann}, {Newcomer}, {van Berg}, {Zhang}, \& {Cortez}}]{hirata88}
{Hirata}, K.~S., {Kajita}, T., {Koshiba}, M., {et~al.} 1988, \prd, 38, 448,
  \dodoi{10.1103/PhysRevD.38.448}

\bibitem[{{Ho} \& {Heinke}(2009)}]{ho09}
{Ho}, W.~C.~G., \& {Heinke}, C.~O. 2009, \nat, 462, 71,
  \dodoi{10.1038/nature08525}

\bibitem[{{Ho} {et~al.}(2008){Ho}, {Potekhin}, \& {Chabrier}}]{ho08b}
{Ho}, W.~C.~G., {Potekhin}, A.~Y., \& {Chabrier}, G. 2008, \apjs, 178, 102,
  \dodoi{10.1086/589238}

\bibitem[{{Hobbs} {et~al.}(2005){Hobbs}, {Lorimer}, {Lyne}, \&
  {Kramer}}]{hobbs05}
{Hobbs}, G., {Lorimer}, D.~R., {Lyne}, A.~G., \& {Kramer}, M. 2005, \mnras,
  360, 974, \dodoi{10.1111/j.1365-2966.2005.09087.x}

\bibitem[{{Houck} \& {Chevalier}(1991)}]{houck91}
{Houck}, J.~C., \& {Chevalier}, R.~A. 1991, \apj, 376, 234,
  \dodoi{10.1086/170272}

\bibitem[{{Hunter}(2007)}]{hunter07}
{Hunter}, J.~D. 2007, Computing in Science and Engineering, 9, 90,
  \dodoi{10.1109/MCSE.2007.55}

\bibitem[{{Indebetouw} {et~al.}(2014){Indebetouw}, {Matsuura}, {Dwek},
  {Zanardo}, {Barlow}, {Baes}, {Bouchet}, {Burrows}, {Chevalier}, {Clayton},
  {Fransson}, {Gaensler}, {Kirshner}, {Laki{\'c}evi{\'c}}, {Long}, {Lundqvist},
  {Mart{\'{\i}}-Vidal}, {Marcaide}, {McCray}, {Meixner}, {Ng}, {Park},
  {Sonneborn}, {Staveley-Smith}, {Vlahakis}, \& {van Loon}}]{indebetouw14}
{Indebetouw}, R., {Matsuura}, M., {Dwek}, E., {et~al.} 2014, \apjl, 782, L2,
  \dodoi{10.1088/2041-8205/782/1/L2}

\bibitem[{{Jerkstrand} {et~al.}(2011){Jerkstrand}, {Fransson}, \&
  {Kozma}}]{jerkstrand11}
{Jerkstrand}, A., {Fransson}, C., \& {Kozma}, C. 2011, \aap, 530, A45,
  \dodoi{10.1051/0004-6361/201015937}

\bibitem[{{Jerkstrand} {et~al.}(2015){Jerkstrand}, {Timmes}, {Magkotsios},
  {Sim}, {Fransson}, {Spyromilio}, {M{\"u}ller}, {Heger}, {Sollerman}, \&
  {Smartt}}]{jerkstrand15}
{Jerkstrand}, A., {Timmes}, F.~X., {Magkotsios}, G., {et~al.} 2015, \apj, 807,
  110, \dodoi{10.1088/0004-637X/807/1/110}

\bibitem[{{Johns} {et~al.}(2012){Johns}, {McCarthy}, {Raybould}, {Bouchez},
  {Farahani}, {Filgueira}, {Jacoby}, {Shectman}, \& {Sheehan}}]{johns12}
{Johns}, M., {McCarthy}, P., {Raybould}, K., {et~al.} 2012, in \procspie, Vol.
  8444, Ground-based and Airborne Telescopes IV, 84441H

\bibitem[{Jones {et~al.}(2001--)Jones, Oliphant, Peterson, {et~al.}}]{jones01}
Jones, E., Oliphant, T., Peterson, P., {et~al.} 2001--

\bibitem[{{Kallivayalil} {et~al.}(2013){Kallivayalil}, {van der Marel},
  {Besla}, {Anderson}, \& {Alcock}}]{kallivayalil13}
{Kallivayalil}, N., {van der Marel}, R.~P., {Besla}, G., {Anderson}, J., \&
  {Alcock}, C. 2013, \apj, 764, 161, \dodoi{10.1088/0004-637X/764/2/161}

\bibitem[{{Kirshner} {et~al.}(1987){Kirshner}, {Sonneborn}, {Crenshaw}, \&
  {Nassiopoulos}}]{kirshner87}
{Kirshner}, R.~P., {Sonneborn}, G., {Crenshaw}, D.~M., \& {Nassiopoulos}, G.~E.
  1987, \apj, 320, 602, \dodoi{10.1086/165579}

\bibitem[{{Kj{\ae}r} {et~al.}(2007){Kj{\ae}r}, {Leibundgut}, {Fransson},
  {Gr{\"o}ningsson}, {Spyromilio}, \& {Kissler-Patig}}]{kjaer07}
{Kj{\ae}r}, K., {Leibundgut}, B., {Fransson}, C., {et~al.} 2007, \aap, 471,
  617, \dodoi{10.1051/0004-6361:20077561}

\bibitem[{{Kj{\ae}r} {et~al.}(2010){Kj{\ae}r}, {Leibundgut}, {Fransson},
  {Jerkstrand}, \& {Spyromilio}}]{kjaer10}
{Kj{\ae}r}, K., {Leibundgut}, B., {Fransson}, C., {Jerkstrand}, A., \&
  {Spyromilio}, J. 2010, \aap, 517, A51, \dodoi{10.1051/0004-6361/201014538}

\bibitem[{{Klochkov} {et~al.}(2015){Klochkov}, {Suleimanov}, {P{\"u}hlhofer},
  {Yakovlev}, {Santangelo}, \& {Werner}}]{klochkov15}
{Klochkov}, D., {Suleimanov}, V., {P{\"u}hlhofer}, G., {et~al.} 2015, \aap,
  573, A53, \dodoi{10.1051/0004-6361/201424683}

\bibitem[{{Kochanek}(2018)}]{kochanek18}
{Kochanek}, C.~S. 2018, \mnras, 473, 1633, \dodoi{10.1093/mnras/stx2423}

\bibitem[{{Kozma} \& {Fransson}(1992)}]{kozma92}
{Kozma}, C., \& {Fransson}, C. 1992, \apj, 390, 602, \dodoi{10.1086/171311}

\bibitem[{{Kuiper} \& {Hermsen}(2015)}]{kuiper15}
{Kuiper}, L., \& {Hermsen}, W. 2015, \mnras, 449, 3827,
  \dodoi{10.1093/mnras/stv426}

\bibitem[{{Laki{\'c}evi{\'c}} {et~al.}(2011){Laki{\'c}evi{\'c}}, {van Loon},
  {Patat}, {Staveley-Smith}, \& {Zanardo}}]{lakicevic11}
{Laki{\'c}evi{\'c}}, M., {van Loon}, J.~T., {Patat}, F., {Staveley-Smith}, L.,
  \& {Zanardo}, G. 2011, \aap, 532, L8, \dodoi{10.1051/0004-6361/201116978}

\bibitem[{{Laki{\'c}evi{\'c}} {et~al.}(2012{\natexlab{a}}){Laki{\'c}evi{\'c}},
  {van Loon}, {Stanke}, {De Breuck}, \& {Patat}}]{lakicevic12b}
{Laki{\'c}evi{\'c}}, M., {van Loon}, J.~T., {Stanke}, T., {De Breuck}, C., \&
  {Patat}, F. 2012{\natexlab{a}}, \aap, 541, L1,
  \dodoi{10.1051/0004-6361/201118661}

\bibitem[{{Laki{\'c}evi{\'c}} {et~al.}(2012{\natexlab{b}}){Laki{\'c}evi{\'c}},
  {Zanardo}, {van Loon}, {Staveley-Smith}, {Potter}, {Ng}, \&
  {Gaensler}}]{lakicevic12}
{Laki{\'c}evi{\'c}}, M., {Zanardo}, G., {van Loon}, J.~T., {et~al.}
  2012{\natexlab{b}}, \aap, 541, L2, \dodoi{10.1051/0004-6361/201218902}

\bibitem[{{Larsson} {et~al.}(2011){Larsson}, {Fransson}, {{\"O}stlin},
  {Gr{\"o}ningsson}, {Jerkstrand}, {Kozma}, {Sollerman}, {Challis}, {Kirshner},
  {Chevalier}, {Heng}, {McCray}, {Suntzeff}, {Bouchet}, {Crotts}, {Danziger},
  {Dwek}, {France}, {Garnavich}, {Lawrence}, {Leibundgut}, {Lundqvist},
  {Panagia}, {Pun}, {Smith}, {Sonneborn}, {Wang}, \& {Wheeler}}]{larsson11}
{Larsson}, J., {Fransson}, C., {{\"O}stlin}, G., {et~al.} 2011, \nat, 474, 484,
  \dodoi{10.1038/nature10090}

\bibitem[{{Larsson} {et~al.}(2013){Larsson}, {Fransson}, {Kjaer}, {Jerkstrand},
  {Kirshner}, {Leibundgut}, {Lundqvist}, {Mattila}, {McCray}, {Sollerman},
  {Spyromilio}, \& {Wheeler}}]{larsson13}
{Larsson}, J., {Fransson}, C., {Kjaer}, K., {et~al.} 2013, \apj, 768, 89,
  \dodoi{10.1088/0004-637X/768/1/89}

\bibitem[{{Larsson} {et~al.}(2016){Larsson}, {Fransson}, {Spyromilio},
  {Leibundgut}, {Challis}, {Chevalier}, {France}, {Jerkstrand}, {Kirshner},
  {Lundqvist}, {Matsuura}, {McCray}, {Smith}, {Sollerman}, {Garnavich}, {Heng},
  {Lawrence}, {Mattila}, {Migotto}, {Sonneborn}, {Taddia}, \&
  {Wheeler}}]{larsson16}
{Larsson}, J., {Fransson}, C., {Spyromilio}, J., {et~al.} 2016, \apj, 833, 147,
  \dodoi{10.3847/1538-4357/833/2/147}

\bibitem[{{Lawrence} {et~al.}(2000){Lawrence}, {Sugerman}, {Bouchet}, {Crotts},
  {Uglesich}, \& {Heathcote}}]{lawrence00}
{Lawrence}, S.~S., {Sugerman}, B.~E., {Bouchet}, P., {et~al.} 2000, \apjl, 537,
  L123, \dodoi{10.1086/312771}

\bibitem[{{Lenzen} {et~al.}(2003){Lenzen}, {Hartung}, {Brandner}, {Finger},
  {Hubin}, {Lacombe}, {Lagrange}, {Lehnert}, {Moorwood}, \&
  {Mouillet}}]{lenzen03}
{Lenzen}, R., {Hartung}, M., {Brandner}, W., {et~al.} 2003, in \procspie, Vol.
  4841, Instrument Design and Performance for Optical/Infrared Ground-based
  Telescopes, ed. M.~{Iye} \& A.~F.~M. {Moorwood}, 944--952

\bibitem[{{Lorimer} \& {Kramer}(2012)}]{lorimer12}
{Lorimer}, D.~R., \& {Kramer}, M. 2012, {Handbook of Pulsar Astronomy}
  (Cambridge, UK: Cambridge University Press)

\bibitem[{{Lucy} {et~al.}(1989){Lucy}, {Danziger}, {Gouiffes}, \&
  {Bouchet}}]{lucy89}
{Lucy}, L.~B., {Danziger}, I.~J., {Gouiffes}, C., \& {Bouchet}, P. 1989, in
  Lecture Notes in Physics, Berlin Springer Verlag, Vol. 350, IAU Colloq. 120:
  Structure and Dynamics of the Interstellar Medium, ed. G.~{Tenorio-Tagle},
  M.~{Moles}, \& J.~{Melnick}, 164

\bibitem[{{Lucy} {et~al.}(1991){Lucy}, {Danziger}, {Gouiffes}, \&
  {Bouchet}}]{lucy91}
{Lucy}, L.~B., {Danziger}, I.~J., {Gouiffes}, C., \& {Bouchet}, P. 1991, in
  Supernovae, ed. S.~E. {Woosley}, 82

\bibitem[{{Lundqvist} {et~al.}(2001){Lundqvist}, {Kozma}, {Sollerman}, \&
  {Fransson}}]{lundqvist01}
{Lundqvist}, P., {Kozma}, C., {Sollerman}, J., \& {Fransson}, C. 2001, \aap,
  374, 629, \dodoi{10.1051/0004-6361:20010725}

\bibitem[{{Manchester} {et~al.}(2005){Manchester}, {Hobbs}, {Teoh}, \&
  {Hobbs}}]{manchester05}
{Manchester}, R.~N., {Hobbs}, G.~B., {Teoh}, A., \& {Hobbs}, M. 2005, \aj, 129,
  1993, \dodoi{10.1086/428488}

\bibitem[{{Matsuura} {et~al.}(2011){Matsuura}, {Dwek}, {Meixner}, {Otsuka},
  {Babler}, {Barlow}, {Roman-Duval}, {Engelbracht}, {Sandstrom},
  {Laki{\'c}evi{\'c}}, {van Loon}, {Sonneborn}, {Clayton}, {Long}, {Lundqvist},
  {Nozawa}, {Gordon}, {Hony}, {Panuzzo}, {Okumura}, {Misselt}, {Montiel}, \&
  {Sauvage}}]{matsuura11}
{Matsuura}, M., {Dwek}, E., {Meixner}, M., {et~al.} 2011, Science, 333, 1258,
  \dodoi{10.1126/science.1205983}

\bibitem[{{Matsuura} {et~al.}(2015){Matsuura}, {Dwek}, {Barlow}, {Babler},
  {Baes}, {Meixner}, {Cernicharo}, {Clayton}, {Dunne}, {Fransson}, {Fritz},
  {Gear}, {Gomez}, {Groenewegen}, {Indebetouw}, {Ivison}, {Jerkstrand},
  {Lebouteiller}, {Lim}, {Lundqvist}, {Pearson}, {Roman-Duval}, {Royer},
  {Staveley-Smith}, {Swinyard}, {van Hoof}, {van Loon}, {Verstappen}, {Wesson},
  {Zanardo}, {Blommaert}, {Decin}, {Reach}, {Sonneborn}, {Van de Steene}, \&
  {Yates}}]{matsuura15}
{Matsuura}, M., {Dwek}, E., {Barlow}, M.~J., {et~al.} 2015, \apj, 800, 50,
  \dodoi{10.1088/0004-637X/800/1/50}

\bibitem[{{Matsuura} {et~al.}(2017){Matsuura}, {Indebetouw}, {Woosley},
  {Bujarrabal}, {Abell{\'a}n}, {McCray}, {Kamenetzky}, {Fransson}, {Barlow},
  {Gomez}, {Cigan}, {De Looze}, {Spyromilio}, {Staveley-Smith}, {Zanardo},
  {Roche}, {Larsson}, {Viti}, {van Loon}, {Wheeler}, {Baes}, {Chevalier},
  {Lundqvist}, {Marcaide}, {Dwek}, {Meixner}, {Ng}, {Sonneborn}, \&
  {Yates}}]{matsuura17}
{Matsuura}, M., {Indebetouw}, R., {Woosley}, S., {et~al.} 2017, \mnras, 469,
  3347, \dodoi{10.1093/mnras/stx830}

\bibitem[{{McCray}(1979)}]{mccray79}
{McCray}, R. 1979, in Active Galactic Nuclei, ed. C.~{Hazard} \& S.~{Mitton}
  (Cambridge University Press), 227--239

\bibitem[{{McCray}(1993)}]{mccray93}
---. 1993, \araa, 31, 175, \dodoi{10.1146/annurev.aa.31.090193.001135}

\bibitem[{{McCray} \& {Fransson}(2016)}]{mccray16}
{McCray}, R., \& {Fransson}, C. 2016, \araa, 54, 19,
  \dodoi{10.1146/annurev-astro-082615-105405}

\bibitem[{{McMullin} {et~al.}(2007){McMullin}, {Waters}, {Schiebel}, {Young},
  \& {Golap}}]{mcmullin07}
{McMullin}, J.~P., {Waters}, B., {Schiebel}, D., {Young}, W., \& {Golap}, K.
  2007, in Astronomical Society of the Pacific Conference Series, Vol. 376,
  Astronomical Data Analysis Software and Systems XVI, ed. R.~A. {Shaw},
  F.~{Hill}, \& D.~J. {Bell}, 127

\bibitem[{{Menon} \& {Heger}(2017)}]{menon17}
{Menon}, A., \& {Heger}, A. 2017, \mnras, 469, 4649,
  \dodoi{10.1093/mnras/stx818}

\bibitem[{{Michael} {et~al.}(2003){Michael}, {McCray}, {Chevalier},
  {Filippenko}, {Lundqvist}, {Challis}, {Sugerman}, {Lawrence}, {Pun},
  {Garnavich}, {Kirshner}, {Crotts}, {Fransson}, {Li}, {Panagia}, {Phillips},
  {Schmidt}, {Sonneborn}, {Suntzeff}, {Wang}, \& {Wheeler}}]{michael03}
{Michael}, E., {McCray}, R., {Chevalier}, R., {et~al.} 2003, \apj, 593, 809,
  \dodoi{10.1086/376725}

\bibitem[{{Mitchell} {et~al.}(2002){Mitchell}, {Baron}, {Branch}, {Hauschildt},
  {Nugent}, {Lundqvist}, {Blinnikov}, \& {Pun}}]{mitchell02}
{Mitchell}, R.~C., {Baron}, E., {Branch}, D., {et~al.} 2002, \apj, 574, 293,
  \dodoi{10.1086/340928}

\bibitem[{{Modigliani} {et~al.}(2007){Modigliani}, {Hummel}, {Abuter}, {Amico},
  {Ballester}, {Davies}, {Dumas}, {Horrobin}, {Neeser}, {Kissler-Patig},
  {Peron}, {Rehunanen}, {Schreiber}, \& {Szeifert}}]{modigliani07}
{Modigliani}, A., {Hummel}, W., {Abuter}, R., {et~al.} 2007, ArXiv Astrophysics
  e-prints

\bibitem[{{Mori} \& {Ho}(2007)}]{mori07}
{Mori}, K., \& {Ho}, W.~C.~G. 2007, \mnras, 377, 905,
  \dodoi{10.1111/j.1365-2966.2007.11663.x}

\bibitem[{{Morris} \& {Podsiadlowski}(2007)}]{morris07}
{Morris}, T., \& {Podsiadlowski}, P. 2007, Science, 315, 1103,
  \dodoi{10.1126/science.1136351}

\bibitem[{{Morris} \& {Podsiadlowski}(2009)}]{morris09}
---. 2009, \mnras, 399, 515, \dodoi{10.1111/j.1365-2966.2009.15114.x}

\bibitem[{{Morrison} \& {McCammon}(1983)}]{morrison83}
{Morrison}, R., \& {McCammon}, D. 1983, \apj, 270, 119, \dodoi{10.1086/161102}

\bibitem[{{Ng} {et~al.}(2009){Ng}, {Gaensler}, {Murray}, {Slane}, {Park},
  {Staveley-Smith}, {Manchester}, \& {Burrows}}]{ng09}
{Ng}, C.-Y., {Gaensler}, B.~M., {Murray}, S.~S., {et~al.} 2009, \apjl, 706,
  L100, \dodoi{10.1088/0004-637X/706/1/L100}

\bibitem[{{Ng} {et~al.}(2008){Ng}, {Gaensler}, {Staveley-Smith}, {Manchester},
  {Kesteven}, {Ball}, \& {Tzioumis}}]{ng08}
{Ng}, C.-Y., {Gaensler}, B.~M., {Staveley-Smith}, L., {et~al.} 2008, \apj, 684,
  481, \dodoi{10.1086/590330}

\bibitem[{{Ng} {et~al.}(2011){Ng}, {Potter}, {Staveley-Smith}, {Tingay},
  {Gaensler}, {Phillips}, {Tzioumis}, \& {Zanardo}}]{ng11}
{Ng}, C.-Y., {Potter}, T.~M., {Staveley-Smith}, L., {et~al.} 2011, \apjl, 728,
  L15, \dodoi{10.1088/2041-8205/728/1/L15}

\bibitem[{{Ng} {et~al.}(2013){Ng}, {Zanardo}, {Potter}, {Staveley-Smith},
  {Gaensler}, {Manchester}, \& {Tzioumis}}]{ng13}
{Ng}, C.-Y., {Zanardo}, G., {Potter}, T.~M., {et~al.} 2013, \apj, 777, 131,
  \dodoi{10.1088/0004-637X/777/2/131}

\bibitem[{{O'Brien}(2015)}]{obrien15}
{O'Brien}, K. 2015, FORS2 User Manual, issue 96.0, VLT-MAN-ESO-13100-1543
  (European Southern Observatory)

\bibitem[{{O'Dell} {et~al.}(2013){O'Dell}, {Swartz}, {Tice}, {Plucinsky},
  {Grant}, {Marshall}, {Vikhlinin}, \& {Tennant}}]{odell13}
{O'Dell}, S.~L., {Swartz}, D.~A., {Tice}, N.~W., {et~al.} 2013, in \procspie,
  Vol. 8859, UV, X-Ray, and Gamma-Ray Space Instrumentation for Astronomy
  XVIII, 88590F

\bibitem[{{Orlando} {et~al.}(2015){Orlando}, {Miceli}, {Pumo}, \&
  {Bocchino}}]{orlando15}
{Orlando}, S., {Miceli}, M., {Pumo}, M.~L., \& {Bocchino}, F. 2015, \apj, 810,
  168, \dodoi{10.1088/0004-637X/810/2/168}

\bibitem[{{{\"O}zel} \& {Freire}(2016)}]{ozel16}
{{\"O}zel}, F., \& {Freire}, P. 2016, \araa, 54, 401,
  \dodoi{10.1146/annurev-astro-081915-023322}

\bibitem[{{Page} {et~al.}(2009){Page}, {Lattimer}, {Prakash}, \&
  {Steiner}}]{page09}
{Page}, D., {Lattimer}, J.~M., {Prakash}, M., \& {Steiner}, A.~W. 2009, \apj,
  707, 1131, \dodoi{10.1088/0004-637X/707/2/1131}

\bibitem[{{Panagia}(1999)}]{panagia99}
{Panagia}, N. 1999, in IAU Symposium, Vol. 190, New Views of the Magellanic
  Clouds, ed. Y.-H. {Chu}, N.~{Suntzeff}, J.~{Hesser}, \& D.~{Bohlender}, 549

\bibitem[{{Panagia} {et~al.}(1991){Panagia}, {Gilmozzi}, {Macchetto}, {Adorf},
  \& {Kirshner}}]{panagia91}
{Panagia}, N., {Gilmozzi}, R., {Macchetto}, F., {Adorf}, H.-M., \& {Kirshner},
  R.~P. 1991, \apjl, 380, L23, \dodoi{10.1086/186164}

\bibitem[{{Park} {et~al.}(2002){Park}, {Burrows}, {Garmire}, {Nousek},
  {McCray}, {Michael}, \& {Zhekov}}]{park02}
{Park}, S., {Burrows}, D.~N., {Garmire}, G.~P., {et~al.} 2002, \apj, 567, 314,
  \dodoi{10.1086/338492}

\bibitem[{{Park} {et~al.}(2010){Park}, {Hughes}, {Slane}, {Mori}, \&
  {Burrows}}]{park10}
{Park}, S., {Hughes}, J.~P., {Slane}, P.~O., {Mori}, K., \& {Burrows}, D.~N.
  2010, \apj, 710, 948, \dodoi{10.1088/0004-637X/710/2/948}

\bibitem[{{Park} {et~al.}(2004){Park}, {Zhekov}, {Burrows}, {Garmire}, \&
  {McCray}}]{park04}
{Park}, S., {Zhekov}, S.~A., {Burrows}, D.~N., {Garmire}, G.~P., \& {McCray},
  R. 2004, \apj, 610, 275, \dodoi{10.1086/421701}

\bibitem[{{Park} {et~al.}(2011){Park}, {Zhekov}, {Burrows}, {Racusin}, {Dewey},
  \& {McCray}}]{park11}
{Park}, S., {Zhekov}, S.~A., {Burrows}, D.~N., {et~al.} 2011, \apjl, 733, L35,
  \dodoi{10.1088/2041-8205/733/2/L35}

\bibitem[{{Patat} \& {Romaniello}(2006)}]{patat06}
{Patat}, F., \& {Romaniello}, M. 2006, \pasp, 118, 146, \dodoi{10.1086/497581}

\bibitem[{{Perego} {et~al.}(2015){Perego}, {Hempel}, {Fr{\"o}hlich}, {Ebinger},
  {Eichler}, {Casanova}, {Liebend{\"o}rfer}, \& {Thielemann}}]{perego15}
{Perego}, A., {Hempel}, M., {Fr{\"o}hlich}, C., {et~al.} 2015, \apj, 806, 275,
  \dodoi{10.1088/0004-637X/806/2/275}

\bibitem[{{Phinney}(1989)}]{phinney89}
{Phinney}, E.~S. 1989, in IAU Symposium, Vol. 136, The Center of the Galaxy,
  ed. M.~{Morris}, 543

\bibitem[{{Posselt} {et~al.}(2013){Posselt}, {Pavlov}, {Suleimanov}, \&
  {Kargaltsev}}]{posselt13}
{Posselt}, B., {Pavlov}, G.~G., {Suleimanov}, V., \& {Kargaltsev}, O. 2013,
  \apj, 779, 186, \dodoi{10.1088/0004-637X/779/2/186}

\bibitem[{{Potter} {et~al.}(2009){Potter}, {Staveley-Smith}, {Ng}, {Ball},
  {Gaensler}, {Kesteven}, {Manchester}, {Tzioumis}, \& {Zanardo}}]{potter09}
{Potter}, T.~M., {Staveley-Smith}, L., {Ng}, C.-Y., {et~al.} 2009, \apj, 705,
  261, \dodoi{10.1088/0004-637X/705/1/261}

\bibitem[{{Primini} {et~al.}(2011){Primini}, {Houck}, {Davis}, {Nowak},
  {Evans}, {Glotfelty}, {Anderson}, {Bonaventura}, {Chen}, {Doe}, {Evans},
  {Fabbiano}, {Galle}, {Gibbs}, {Grier}, {Hain}, {Hall}, {Harbo}, {(Helen He},
  {Karovska}, {Kashyap}, {Lauer}, {McCollough}, {McDowell}, {Miller},
  {Mitschang}, {Morgan}, {Mossman}, {Nichols}, {Plummer}, {Refsdal}, {Rots},
  {Siemiginowska}, {Sundheim}, {Tibbetts}, {Van Stone}, {Winkelman}, \&
  {Zografou}}]{primini11}
{Primini}, F.~A., {Houck}, J.~C., {Davis}, J.~E., {et~al.} 2011, \apjs, 194,
  37, \dodoi{10.1088/0067-0049/194/2/37}

\bibitem[{{Rees}(1988)}]{rees88}
{Rees}, M.~J. 1988, \nat, 333, 523, \dodoi{10.1038/333523a0}

\bibitem[{{Reynolds} {et~al.}(1995){Reynolds}, {Jauncey}, {Staveley-Smith},
  {Tzioumis}, {de Vegt}, {Zacharias}, {Perryman}, {van Leeuwen}, {King},
  {McCulloch}, {Russell}, {Johnston}, {Hindsley}, {Malin}, {Argue},
  {Manchester}, {Kesteven}, {White}, \& {Jones}}]{reynolds95}
{Reynolds}, J.~E., {Jauncey}, D.~L., {Staveley-Smith}, L., {et~al.} 1995, \aap,
  304, 116

\bibitem[{{Rousset} {et~al.}(2003){Rousset}, {Lacombe}, {Puget}, {Hubin},
  {Gendron}, {Fusco}, {Arsenault}, {Charton}, {Feautrier}, {Gigan}, {Kern},
  {Lagrange}, {Madec}, {Mouillet}, {Rabaud}, {Rabou}, {Stadler}, \&
  {Zins}}]{rousset03}
{Rousset}, G., {Lacombe}, F., {Puget}, P., {et~al.} 2003, in \procspie, Vol.
  4839, Adaptive Optical System Technologies II, ed. P.~L. {Wizinowich} \&
  D.~{Bonaccini}, 140--149

\bibitem[{{Ruiz-Lapuente} \& {Spruit}(1998)}]{ruiz_lapuente98}
{Ruiz-Lapuente}, P., \& {Spruit}, H.~C. 1998, \apj, 500, 360,
  \dodoi{10.1086/305697}

\bibitem[{{Salaris} \& {Cassisi}(2005)}]{salaris05}
{Salaris}, M., \& {Cassisi}, S. 2005, {Evolution of Stars and Stellar
  Populations} (Wiley-VCH), 400

\bibitem[{{Sanduleak}(1970)}]{sanduleak70}
{Sanduleak}, N. 1970, Contributions from the Cerro Tololo Inter-American
  Observatory, 89

\bibitem[{{Schreiber} {et~al.}(2004){Schreiber}, {Thatte}, {Eisenhauer},
  {Tecza}, {Abuter}, \& {Horrobin}}]{schreiber04}
{Schreiber}, J., {Thatte}, N., {Eisenhauer}, F., {et~al.} 2004, in Astronomical
  Society of the Pacific Conference Series, Vol. 314, Astronomical Data
  Analysis Software and Systems (ADASS) XIII, ed. F.~{Ochsenbein}, M.~G.
  {Allen}, \& D.~{Egret}, 380

\bibitem[{{Scuderi} {et~al.}(1996){Scuderi}, {Panagia}, {Gilmozzi}, {Challis},
  \& {Kirshner}}]{scuderi96}
{Scuderi}, S., {Panagia}, N., {Gilmozzi}, R., {Challis}, P.~M., \& {Kirshner},
  R.~P. 1996, \apj, 465, 956, \dodoi{10.1086/177480}

\bibitem[{{Serafimovich} {et~al.}(2004){Serafimovich}, {Shibanov}, {Lundqvist},
  \& {Sollerman}}]{serafimovich04}
{Serafimovich}, N.~I., {Shibanov}, Y.~A., {Lundqvist}, P., \& {Sollerman}, J.
  2004, \aap, 425, 1041, \dodoi{10.1051/0004-6361:20040499}

\bibitem[{{Shapiro}(1973)}]{shapiro73}
{Shapiro}, S.~L. 1973, \apj, 180, 531, \dodoi{10.1086/151982}

\bibitem[{{Shapiro} \& {Teukolsky}(1983)}]{shapiro83}
{Shapiro}, S.~L., \& {Teukolsky}, S.~A. 1983, {Black holes, white dwarfs, and
  neutron stars: The physics of compact objects} (Wiley-Interscience)

\bibitem[{{Shternin} \& {Yakovlev}(2008)}]{shternin08}
{Shternin}, P.~S., \& {Yakovlev}, D.~G. 2008, Astronomy Letters, 34, 675,
  \dodoi{10.1134/S1063773708100034}

\bibitem[{{Shtykovskiy} {et~al.}(2005){Shtykovskiy}, {Lutovinov}, {Gilfanov},
  \& {Sunyaev}}]{shtykovskiy05}
{Shtykovskiy}, P.~E., {Lutovinov}, A.~A., {Gilfanov}, M.~R., \& {Sunyaev},
  R.~A. 2005, Astronomy Letters, 31, 258, \dodoi{10.1134/1.1896069}

\bibitem[{{Skrutskie} {et~al.}(2006){Skrutskie}, {Cutri}, {Stiening},
  {Weinberg}, {Schneider}, {Carpenter}, {Beichman}, {Capps}, {Chester},
  {Elias}, {Huchra}, {Liebert}, {Lonsdale}, {Monet}, {Price}, {Seitzer},
  {Jarrett}, {Kirkpatrick}, {Gizis}, {Howard}, {Evans}, {Fowler}, {Fullmer},
  {Hurt}, {Light}, {Kopan}, {Marsh}, {McCallon}, {Tam}, {Van Dyk}, \&
  {Wheelock}}]{skrutskie06}
{Skrutskie}, M.~F., {Cutri}, R.~M., {Stiening}, R., {et~al.} 2006, \aj, 131,
  1163, \dodoi{10.1086/498708}

\bibitem[{{Smartt} {et~al.}(2009){Smartt}, {Eldridge}, {Crockett}, \&
  {Maund}}]{smartt09b}
{Smartt}, S.~J., {Eldridge}, J.~J., {Crockett}, R.~M., \& {Maund}, J.~R. 2009,
  \mnras, 395, 1409, \dodoi{10.1111/j.1365-2966.2009.14506.x}

\bibitem[{{Spitkovsky}(2006)}]{spitkovsky06}
{Spitkovsky}, A. 2006, \apjl, 648, L51, \dodoi{10.1086/507518}

\bibitem[{{Steiner} {et~al.}(2013){Steiner}, {Lattimer}, \&
  {Brown}}]{steiner13}
{Steiner}, A.~W., {Lattimer}, J.~M., \& {Brown}, E.~F. 2013, \apjl, 765, L5,
  \dodoi{10.1088/2041-8205/765/1/L5}

\bibitem[{{Stetson}(1987)}]{stetson87}
{Stetson}, P.~B. 1987, \pasp, 99, 191, \dodoi{10.1086/131977}

\bibitem[{{Sukhbold} {et~al.}(2016){Sukhbold}, {Ertl}, {Woosley}, {Brown}, \&
  {Janka}}]{sukhbold16}
{Sukhbold}, T., {Ertl}, T., {Woosley}, S.~E., {Brown}, J.~M., \& {Janka}, H.-T.
  2016, \apj, 821, 38, \dodoi{10.3847/0004-637X/821/1/38}

\bibitem[{{Suntzeff} {et~al.}(1992){Suntzeff}, {Phillips}, {Elias}, {Walker},
  \& {Depoy}}]{suntzeff92}
{Suntzeff}, N.~B., {Phillips}, M.~M., {Elias}, J.~H., {Walker}, A.~R., \&
  {Depoy}, D.~L. 1992, \apjl, 384, L33, \dodoi{10.1086/186256}

\bibitem[{{Szary} {et~al.}(2014){Szary}, {Zhang}, {Melikidze}, {Gil}, \&
  {Xu}}]{szary14}
{Szary}, A., {Zhang}, B., {Melikidze}, G.~I., {Gil}, J., \& {Xu}, R.-X. 2014,
  \apj, 784, 59, \dodoi{10.1088/0004-637X/784/1/59}

\bibitem[{{Taylor}(2013)}]{taylor13}
{Taylor}, A.~R. 2013, in IAU Symposium, Vol. 291, Neutron Stars and Pulsars:
  Challenges and Opportunities after 80 years, ed. J.~{van Leeuwen}, 337--341

\bibitem[{{Tziamtzis} {et~al.}(2011){Tziamtzis}, {Lundqvist},
  {Gr{\"o}ningsson}, \& {Nasoudi-Shoar}}]{tziamtzis11}
{Tziamtzis}, A., {Lundqvist}, P., {Gr{\"o}ningsson}, P., \& {Nasoudi-Shoar}, S.
  2011, \aap, 527, A35, \dodoi{10.1051/0004-6361/201015576}

\bibitem[{{Utrobin} {et~al.}(2015){Utrobin}, {Wongwathanarat}, {Janka}, \&
  {M{\"u}ller}}]{utrobin15}
{Utrobin}, V.~P., {Wongwathanarat}, A., {Janka}, H.-T., \& {M{\"u}ller}, E.
  2015, \aap, 581, A40, \dodoi{10.1051/0004-6361/201425513}

\bibitem[{{van der Marel} \& {Kallivayalil}(2014)}]{van_der_marel14}
{van der Marel}, R.~P., \& {Kallivayalil}, N. 2014, \apj, 781, 121,
  \dodoi{10.1088/0004-637X/781/2/121}

\bibitem[{{van der Marel} \& {Sahlmann}(2016)}]{van_der_marel16}
{van der Marel}, R.~P., \& {Sahlmann}, J. 2016, \apjl, 832, L23,
  \dodoi{10.3847/2041-8205/832/2/L23}

\bibitem[{van~der Walt {et~al.}(2011)van~der Walt, Colbert, \&
  Varoquaux}]{van_der_walt11}
van~der Walt, S., Colbert, S.~C., \& Varoquaux, G. 2011, Computing in Science
  Engineering, 13, 22, \dodoi{10.1109/MCSE.2011.37}

\bibitem[{{Walborn} {et~al.}(1987){Walborn}, {Lasker}, {Laidler}, \&
  {Chu}}]{walborn87}
{Walborn}, N.~R., {Lasker}, B.~M., {Laidler}, V.~G., \& {Chu}, Y.-H. 1987,
  \apjl, 321, L41, \dodoi{10.1086/185002}

\bibitem[{{Walborn} {et~al.}(1993){Walborn}, {Phillips}, {Walker}, \&
  {Elias}}]{walborn93}
{Walborn}, N.~R., {Phillips}, M.~M., {Walker}, A.~R., \& {Elias}, J.~H. 1993,
  \pasp, 105, 1240, \dodoi{10.1086/133302}

\bibitem[{{Walker} \& {Suntzeff}(1990)}]{walker90}
{Walker}, A.~R., \& {Suntzeff}, N.~B. 1990, \pasp, 102, 131,
  \dodoi{10.1086/132618}

\bibitem[{{Wang} {et~al.}(1996){Wang}, {Wheeler}, {Kirshner}, {Challis},
  {Filippenko}, {Fransson}, {Panagia}, {Phillips}, \& {Suntzeff}}]{wang96}
{Wang}, L., {Wheeler}, J.~C., {Kirshner}, R.~P., {et~al.} 1996, \apj, 466, 998,
  \dodoi{10.1086/177570}

\bibitem[{{Wesson} {et~al.}(2015){Wesson}, {Barlow}, {Matsuura}, \&
  {Ercolano}}]{wesson15}
{Wesson}, R., {Barlow}, M.~J., {Matsuura}, M., \& {Ercolano}, B. 2015, \mnras,
  446, 2089, \dodoi{10.1093/mnras/stu2250}

\bibitem[{{West} {et~al.}(1987){West}, {Lauberts}, {Schuster}, \&
  {Jorgensen}}]{west87}
{West}, R.~M., {Lauberts}, A., {Schuster}, H.-E., \& {Jorgensen}, H.~E. 1987,
  \aap, 177, L1

\bibitem[{{White} \& {Malin}(1987)}]{white87}
{White}, G.~L., \& {Malin}, D.~F. 1987, \nat, 327, 36, \dodoi{10.1038/327036a0}

\bibitem[{{White}(1994)}]{white94}
{White}, R.~L. 1994, in The Restoration of HST Images and Spectra - II, ed.
  R.~J. {Hanisch} \& R.~L. {White}, 104

\bibitem[{{Willingale} {et~al.}(2001){Willingale}, {Aschenbach}, {Griffiths},
  {Sembay}, {Warwick}, {Becker}, {Abbey}, \& {Bonnet-Bidaud}}]{willingale01}
{Willingale}, R., {Aschenbach}, B., {Griffiths}, R.~G., {et~al.} 2001, \aap,
  365, L212, \dodoi{10.1051/0004-6361:20000114}

\bibitem[{{Willingale} {et~al.}(2013){Willingale}, {Starling}, {Beardmore},
  {Tanvir}, \& {O'Brien}}]{willingale13}
{Willingale}, R., {Starling}, R.~L.~C., {Beardmore}, A.~P., {Tanvir}, N.~R., \&
  {O'Brien}, P.~T. 2013, \mnras, 431, 394, \dodoi{10.1093/mnras/stt175}

\bibitem[{{Wilms} {et~al.}(2000){Wilms}, {Allen}, \& {McCray}}]{wilms00}
{Wilms}, J., {Allen}, A., \& {McCray}, R. 2000, \apj, 542, 914,
  \dodoi{10.1086/317016}

\bibitem[{{Wongwathanarat} {et~al.}(2013){Wongwathanarat}, {Janka}, \&
  {M{\"u}ller}}]{wongwathanarat13}
{Wongwathanarat}, A., {Janka}, H.-T., \& {M{\"u}ller}, E. 2013, \aap, 552,
  A126, \dodoi{10.1051/0004-6361/201220636}

\bibitem[{{Wongwathanarat} {et~al.}(2015){Wongwathanarat}, {M{\"u}ller}, \&
  {Janka}}]{wongwathanarat15}
{Wongwathanarat}, A., {M{\"u}ller}, E., \& {Janka}, H.-T. 2015, \aap, 577, A48,
  \dodoi{10.1051/0004-6361/201425025}

\bibitem[{{Wooden} {et~al.}(1993){Wooden}, {Rank}, {Bregman}, {Witteborn},
  {Tielens}, {Cohen}, {Pinto}, \& {Axelrod}}]{wooden93}
{Wooden}, D.~H., {Rank}, D.~M., {Bregman}, J.~D., {et~al.} 1993, \apjs, 88,
  477, \dodoi{10.1086/191830}

\bibitem[{{Woosley} {et~al.}(1988){Woosley}, {Pinto}, \& {Ensman}}]{woosley88}
{Woosley}, S.~E., {Pinto}, P.~A., \& {Ensman}, L. 1988, \apj, 324, 466,
  \dodoi{10.1086/165908}

\bibitem[{{Yakovlev} {et~al.}(2011){Yakovlev}, {Ho}, {Shternin}, {Heinke}, \&
  {Potekhin}}]{yakovlev11}
{Yakovlev}, D.~G., {Ho}, W.~C.~G., {Shternin}, P.~S., {Heinke}, C.~O., \&
  {Potekhin}, A.~Y. 2011, \mnras, 411, 1977,
  \dodoi{10.1111/j.1365-2966.2010.17827.x}

\bibitem[{{Yakovlev} \& {Pethick}(2004)}]{yakovlev04}
{Yakovlev}, D.~G., \& {Pethick}, C.~J. 2004, \araa, 42, 169,
  \dodoi{10.1146/annurev.astro.42.053102.134013}

\bibitem[{{Zanardo} {et~al.}(2013){Zanardo}, {Staveley-Smith}, {Ng},
  {Gaensler}, {Potter}, {Manchester}, \& {Tzioumis}}]{zanardo13}
{Zanardo}, G., {Staveley-Smith}, L., {Ng}, C.-Y., {et~al.} 2013, \apj, 767, 98,
  \dodoi{10.1088/0004-637X/767/2/98}

\bibitem[{{Zanardo} {et~al.}(2014){Zanardo}, {Staveley-Smith}, {Indebetouw},
  {Chevalier}, {Matsuura}, {Gaensler}, {Barlow}, {Fransson}, {Manchester},
  {Baes}, {Kamenetzky}, {Laki{\'c}evi{\'c}}, {Lundqvist}, {Marcaide},
  {Mart{\'{\i}}-Vidal}, {Meixner}, {Ng}, {Park}, {Sonneborn}, {Spyromilio}, \&
  {van Loon}}]{zanardo14}
{Zanardo}, G., {Staveley-Smith}, L., {Indebetouw}, R., {et~al.} 2014, \apj,
  796, 82, \dodoi{10.1088/0004-637X/796/2/82}

\bibitem[{{Zhang} {et~al.}(2018){Zhang}, {Dai}, {Hobbs}, {Staveley-Smith},
  {Manchester}, {Russell}, {Zanardo}, \& {Wu}}]{zhang18}
{Zhang}, S.-B., {Dai}, S., {Hobbs}, G., {et~al.} 2018, \mnras, 479, 1836,
  \dodoi{10.1093/mnras/sty1573}

\end{thebibliography}
\end{document}